\documentclass[printer]{aa}  
\usepackage{natbib}
\bibpunct{(}{)}{;}{a}{}{,} 

\usepackage{graphicx}
\usepackage{txfonts}
\usepackage{amstext}

\begin{document}

   \title{Comparative testing of dark matter models\\ with 15 HSB and 15 LSB galaxies}

   \author{E. Kun\inst{1}, Z. Keresztes\inst{1}, A. Simk\'o\inst{2}, G. Sz\H{u}cs\inst{3}, L. \'{A}. Gergely\inst{1}}

   \institute{Physics Institute, University of Szeged, D\'om t\'er 9, H-6720 Szeged, Hungary\\
              \email{kun@titan.physx.u-szeged.hu}
                   \and
                   Department of Computing Science, Umeå University, SE-901 87 Umeå, Sweden
                   \and
             Bolyai Institute, University of Szeged, Aradi v\'ertan\'uk tere 1., H-6720 Szeged, Hungary
               }
\titlerunning{Testing dark matter galaxy models}

   \date{2017}

 
  \abstract
   {We assemble a database of $15$ high surface brightness (HSB) and $15$ low surface brightness (LSB) galaxies, for which surface brightness density and spectroscopic rotation curve data are both available and representative for various morphologies. We use this dataset to test the Navarro-Frenk-White, the Einasto, and the pseudo-isothermal sphere dark matter models.}
  {We investigate the compatibility of the pure baryonic model and baryonic plus one of the three dark matter models with observations on the assembled galaxy database. When a dark matter component improves the fit with the spectroscopic rotational curve, we rank the models according to the goodness of fit to the datasets.}
{We constructed the spatial luminosity density of the baryonic component based on the surface brightness profile of the
galaxies. We estimated the mass-to-light ($M/L$) ratio of the stellar component through a previously proposed color--mass-to-light ratio relation (CMLR), which yields stellar masses independent of the photometric band. We assumed an axissymetric baryonic mass model with variable axis ratios together with one of the three dark matter models to provide the theoretical rotational velocity curves, and we compared them with the dataset. In a second attempt, we addressed the question whether the dark component could be replaced by a pure baryonic model with fitted $M/L$ ratios, varied over ranges consistent with CMLR relations derived from the available stellar population models. We employed the Akaike information criterion to establish the performance of the best-fit models.}
{For 7 galaxies (2 HSB, and 5 LSB), neither model fits the dataset within the $1\sigma$ confidence level. For the other $23$ cases, one of the models with dark matter explains the rotation curve data best. According to the Akaike information criterion, the pseudo-isothermal sphere emerges as most favored in $14$ cases, followed by the Navarro-Frenk-White ($6$ cases) and the Einasto ($3$ cases) dark matter models. We find that the pure baryonic model with fitted $M/L$ ratios falls within the $1\sigma$ confidence level for $10$ HSB and $2$ LSB galaxies, at the price of growing the $M/L$s on average by a factor of two, but the fits are inferior compared to the best-fitting dark matter model.}
   {}


   \maketitle
%

\section{Introduction}

Recent cosmological data suggest that $26.8$ percent of the energy content of the Universe is in the form of dark matter \citep{Planck2015}, but no compelling understanding of its nature has been possible
so far. Galaxy-structure studies provide an efficient test of various dark matter candidates, as dark matter plays a key role on this scale. The understanding of the kinematics of galaxy clusters \citep[][]{Zwicky1937} and galactic rotation curves \citep[e.g.,][]{Rubin1978,Rubin1985,Mathewson1992,Prugniel1998} all require dark matter. When we compare different dark matter models with galactic rotation curves, it is crucuial to estimate the mass of the baryonic (luminous) component accurately.

The baryonic mass density of the galaxy can be calculated from the luminosity distribution, assuming a certain mass-to-light ($M/L$) ratio. There are five basic techniques employed to estimate this ratio: a) using tabulated relations between color and $M/L$ \citep[e.g.,][]{Bell2001}, b) modeling broadband photometry \citep[e.g.,][]{Sawicki1998}, c) modeling moderate-resolution spectra \citep[e.g.,][]{Giallongo1998}, d) the analysis of CMDs in nearby galaxies with resolved stellar populations \citep[e.g.][]{Dalcanton2012}, and e) dynamical modeling via the Jeans equation in early-type galaxies \citep[e.g.,][]{Cappellari2013} or in
intermediate- and late-type disks \citep[e.g.,][]{Bershady2010,Martinsson2013}. In this paper we employ the first method.

The accuracy of the color--mass-to-light relations \citep[CMLR, e.g.,][]{Bruzual2003,McGaugh2014} highly depends on the assumed initial mass function (IMF) of the employed stellar population synthesis model, the variations in the star formation histories of galaxies, the distribution of stellar metallicities, and the contribution of stars in bright but short-lived phases of evolution (e.g., TP-AGB stars). According to \citet{McGaugh2014}, the semi-empirical stellar population synthesis model of \citet{Bell2003} provides the most identical stellar masses in different photometric bands. They modified the according CMLR relation to achieve the best match of the calculated stellar masses in different photometric bands.

When the Newtonian law of gravity is employed to deduce the rotational curve of the luminous component, the model curve is qualitatively different from the curve emerging from spectroscopic measurements (Doppler shift of the spectral lines). Either gravity is not well understood and needs refinement on galactic scales, or there is an invisible contribution to the mass of the galaxy, which interacts only gravitationally. In this paper we compare the compatibility with galactic rotation curves of three frequently used dark matter models and of the pure baryonic model.

The Navarro-Frenk-White (hereafter NFW) dark matter model is motivated by cold dark matter simulations. \citet{NFW1997} used high-resolution N-body simulations to study the equilibrium density profiles of dark matter, and found that their halos have the same shape regardless of the halo mass, initial density fluctuations, and cosmological parameters. The NFW model has a divergent central density, and it is cuspy, the density scaling as $r^{-1}$ ($r$ being the radial distance).

\citet{Einasto1965} proposed the density $\rho \sim \exp (-A r^{\alpha})$ for a spherical stellar system, able to model both steep and shallow profiles. The Einasto model is formally similar to Sersic's law, but it is fit to the space density as compared to the projected surface density for the latter. \citet{Merritt2006} pointed out that Sersic's law is also an excellent description for $N$-body dark matter halos (see references therein).

The pseudo-isothermal sphere (hereafter PSE) halo has no cosmological motivation, but it often fits the rotational curves better than NFW \citep[][]{deBlok2002,Kuzio2008}, as the PSE profile exhibits finite density at the center of the halo \citep[e.g.,][]{Chemin2011}.

For each density profile, the rotational velocity of the halo can be fit to the spectroscopically measured curves. We did this for either the pure baryonic model or for the three frequently used dark matter profiles (NFW, Einasto, and PSE). We have chosen $15$ high surface brightness (HSB) and $15$ low surface brightness
(LSB) galaxies for this purpose.

In Section \ref{galaxy_rotation_curves} we generate the spatial luminosity density of the baryonic component from the projected surface brightness profile of the galaxies. We also summarize the contributions of the baryonic and the dark matter components to the rotational velocity. In Section \ref{section_conf} we present the model fit results regarding the spatial luminosity density of the baryonic components. In Section \ref{bestfitrot} we compare the rotational velocity models with the spectroscopic rotational curves. In Section \ref{stat_ranking} we investigate the relevance of the dark matter models. In Section \ref{summary} we summarize the results.

The $\Lambda$CDM cosmological model is adopted throughout the paper, with the Hubble constant $H_0 = 67.8 km s^{-1} Mpc^{-1}$ and (baryonic+dark) matter density $\Omega_m = 0.308$ \citep{Planck2015}.

\section{Galactic rotation curves}
\label{galaxy_rotation_curves}

\subsection{Contribution of baryonic matter}
The baryonic rotational curves are derived based on the distribution of the luminous matter, which is deduced from the surface brightness of the galaxy.

The surface brightness $S$ is the radiative flux $F$ per solid angle $\Delta \Omega$ of the image such that $S \approx F /\Delta \Omega$, which is a function of the redshift, and it is independent of the distance $D$ of the emitting surface in a Friedmann universe. The observed surface brightness $S_\mathrm{obs}$ in units of $L_\odot/kpc^2$ can alternatively be expressed as the quantity $\mu$ in units of $mag/arcsec^2$:
\begin{equation}
S_\mathrm{obs}(R)=4.255\times 10^{14}\times 10^{(0.4(\mathcal{M}_{\odot}-\mu(R)))},
\label{eq:mutrafo}
\end{equation}
where $R$ is the distance measured from the center of the galaxy in the galaxy plane, and $\mathcal{M}_{\odot}$ is the absolute brightness of the Sun in units of $mag$. We translated $\mu(R)$ into $ S_\mathrm{obs}(R)$ using Eq. (\ref{eq:mutrafo}), which is valid in the local Universe ($z\ll1$), having an $(1+z)^3$ factor suppressed.

We followed \cite{Tempel2006} to derive the surface brightness, assuming that the spatial luminosity density distribution of each visible component is given by
\begin{equation}
l(a)=l(0)\exp\left[ -\left( \frac{a}{ka_0}\right)^{{1/N}} \right].
\end{equation}
Here $l(0)=hL/(4\pi q a_0^3)$ is the central density, where $a_0$ characterizes the harmonic mean radius of the respecting component, and $k$ and $h$ are scaling parameters. Furthermore, $a=\sqrt{R^2+z^2/q^2}$, where $q$ is the axis ratio, and $R$ and $z$ are cylindrical coordinates.
From the measurements the projection of $l(a)$ onto the plane of the sky perpendicular to the line of sight is derived:
\begin{equation}
S(R)=2 \sum_i^n q_i \int_R^\infty \frac{l_i(a) a}{\sqrt{a^2-R^2}}da.
\label{eq:sr}
\end{equation}
Here $S(R)$ arises as a sum for $n$ visible components, and we assumed constant axis ratios $q_i$. Equation (\ref{eq:sr}) was fit to the surface brightness of the galaxies $\mu(R)$, in order to reveal the spatial luminosity density $l(a)$. 

We decomposed the baryonic model into two components, a bulge and a disk. Therefore the mass density is 
\begin{equation}
\rho(a)=\sigma l_b(a)+\tau l_d(a),
\end{equation}
where $l_b(a)$ and $l_d(a)$ are the spatial luminosity density of the bulge and of the disk, respectively, and $\sigma$ and $\tau$ are the corresponding mass-to-light ($M/L$) ratios (both are given in solar units).

It follows from the Poisson equation that for spheroidal shape matter, the rotational velocity in the galactic plane induced by the $i$th baryonic component is given by \citep{Tamm2005}
\begin{equation}
V_i^2(R)=4 \pi q_i G \int_0^R \frac{\rho_i(a) a^2}{(R^2-e_i^2 a^2)^{1/2}} da,
\end{equation}
where $G$ is the gravitational constant, $e_i=(1-q_i^2)^{1/2}$ is the eccentricity of the $i$th component, and $\rho_i(a)$ is its mass density. Then a summation of $V_i^2(R)$ over all visible components gives the square of the rotational velocity of the baryonic model.

\subsection{Contribution of the dark matter}

For a spherically symmetric dark matter halo, the rotational velocity square is
\begin{equation}
V^2_\mathrm{DM}(r)=\frac{G M_\mathrm{DM}(r)}{r},
\end{equation}
with the spherical radial coordinate $r$, and cumulative mass within a sphere of $r$ radius
\begin{equation}
M_\mathrm{DM}(r)=4 \pi \int_0^{r} \rho_\mathrm{DM}(r') r^{'2} dr'.
\end{equation}

The NFW dark matter density profile is \citep{NFW1997}
\begin{equation}
\rho_\mathrm{NFW}(r)=\frac{\rho_s}{\left( \frac{r}{r_s}\right)  \left( 1 + \frac{r}{r_s} \right)^2},
\end{equation}
where $\rho_s$ and $r_s$ are the characteristic density and scale distance. The contribution of this dark matter to the rotational velocity squared at radial distance $r$ is
\begin{equation}
V^2_\mathrm{NFW}(r)=4\pi G \rho\frac{r_s^3}{r} \left[ \ln \left(1+\frac{r}{r_s}\right)-\frac{r}{r_s} \left(\frac{1}{1+\frac{r}{r_s}}\right) \right].
\end{equation}

The Einasto dark matter profile is described by \citep[e.g.,][]{Merritt2006}
\begin{equation}
\rho_\mathrm{E}(r)=\rho_e \exp \left\lbrace - d_n \left[ \left(\frac{r}{r_e}\right)^{1/n} -1 \right] \right\rbrace,
\end{equation}
where $n$ is a positive parameter. The term $d_n$ is a function of $n$ with the property that $\rho_e$ is the density at $r_e$ defining a half-mass radius. An empirical relation between $d_n$ and $n$ is \citep[][]{Merritt2006}
\begin{equation}
d_n\approx 3n-\frac{1}{3}+\frac{0.0079}{n}.
\end{equation}
The total dark matter mass is
\begin{equation}
M_\mathrm{E,tot}=4 \pi \rho_\mathrm{E,0} h^3 n \Gamma(3n),
\end{equation}
with the central density $\rho_\mathrm{E,0}=\rho_e e^\mathrm{d_n}$, complete Gamma function $\Gamma(3n)$, and $h=r_e/d_n^n$ \citep{Retana2012}.
This dark matter contributes to the rotational velocity squared as
\begin{equation}
V^2_\mathrm{E}(r)=\frac{G M_{\mathrm{E,tot}}}{r} \left[ 1-\frac{\Gamma(3n,s^{1/n})}{\Gamma(3n)} \right],
\end{equation}
with the incomplete Gamma function $\Gamma(3n,s^{1/n})$, and $s=d_n^n r/r_e$.

The PSE density profile is given by \citep[e.g.,][]{Jimenez2003}
\begin{equation}
\rho_\mathrm{P}(r)=\rho_\mathrm{P,0} \left[ 1+\left(\frac{r}{r_c}\right)^2\right]^{-1},
\end{equation}
where $\rho_\mathrm{P,0}$ is the central density, and $r_c$ scales the size of the core.
For this model the contribution to the square of the rotational velocity reads
\begin{equation}
V^2_\mathrm{P}(r)=4 \pi G \rho_\mathrm{P,0} r_c^2 \left[1- \frac{r_c}{r} \arctan \left( \frac{r}{r_c}\right)\right].
\end{equation}

Taking into account both visible and dark matter, the rotational velocity squared in the galactic plane becomes: 
\begin{equation}
V^2(R)=V^2_\mathrm{b}(R)+V^2_\mathrm{d}(R)+V_\mathrm{DM}^2 (R,z=0).
\end{equation}
This was fit to the spectroscopic rotational curves for each of the dark matter models (NFW, Einasto, and PSE).

\begin{table*}
\centering
\caption{Best-fit parameters describing the luminosity density distribution of the baryonic matter of $15$ HSB spiral and $15$ LSB galaxies. The surface brightness photometry of the galaxies
that were fit with the models are taken from $^{1}$\citet{Palunas2000}, $^2$\citet{vanderHulst1993}, $^{3}$\citet{deBlok1996}, $^{4}$\citet{deBlok1995}, $^5$\citet{Kim2007}, and $^{6}$\citet{deBlok2008}. The superscript asterisk indicates$^\text{}$  galaxies for which a bulge component fully describes the surface brightness density.}
\label{table:gx_phot}
\resizebox{\textwidth}{!}{\begin{tabular}{lcccccccccc}
\hline
\hline
ID & $z$ & $l(0)_b$ & $ka_{0,b}$ & $N_b$ & $q_b$ & $l(0)_{d}$ & $ka_{0,d}$ & $N_d$ & $q_d$\\
 & & $10^9 L_\odot /kpc^3$ & $kpc$ &  &  & $10^7 L_\odot /kpc^3$ & $kpc$ &  &\\
\hline
ESO215G39$^1$ & $0.014$ & $0.349 \pm 0.003$ & $0.404 \pm 0.013$ & $0.956 \pm 0.042$ & $0.742\pm0.040$ & $3.113 \pm 0.007$ & $6.082 \pm 0.002$ & $0.432 \pm 0.003$ & $0.098\pm0.010$\\
ESO322G76$^1$ & $0.015$ & $1.3 \pm 0.003$ & $0.720 \pm 0.011$ & $0.780 \pm 0.022$ & $0.810\pm0.018$ & $35.62 \pm 0.07$ & $3.5 \pm 0.002$ & $0.87 \pm 0.003$ & $0.100\pm0.002$\\
ESO322G77$^1$ & $0.008$ & $7.3 \pm 0.006$ & $0.140 \pm 0.001$ & $1.300 \pm 0.002$ & $0.640\pm0.006$ & $42 \pm 0.02$ & $2.800 \pm 0.002$ & $0.760 \pm 0.011$ & $0.140\pm0.003$\\
ESO322G82$^1$ & $0.015$ &  $4.3 \pm 0.001$ & $0.21 \pm 0.005$ & $1.66 \pm 0.02$ & $0.66\pm0.080$ & $12.6 \pm 0.01$ & $8.58 \pm 0.056$ & $0.659 \pm 0.006$ & $0.08\pm0.034$\\
ESO323G25$^1$ & $0.014$ &  $2.309 \pm 0.005$ & $0.458 \pm 0.005$ & $0.535 \pm 0.006$ & $0.462\pm0.008$ & $56.58 \pm 0.04$ & $2.467 \pm 0.002$ & $1.055 \pm 0.008$ & $0.154\pm0.002$\\
ESO374G02$^1$ & $0.009$ & $58.4 \pm 0.001$ & $0.07 \pm 0.004$ & $1.92 \pm 0.04$ & $0.66\pm0.04$ & $158.0 \pm 0.01$ & $1.48 \pm 0.04$ & $1.40 \pm 0.018$ & $0.08\pm0.02$\\
ESO375G12$^1$ & $0.010$ & $55.4 \pm 0.001$ & $0.08 \pm 0.003$ & $1.80 \pm 0.03$ & $0.67\pm0.05$ & $88.3 \pm 0.01$ & $2.70 \pm 0.029$ & $1.33 \pm 0.008$ & $0.08\pm0.01$\\
ESO376G02$^1$ & $0.014$ & $28.3 \pm 0.01$ & $0.095 \pm 0.002$ & $1.64 \pm 0.02$ & $0.67\pm0.41$ & $51.0 \pm 0.01$ & $3.20 \pm 0.087$ & $1.01 \pm 0.02$ & $0.08\pm0.17$\\
ESO383G02$^1$ & $0.021$ &  $56.056 \pm 0.05$ & $0.130 \pm 0.002$ & $1.20 \pm 0.001$ & $0.43\pm0.04$ & $16 \pm 0.05$ & $ 2.6\pm 0.003$ & $1.200 \pm 0.003$ & $0.200\pm0.005$\\
ESO383G88$^1$ & $0.014$ &  $3.89 \pm 0.05$ & $0.100 \pm 0.012$ & $1.61 \pm 0.089$ & $0.68\pm0.09$ & $17.2 \pm 0.00$ & $ 5.61\pm 0.03$ & $0.776 \pm 0.005$ & $0.08\pm0.04$\\
ESO445G19$^1$ & $0.016$ &  $6.048 \pm 0.004$ & $0.129 \pm 0.003$ & $1.519 \pm 0.029$ & $0.741\pm0.054 $ & $12.40 \pm 0.08$ & $5.631 \pm 0.008$ & $0.744 \pm 0.003$ & $0.148\pm0.002$\\
ESO446G01$^1$ & $0.023$ & $2.2 \pm 0.001$ & $0.740 \pm 0.007$ & $1.100 \pm 0.017$ & $0.42\pm0.020$ & $8.000 \pm 0.006$ & $4.4 \pm 0.001$ & $1.1 \pm 0.007$ & $0.19\pm0.008$\\
ESO502G02$^1$ & $0.013$ & $23.5 \pm 0.003$ & $0.09 \pm 0.001$ & $1.60 \pm 0.01$ & $0.80\pm0.043$ & $106.8 \pm 0.01$ & $2.16 \pm 0.023$ & $1.06 \pm 0.01$ & $0.08\pm0.010$\\
ESO509G80$^1$ & $0.022$ & $0.972 \pm 0.001$ & $0.666 \pm 0.022$ & $0.963 \pm 0.031$ & $0.986\pm0.023$ & $2.036 \pm 0.001$ & $11.304 \pm 0.001$ & $0.564 \pm 0.005$ & $0.265\pm0.005$\\
ESO569G17$^1$ & $0.013$ & $3.815 \pm 0.003$ & $0.385 \pm 0.023$ & $0.590 \pm 0.06$ & $0.713\pm0.063$ & $174 \pm 2$ & $1.643 \pm 0.067$ & $0.902 \pm 0.018$ & $0.146\pm0.011$\\
\hline
ID & $z$ & $l(0)_b$ & $ka_{0,b}$ & $N_b$ & $q_b$ & $l(0)_{d}$ & $ka_{0,d}$ & $N_d$ & $q_d$\\
 & & $10^7 L_\odot /kpc^3$ & $kpc$ &  &  & $10^6 L_\odot /kpc^3$ & $kpc$ &  &\\
\hline
F561-1$^2$ & $0.016$ & $2.235\pm0.002$ & $0.877\pm0.098$ & $1.045\pm0.088$ & $0.894\pm0.085$ & $1.731\pm0.001$ & $9.482\pm0.036$ & $0.138\pm0.048$ & $0.292\pm0.016$\\
F563-1$^{3 \star}$      & $0.012$ & $61.08\pm0.20$ & $0.174\pm0.015$ & $2.128\pm0.019$ & $0.855\pm0.019$ &   -               &       -        & -                &  -\\
F568-3$^4$      & $0.019$ & $1.561\pm0.007$ & $2.290\pm0.014$ & $0.649\pm0.018$ & $0.936\pm0.028$ & $6.368\pm0.05$ & $11.087\pm0.02$ & $0.251\pm0.002$ & $0.100\pm0.002$\\
F579-V1$^3$      & $0.021$ & $1.639\pm0.002$ & $1.283\pm0.026$ & $0.574\pm0.054$ & $0.888\pm0.051$ & $7.342\pm0.01$ & $6.741\pm0.004$ & $0.601\pm0.005$  & $0.262\pm0.007$\\
F583-1$^{3 \star}$      & $0.008$ & $6.059\pm0.008$ & $0.390\pm0.004$ & $1.629\pm0.007$ & $0.625\pm0.006$ &   -          &    -             &  -                & - &\\
F730-V1$^5$ & $0.036$ & $4.351\pm0.01$ & $1.120\pm0.025$ & $1.217\pm0.031$ & $0.816\pm0.018$ & $5.434\pm0.02$ & $9.404\pm0.004$ & $0.43\pm0.076$ & $0.11\pm0.014$\\
UGC128$^{4}$ & $0.015$ & $9.360\pm0.005$ & $0.356\pm0.014$ & $1.595\pm0.026$ & $0.869\pm0.034$ & $2.983\pm0.051$ & $11.362\pm0.005$ & $0.690\pm0.08$ & $0.190\pm0.026$\\
UGC1230$^2$ & $0.012$ & $1.9\pm0.05$ & $0.818\pm0.028$ & $1.002\pm0.027$ & $0.60\pm0.018$ & $9.7\pm0.004$ & $4.3\pm0.005$ & $1.0\pm0.045$ & $0.11\pm0.011$\\
UGC5750$^{4}$ & $0.014$ & $12\pm0.05$ & $0.270\pm0.004$ & $1.5\pm0.021$ & $1.0\pm0.024$ & $6.6\pm0.003$ & $8.8\pm0.006$ & $0.430\pm0.051$ & $0.13\pm0.009$\\
UGC6614$^5$ & $0.021$ & $167.8\pm0.5$ & $0.270\pm0.005$ & $1.47\pm0.016$ & $0.65\pm0.05$ & $5.7\pm0.03$ & $15.8\pm0.008$ & $0.752\pm0.069$ & $0.08\pm0.02$\\
UGC10310$^5$ & $0.002$ & $9.4\pm0.01$ & $0.44\pm0.014$ & $0.88\pm0.044$ & $0.71\pm0.01$ & $17.5\pm0.001$ & $1.4\pm0.12$ & $1.1\pm0.06$ & $0.12\pm0.01$\\
UGC11454$^5$ & $0.022$ & $127.5\pm5$ & $0.206\pm0.011$ & $1.4\pm0.062$ & $0.80\pm0.02$ & $64.4\pm0.001$ & $3.47\pm0.099$ & $1.09\pm0.031$ & $0.09\pm0.01$\\
UGC11616$^5$ & $0.017$ & $83.1\pm0.01$ & $0.545\pm0.043$ & $1.23\pm0.096$ & $0.66\pm0.04$ & $43.8\pm0.002$ & $5.5\pm0.31$ & $0.844\pm0.069$ & $0.08\pm0.02$\\
UGC11748$^5$ & $0.018$ & $114.4\pm5$ & $0.980\pm0.010$ & $0.847\pm0.014$ & $0.80\pm0.04$ & $286.1\pm0.5$ & $3.1\pm0.17$ & $1.19\pm0.046$ & $0.08\pm0.03$\\
UGC11819$^5$ & $0.014$ & $88.01\pm0.1$ & $0.197\pm0.002$ & $1.081\pm0.013$ & $0.991\pm0.013$ & $130\pm1$ & $5.789\pm0.002$ & $0.753\pm0.015$ & $0.113\pm0.001$ \\
\hline
\end{tabular}}
\end{table*}

\section{Best-fit surface brightness profile of the galaxies}
\label{section_conf}

We calculated the $S_\mathrm{obs}(R)$ of the galaxies from $\mu(R)$, given in the literature. The ESO HSB galaxies were imaged in I band \citep{Palunas2000}. Eight of the $15$ LSB galaxies$\text{}$ were detected in R band \citep[F561-1, UGC1230,][]{vanderHulst1993}, \citep[F563-1, F579-V1, F583-1,][]{deBlok1996}, and \citep[F568-3, UGC128, UGC5750,][]{deBlok1995}, and $7$ in V band \citep[F730-V1, UGC10310, UGC11454, UGC11616, UGC11748, UGC11819, and UGC6614][]{Kim2007}. The absolute brightness of the Sun ($\mathcal{M}_{\odot}$) was substituted into Eq. (\ref{eq:mutrafo}) according to the observing photometric filter, assuming the values of \citet{Binney1998} (Table 2.1) as $M_{I,\odot}=4.08^\mathbf{m}$, $M_{R,\odot}=4.42^\mathbf{m}$,  and $M_{V,\odot}=4.83^\mathbf{m}$.

Then by using Eq. (\ref{eq:sr}) and varying the parameters $q$, $l(0)$, $ka_0$, and $N$, the spatial luminosity densities were fit to the surface brightness of the galaxies by nonlinear least-squares fitting methods. We set the initial value of the axial ratio of the components as $q=0.7$ for the bulge and $q=0.1$ for the disk, which were shown to be the most frequently used values for the nearby galaxies \citep{Tamm2005}. We set a lower limit of $0.4$ for the bulge and an upper limit of $0.3$ for the disk based on SDSS results \citep{Padilla2008,Rodriguez2013}. Using the photometry of the $15$ HSB and $15$ LSB galaxies, we decomposed the surface brightness profile into bulge and disk components. The fit results are given in Table \ref{table:gx_phot} and shown in Figures \ref{fig:bright_fit_plots1} and \ref{fig:bright_fit_plots2} in the Appendix. In Table \ref{table:gx_phot} we also list the redshift $z$ of the galaxies from the NASA/IPAC Extragalactic Database, confirming that all galaxies belong to the local Universe. In the case of the LSB galaxies F563-1 and F583-1, the surface brightness distribution indicates only a bulge component. Knowing the spatial distribution of the luminous matter $l(a)$, we are able to construct the mass model, and consequently the rotational velocity curve of the baryonic component. In the next subsections we present the fitting results with the baryonic matter and three different dark matter density profiles.


\section{Best-fit galaxy rotational curves}
\label{bestfitrot}
We proceeded in a similar way for the HSB and LSB galaxies. First we fit the surface brightness distributions with the model described in Section \ref{galaxy_rotation_curves}, inferring the luminosity distribution of the baryonic component. When we fit the dark matter models, we included the baryonic component, with fixed $M/L$ ratios as described in the next subsection. We applied
the nonlinear least-squares method to perform the fit with 1/error$^2$ weights, minimizing the residual sum of squares ($\chi^2$) between the data and the model.

\subsection{Estimated mass-to-light ratio of the bulge and the disk}
\label{estimated_ml}

\begin{table*}
\centering
\caption{Coefficients of the color--mass-to-light ratios employed in this paper.}
\label{table:cmlrs}
\begin{tabular}{lccccccc}
\hline
Ref & Color index & $\alpha_V$ & $\beta_V$ & $\alpha_R$ & $\beta_R$ & $\alpha_I$ & $\beta_I$\\
\hline
\cite{Bell2003}& $B-V$ & -0.628 & 1.305 & -0.520 & 1.094 & -0.399 & 0.824\\
& $B-R$ & -0.633& 0.816 & -0.523 & 0.683 & -0.405 & 0.518  \\
\hline
\cite{McGaugh2014}& $B-V$ & -0.628 & 1.305 & - & - & -0.275 & 0.615  \\
\hline
\end{tabular}
\end{table*}

\begin{table*}
\centering
\caption{Color indices of the chosen galaxies (\citet{deVaucouleurs1992}$^1$, \citet{Lauberts1989}$^2$, \citet{McGaugh1997}$^3$, \citet{SDSSDR6}$^4$, \citet{Stark2009}$^5$, \citet{Kim2007}$^6$), their gas mass fraction ($f_g$), total luminosity in $i$ photometric band ($L_i$), stellar ($M_\star$) and gas masses ($M_\mathrm{gas}$), and estimated $M/L$ ratios for the bulge ($\Upsilon_b$), and for the disk ($\Upsilon_d$).}
\label{table:mtolight}
\begin{tabular}{lccccccccccc}
\hline
\hline
ID & $B-V$ & $B-R$ & Ref. & $f_g$  & $L_{i}$ & $M_{\star}$ & $M_\mathrm{gas}$ &  $\Upsilon_b$ & $\Upsilon_d$\\
 & ($^{m}$) & ($^{m}$) &  &  & $\times10^8 L\odot$ & $\times10^8 M_\odot$ & $\times10^8 M_\odot$ &($M_{\odot}/L_\odot$) &  ($M_{\odot}/L_\odot$)\\
\hline
ESO215G39 & 0.54 &- & 1 & 0.38 & 26.3 & 29.9 & 18.5 & 1.14 & 1.84\\
ESO322G76 &- & 0.13 & 2 & 0.26 & 296 & 136 & 48.4 & 0.46 & 0.62\\
ESO322G77 &- & 0.72 & 2 & 0.52 & 9.11 & 8.48 & 9.34 & 0.93 & 1.96\\
ESO322G82 &- & 0.99 & 2 & 0.24 & 10.2 & 13.1 & 4.24 & 1.28 & 1.69\\
ESO323G25 &- & 1.10 & 2 & 0.28 & 120 & 176 & 70 & 1.47 & 2.05\\
ESO374G02 &- & 0.64 & 2 & 0.27 & 9.01 & 7.58 & 2.77 & 0.84 & 1.15\\
ESO375G12 & 0.61 &- & 1 & 0.20 & 13.4 & 16.8 & 4.31 & 1.25 & 1.58\\
ESO376G02 & 0.40 &- & 1 & 0.37 & 7.31 & 6.82 & 3.98 & 0.93 & 1.48\\
ESO383G02 &- & 0.84 & 2 & 0.31 & 96.1 & 103 & 46.7 & 1.08 & 1.56\\
ESO383G88 &- & 1.08 & 2 & 0.38 & 5.32 & 7.61 & 4.73 & 1.43 & 2.32\\
ESO445G19 &- & 0.71 & 2 & 0.38 & 5.4 & 4.94 & 2.98 & 0.92 & 1.47\\
ESO446G01 &- & 0.57 & 2 & 0.26 & 231 & 180 & 63.2 & 0.78 & 1.05\\
ESO502G02 &- & 1.24 & 2 & 0.35 & 6.21 & 10.7 & 5.71 & 1.72 & 2.64\\
ESO509G80 &- & 1.04 & 2 & 0.29 & 266 & 363 & 152 & 1.37 & 1.94\\
ESO569G17 &- & 0.53 & 2 & 0.35 & 391 & 291 & 159 & 0.74 & 1.15\\
F561-1 & 0.41 &- & 3 & 0.46 & 159 & 135.00 & 115 & 0.85 & 1.57\\
F563-1 & 0.65 &- & 3 &- & 2.89 & 4.48 &- & 1.55 & -\\
F568-3 & 0.55 &- & 3 & 0.55 & 320 & 386 & 472 & 1.21 & 2.68\\
F579-V1 & 0.76 &- & 4 & 0.34 & 74.40 & 151 & 77.9 & 2.03 & 3.08\\
F583-1 & 0.39 &- & 5 &- & 0.72 & 0.58 &- & 0.81 & -\\
F730-V1 & 0.54 &- & 6 & 0.57 & 11.60 & 13.9 & 18.5 & 1.20 & 2.80\\
UGC128 & 0.60 &- & 6 & 0.72 & 1.43 & 1.96 & 5.04 & 1.37 & 4.89\\
UGC1230 & 0.52 &- & 3 & 0.80 & 11.3 & 12.7 & 50.7 & 1.12 & 5.60\\
UGC5750 & 0.53 &- & 4 & 0.67 & 1.08 & 1.25 & 2.54 & 1.16 & 3.51\\
UGC6614 & 0.72 &- & 3 & 0.45 & 2.53 & 5.18 & 4.23 & 2.05 & 3.73\\
UGC10310 & 0.42 &- & 1 & 0.74 & 3.3 & 2.75 & 7.9 & 0.83 & 3.22\\
UGC11454 & 0.47 &- & 6 & 0.67 & 5.61 & 5.46 & 11 & 1.1 & 0.82\\
UGC11616 & 0.36 &- & 6 & 0.83 & 21.7 & 15 & 71.4 & 0.69 & 3.99\\
UGC11748 & 0.38 &- & 6 & 0.80 & 78.5 & 58.3 & 226 & 0.74 & 3.62\\
UGC11819 & 0.60 &- & 6 & 0.29 & 65.2 & 93.4 & 37.9 & 1.43 & 2.01\\
\hline
\end{tabular}
\end{table*}

To estimate the $M/L$ ratios, we employed color--to-mass-to-light ratio relations (CMLR). The relation between the stellar $M/L$ ratio and the color index is $\log \Upsilon_\star=\alpha_i+\beta_i(m_i-m_j)$, where $\Upsilon_\star$ is the stellar $M/L$ ratio, and $m_i-m_j$ is the color index calculated from $i$ and $j$ photometric bands.

 \citet{McGaugh2014} combined Spitzer $3.6 \mu m$ infrared observations of a sample of disk galaxies with optical luminosities to test four different population synthesis prescriptions for computing stellar mass. In their analysis the bulge and disk were not distinguished. They found that the semi-empirical stellar population synthesis model of \citet{Bell2003} is self-consistent, and the revised CMLR based on the model of \citet{Bell2003} has the least scatter (the self-consistency satisfies that different photometric bands provide the same stellar mass). This model is one of the two population synthesis models that required the smallest corrections to the Spitzer data. Therefore we calculated $\Upsilon_\star$ considering the revised CMLR of \citet{McGaugh2014}, which explores the color index $B-V$. This model was tested by circular arguments and assumes that the scatter in the baryonic Tully-Fisher relation should be minimized for all galaxies. For the HSB galaxies with known color index $B-R$, we used the original CMLR of \citet{Bell2003}. We summarize the coefficients of these CMLR-s in Table \ref{table:cmlrs}.

The color indices were collected from the literature as indicated in Table \ref{table:mtolight}. When transforming the SDSS colors to $B-V$ color index, we applied the relation $B-V=0.98(g-r)+0.22$ given by \citet{Jester2005}. Where the color indices were not corrected for extinction \citep{SDSSDR6,Kim2007}, we have done it using Landolt standard-fields.

We estimate the mass of the gaseous disk $M_\mathrm{gas}$ employing the gas mass fraction
\begin{flalign}
f_g=\frac{M_\mathrm{gas}}{M_\mathrm{gas}+M_{\star}},
\end{flalign}
where $M_{\star}$ is the stellar mass of the galaxy. There are empirical relations between the gas mass fraction, the brightness, and the color index of the galaxies. \citet{McGaugh1997} found that in general, $f_g$ computed from $B$-band and $I$-band data of spiral galaxies correspond closely. More precisely, for the $I$-band HSB galaxies $f_g=0.12(M_B+23)$ holds for the gas mass fraction and the absolute magnitude of the galaxy in $B$ band \citep{McGaugh1997}. The absolute $B$ magnitudes are derived based on the galaxy distances collected from \citet{Palunas2000} and their apparent $B$ magnitudes collected from the NASA/IPAC Extragalactic Database \citep{Lauberts1989}. \citet{McGaugh1997} gave a relation between $f_g$ and the $B-V$ color index, $f_g=-1.4[(B-V)-0.95]$. We used this equation to calculate $f_g$ for the LSB galaxies with available $B-V$ indices: F730-V1, UGC10310, UGC11454, UGC11616, UGC11748, and UGC11819. For the other galaxies, $f_g$ is available directly from the literature: F561-1, F568-3, UGC128, UGC1230, UGC6614 \citep{McGaugh1997}, F579-V1, and UGC5750 \citep{Schombert2014}.

The stellar mass-to-light ratio $\Upsilon_\star$ was taken as the bulge $M/L$, $\Upsilon_b$. Then the stellar mass, encompassed by the bulge and the disk, is $M_\star=\Upsilon_\star L$, where $L=L_b+L_d$ is the total luminosity of the galaxy, calculated from the best-fit central luminosity density of the bulge ($L_b$) and the disk ($L_d$). The corresponding scaling parameters, $k=\Gamma(2N)/\Gamma(3N)$ and $h=\Gamma ^2(3N)/[N\Gamma^3(2N)]$ \citep{Tamm2012} were calculated based on the best-fit $N$ and $ka_0$ of the bulge and disk components. The gas mass $M_\mathrm{gas}$ was derived based on $f_g$ and $M_\star$. Then the $M/L$ ratio of the disk is $\Upsilon_{d}=(M_\star+M_\mathrm{gas})/L$. 

We summarize the photometric informations, $M/L$ ratios, and
total luminosities and masses in Table \ref{table:mtolight}. These $M/L$ ratios are assumed constant when fitting the baryonic + dark matter models to the rotation curve data.

\subsection{HSB galaxies}

\begin{figure*}
\centering
\includegraphics[width=170pt,height=120pt]{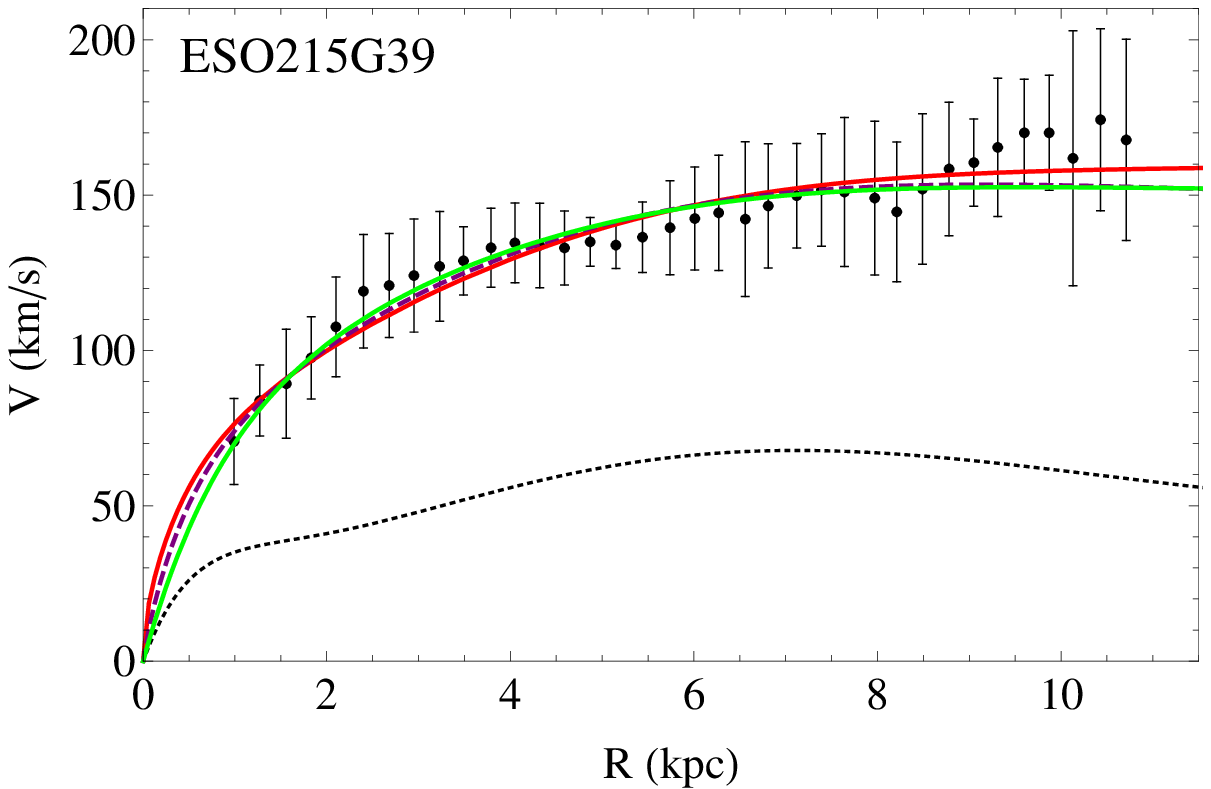}
\includegraphics[width=170pt,height=120pt]{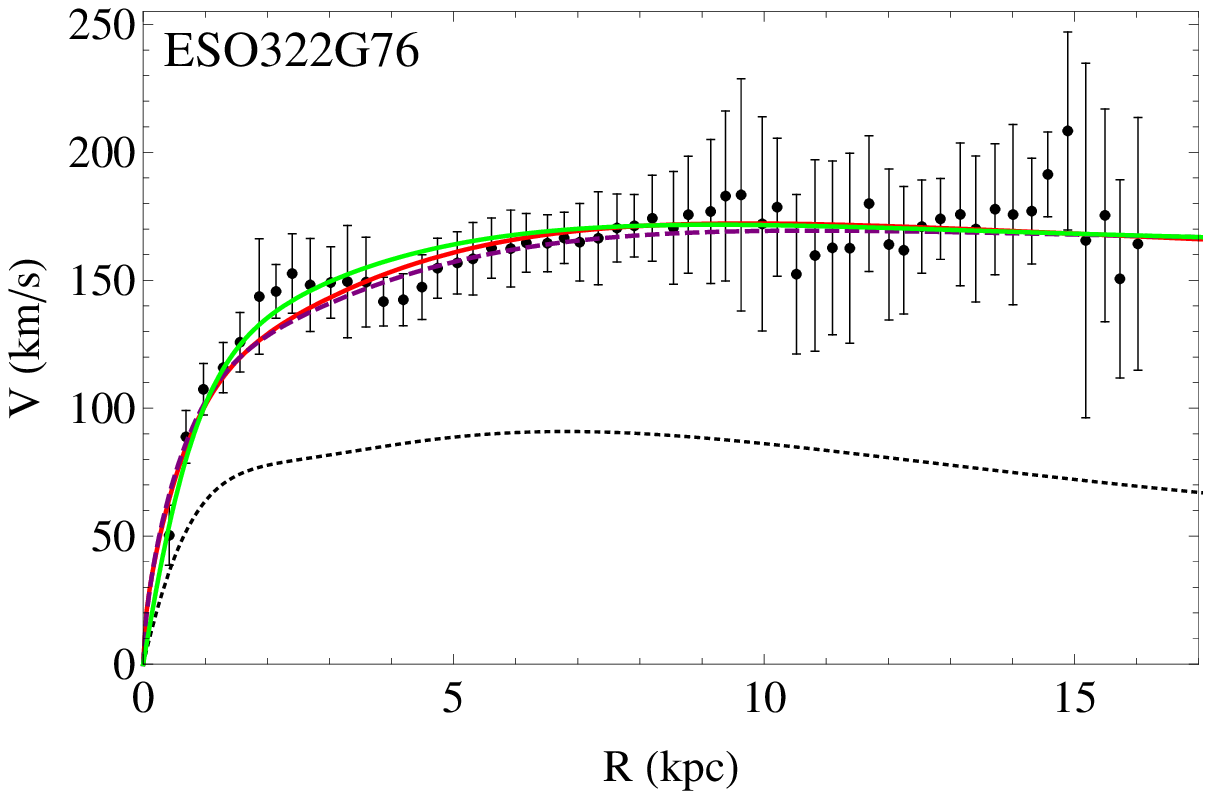}
\includegraphics[width=170pt,height=120pt]{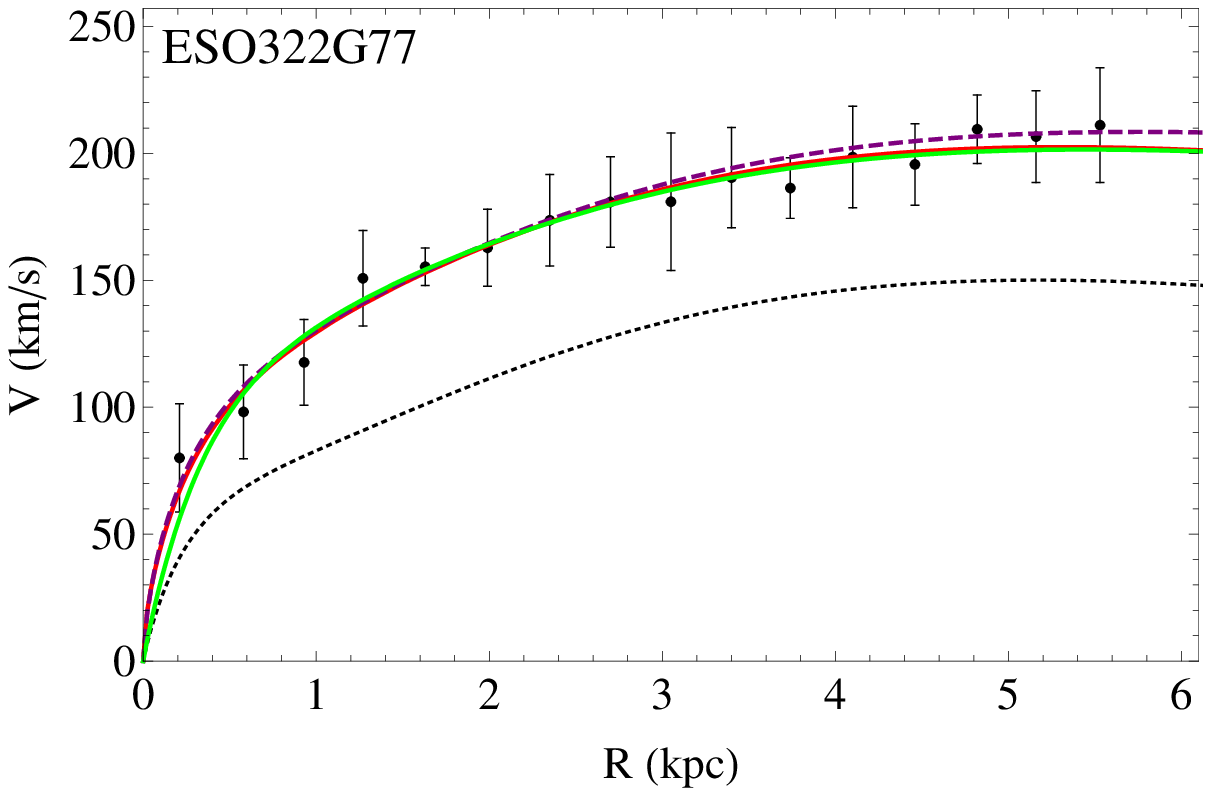}\newline
\includegraphics[width=170pt,height=120pt]{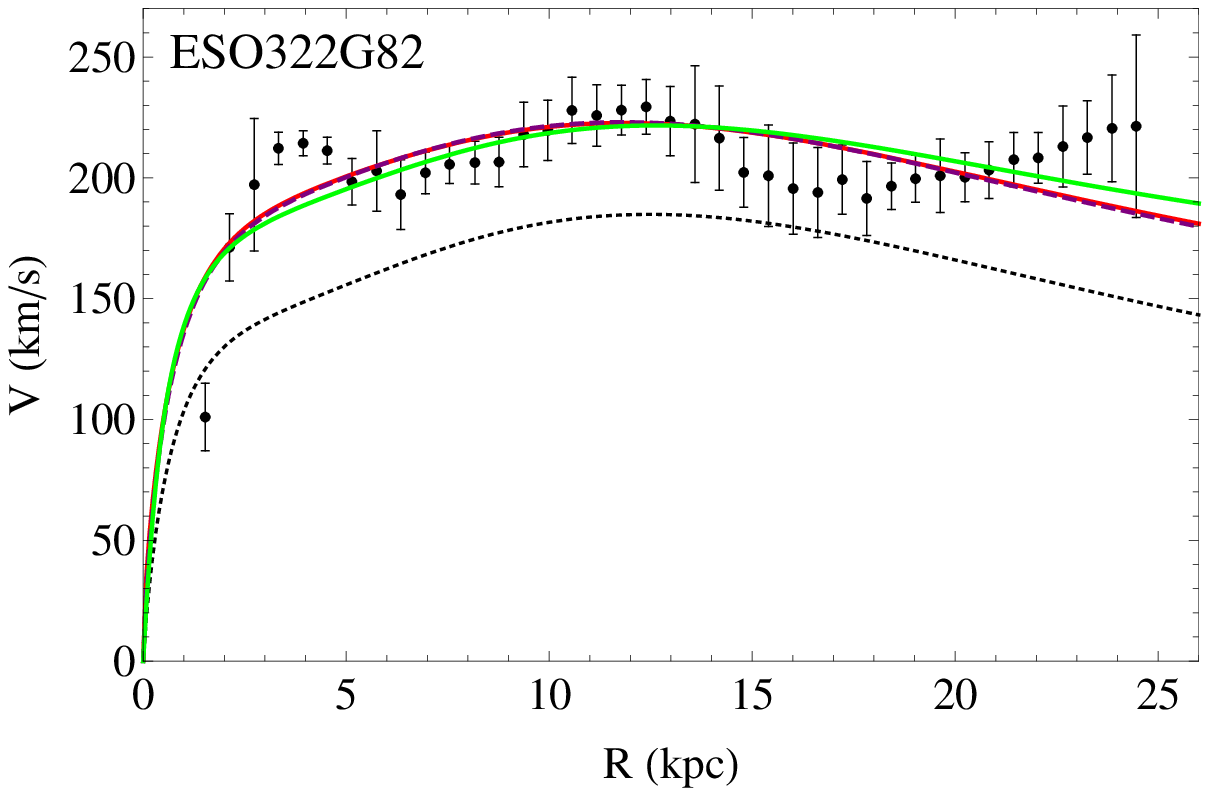}
\includegraphics[width=170pt,height=120pt]{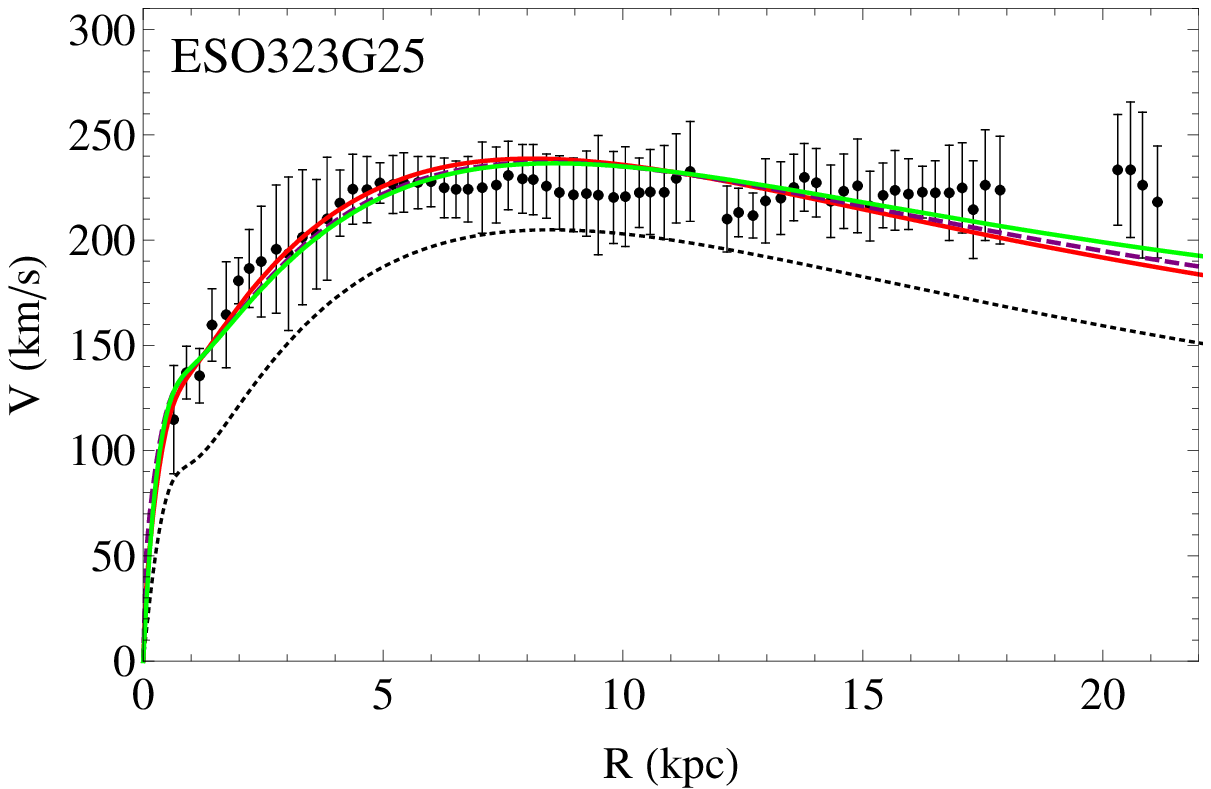}
\includegraphics[width=170pt,height=120pt]{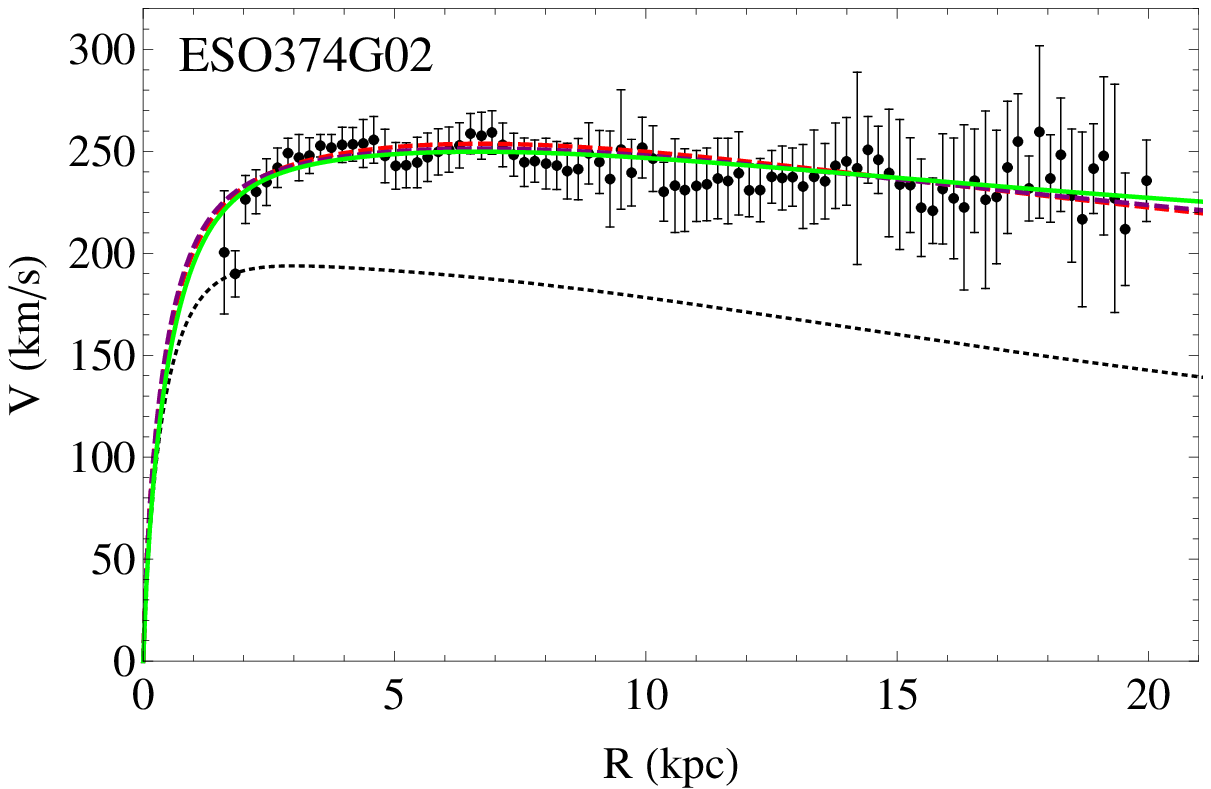}\newline
\includegraphics[width=170pt,height=120pt]{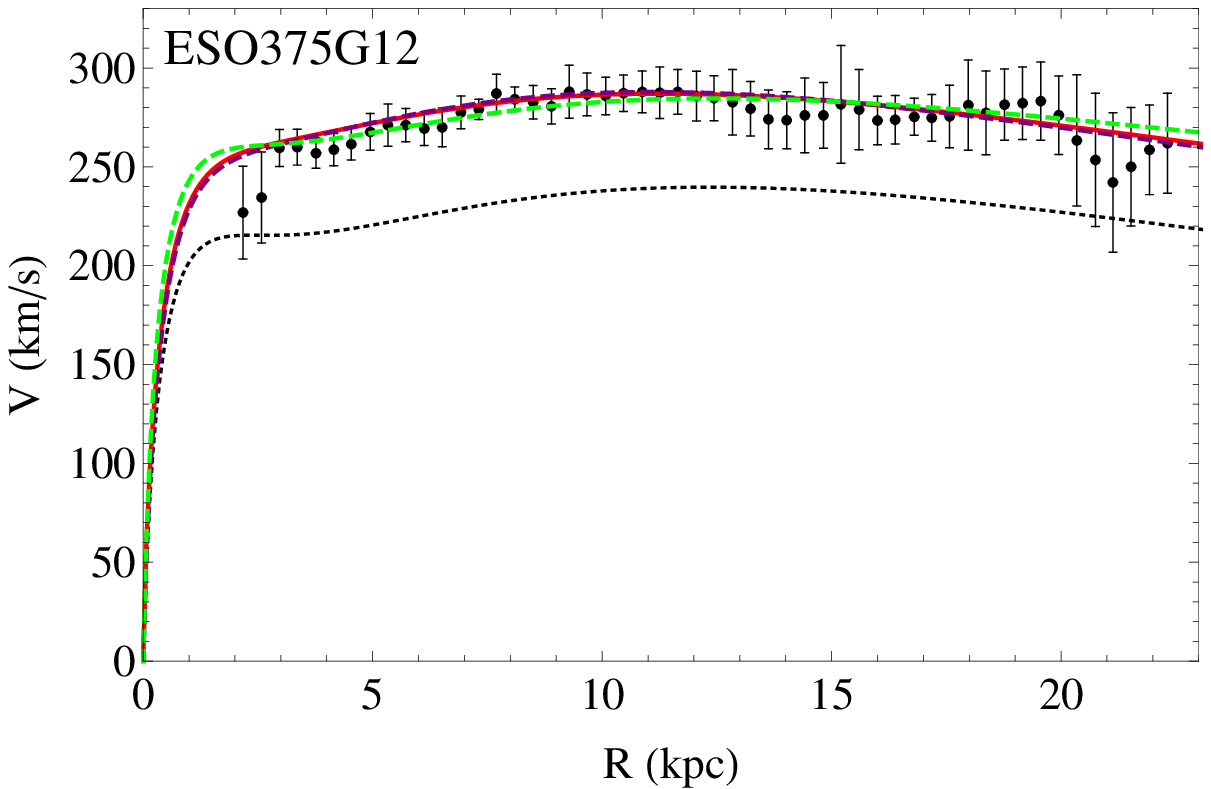}
\includegraphics[width=170pt,height=120pt]{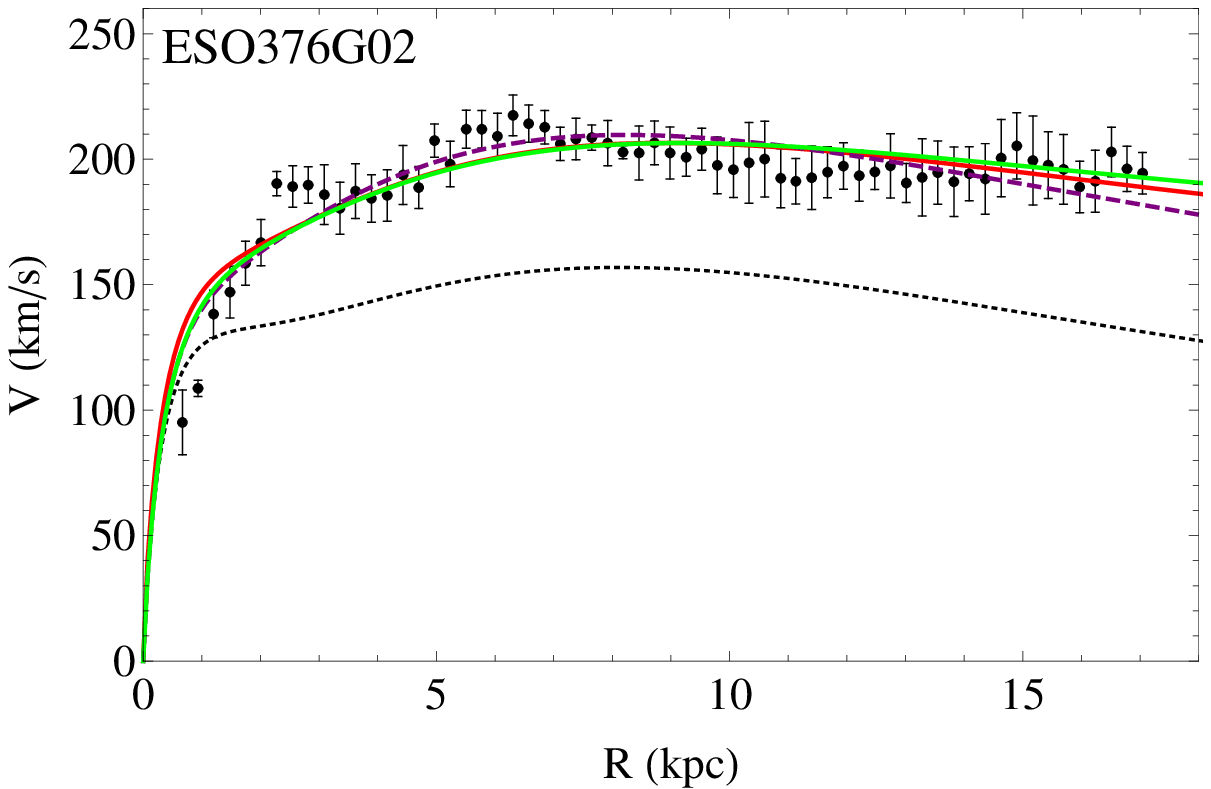}
\includegraphics[width=170pt,height=120pt]{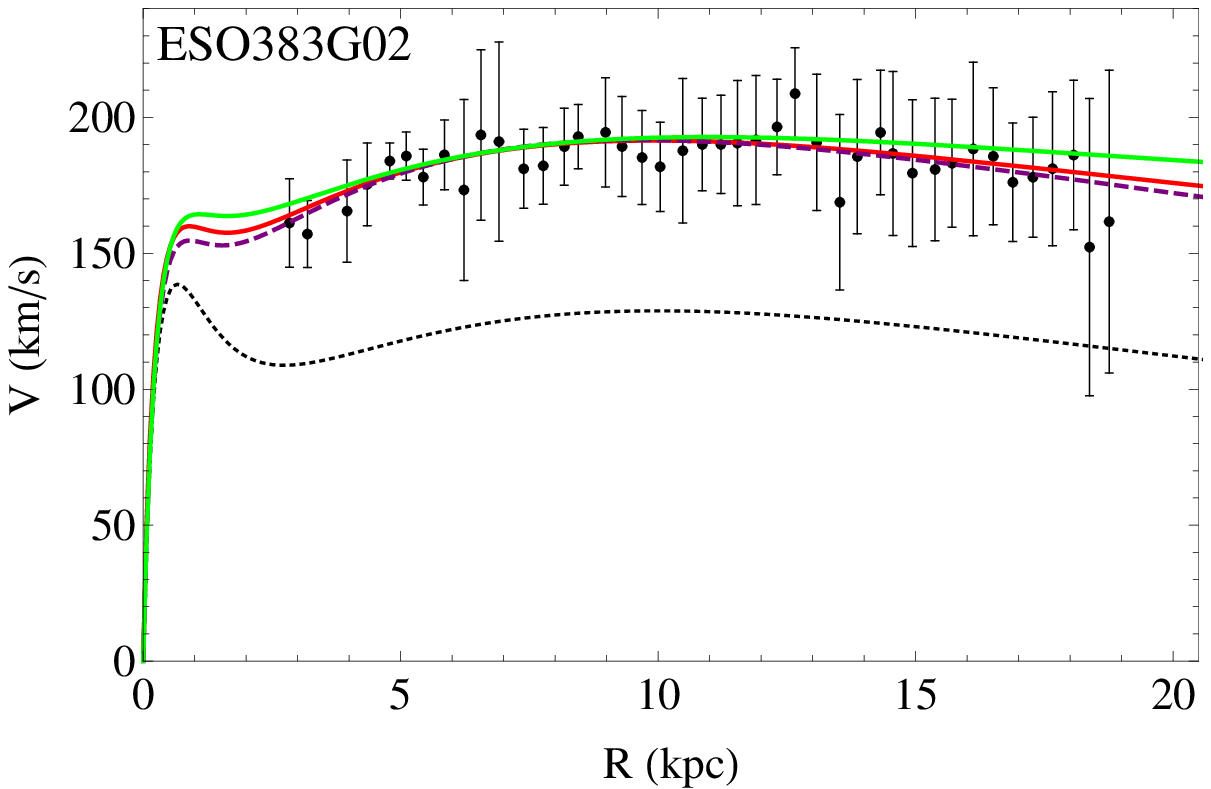}\newline
\includegraphics[width=170pt,height=120pt]{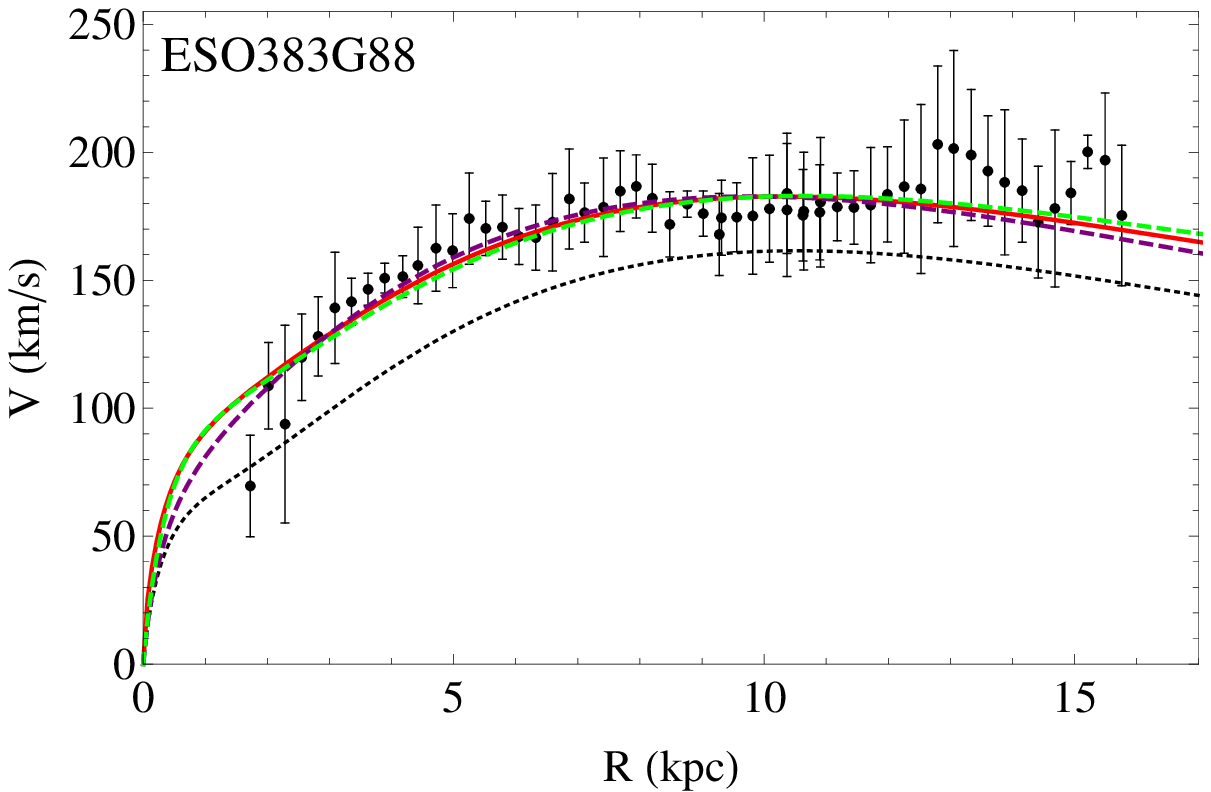}
\includegraphics[width=170pt,height=120pt]{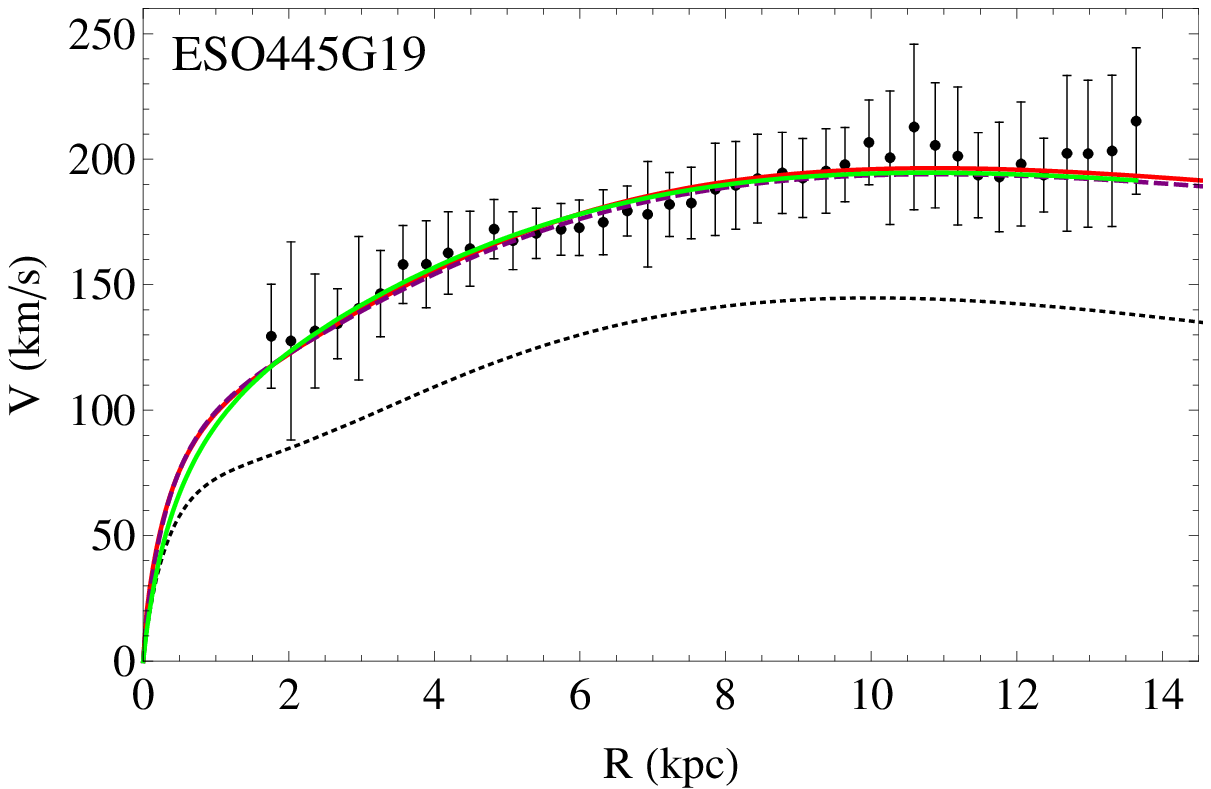}
\includegraphics[width=170pt,height=120pt]{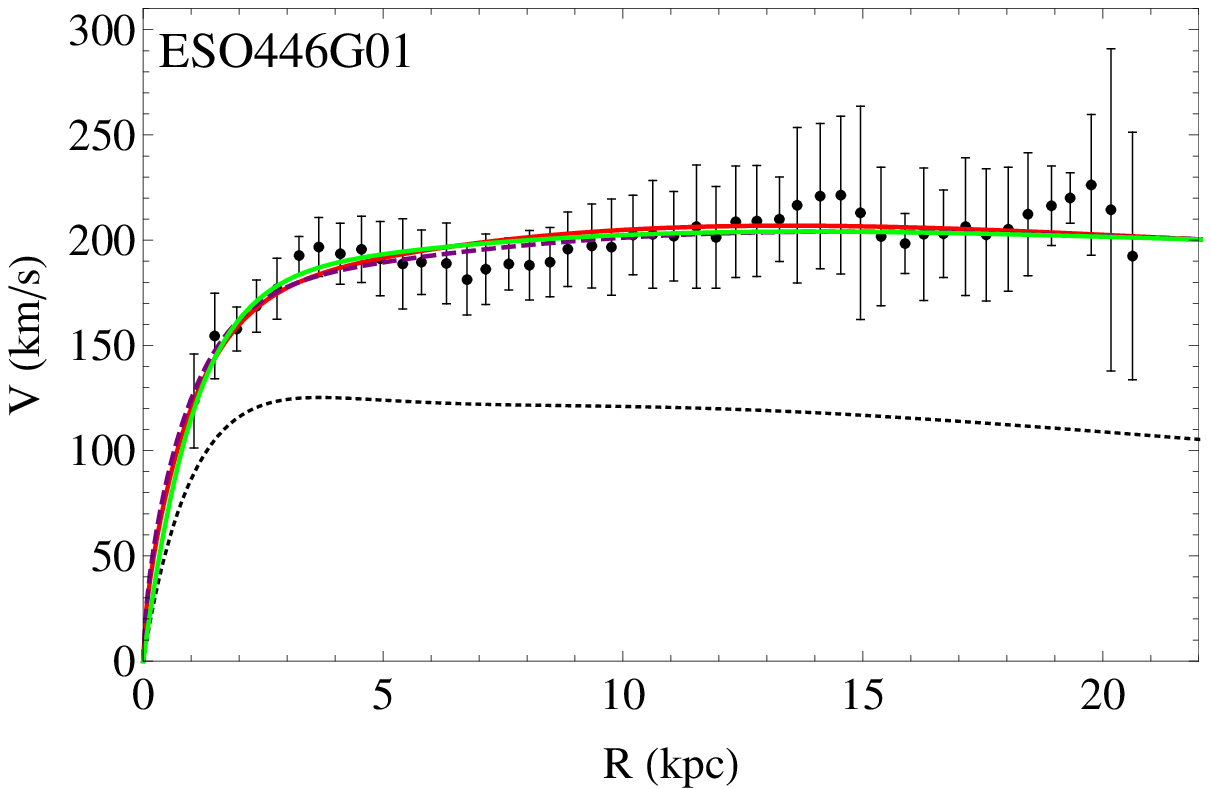}\newline
\includegraphics[width=170pt,height=120pt]{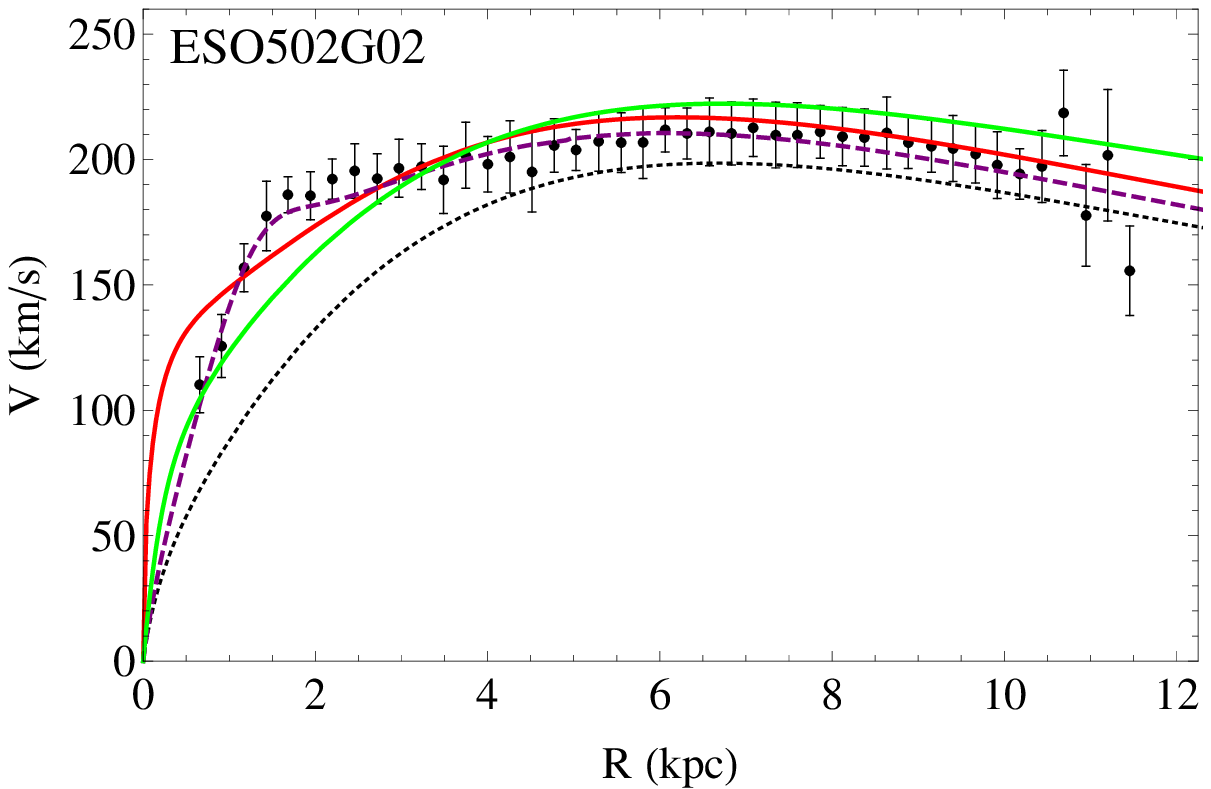}
\includegraphics[width=170pt,height=120pt]{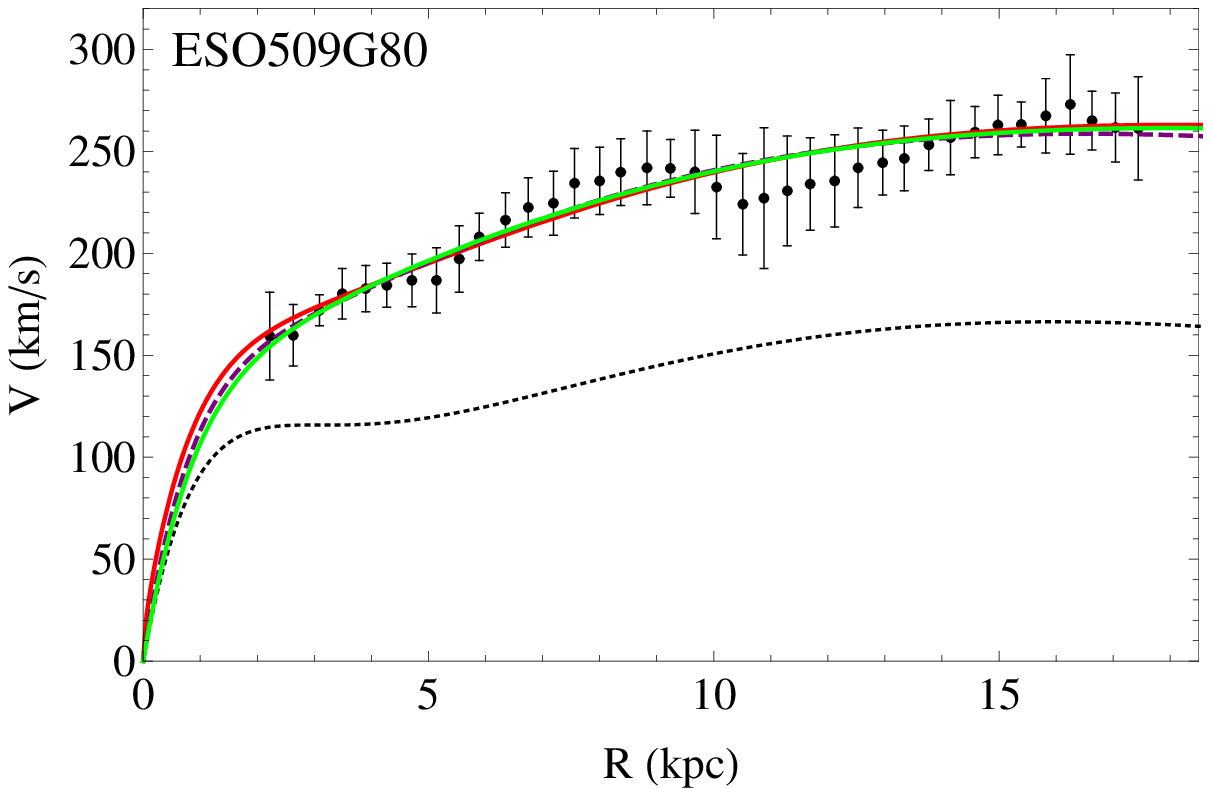}
\includegraphics[width=170pt,height=120pt]{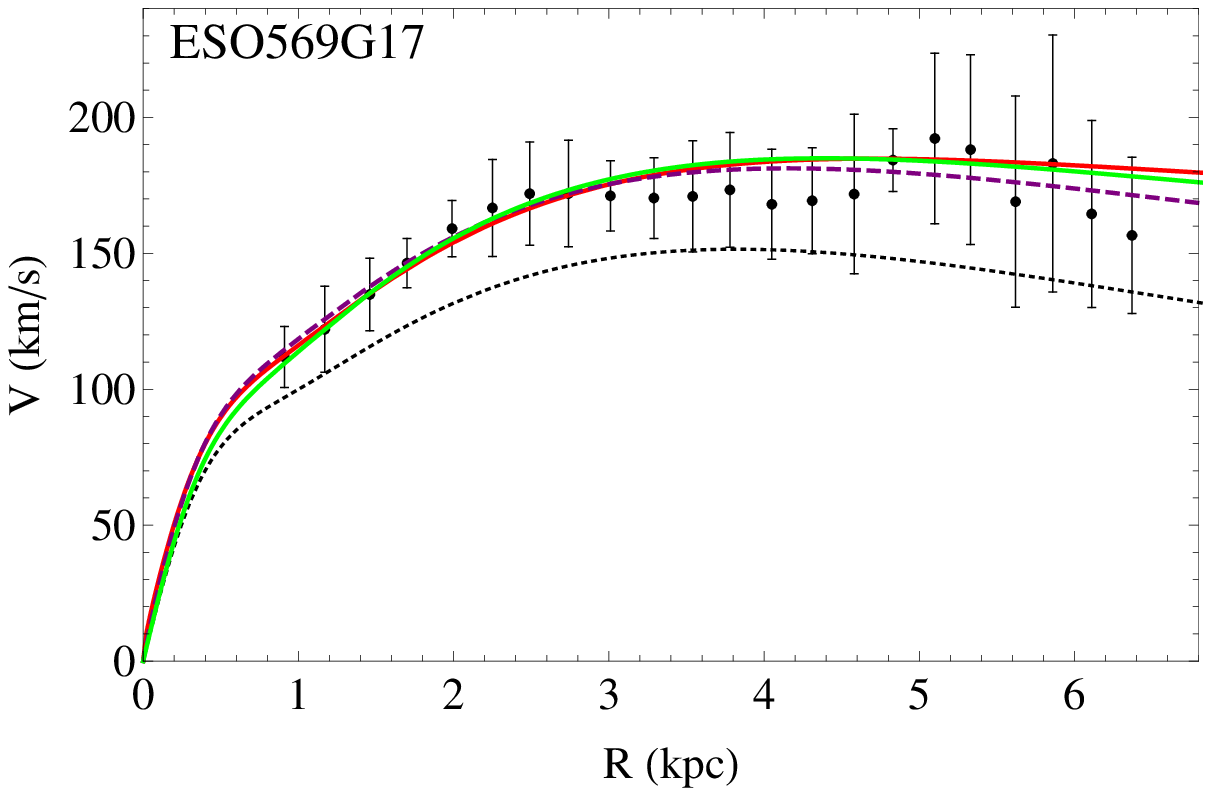}\newline
\caption{Best-fit rotational curves for the HSB galaxy sample. The dots with error bars denote archive rotational velocity curves derived from spectroscopic data. The fitted models are pure baryonic (black short-dashed curve), baryonic + NFW (red continuous curve), baryonic + Einasto (purple dashed curve), and baryonic + PSE (green continuous curve).}
\label{fig:hsb_vrot}
\end{figure*}

We collected the rotational velocity data of the HSB galaxy sample from \citet{Palunas2000}, who presented maximum disk models for a sample of $74$ field and cluster spiral galaxies located in the vicinity of the Hydra-Centaurus cluster. For each galaxy they had $I$-band CCD images and two-dimensional (2D) H$\alpha$ velocity maps, from which the surface brightness distribution and rotational velocity curve of each galaxy were produced. From this sample $15$ galaxies were selected such that these galaxies did not show bars, or rings, which might contradict the assumption of axisymmetry. We summarize the best-fit parameters in Table \ref{table:hsb_vrot}. We present the best-fit rotational velocity models and the galaxy velocity curves on Fig. \ref{fig:hsb_vrot}. We note that the error bars on the individual data points are quite large compared to the scatter in the mean data values, suggesting that the error
bars are slightly overestimated.

\subsection{LSB galaxies}

\begin{figure*}
\centering
\includegraphics[width=170pt,height=120pt]{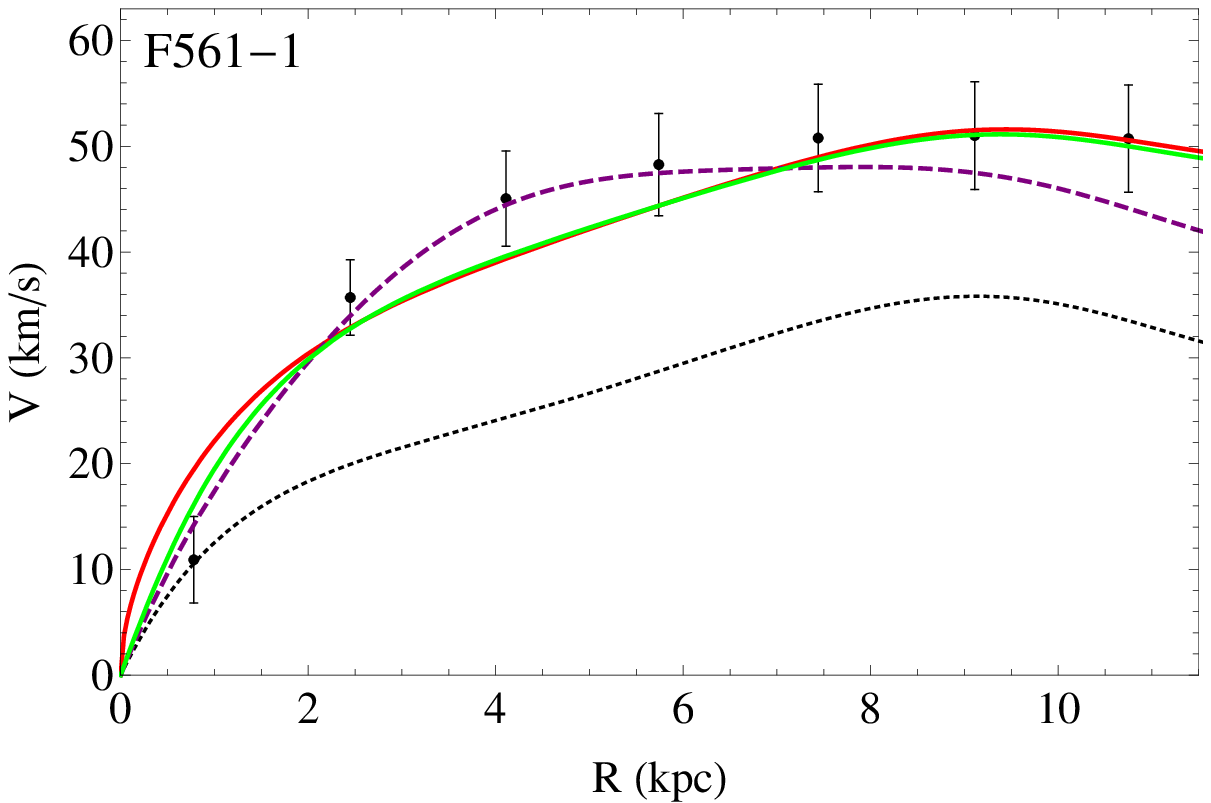}
\includegraphics[width=170pt,height=120pt]{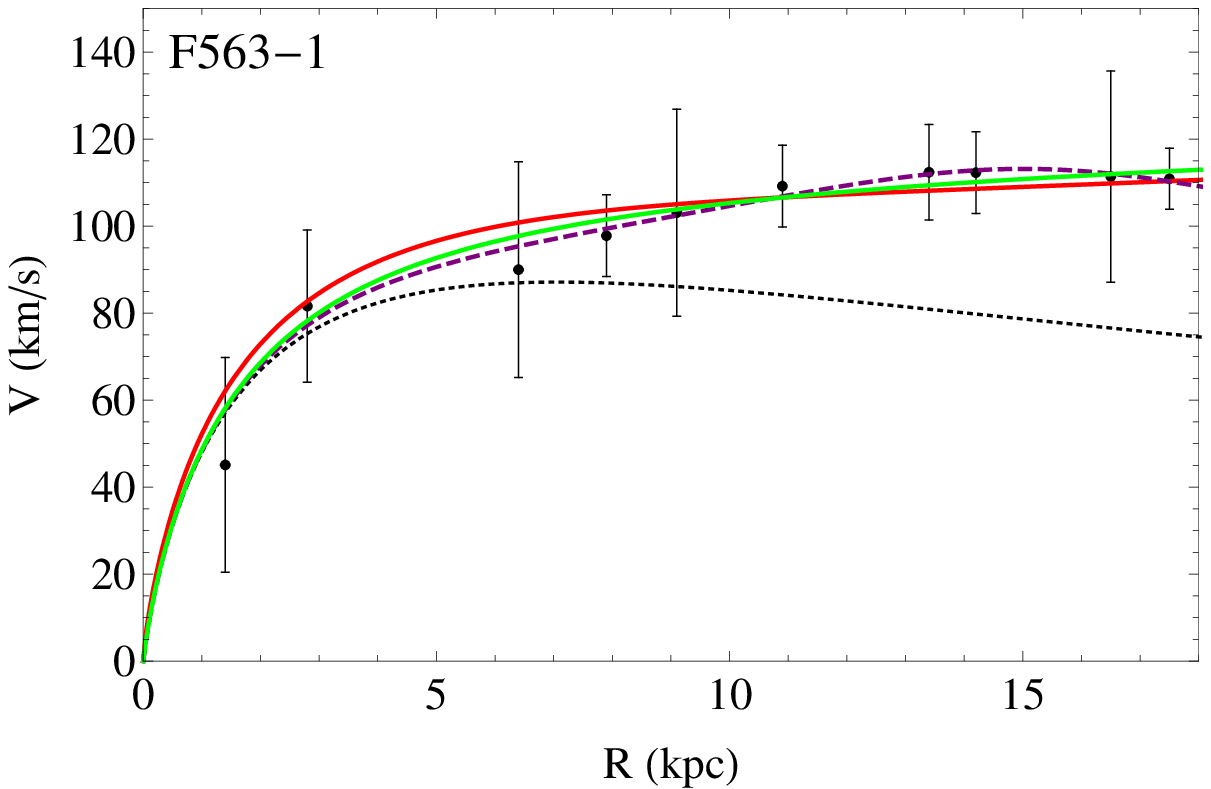}
\includegraphics[width=170pt,height=120pt]{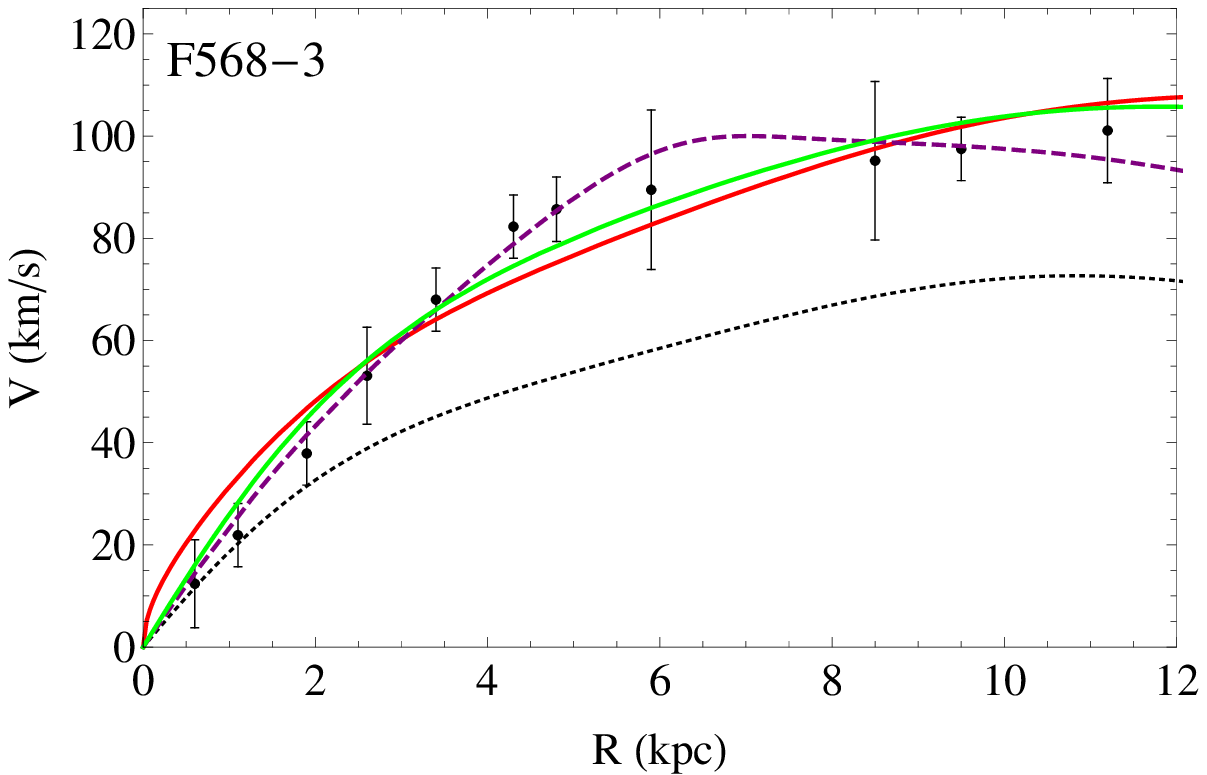}\newline
\includegraphics[width=170pt,height=120pt]{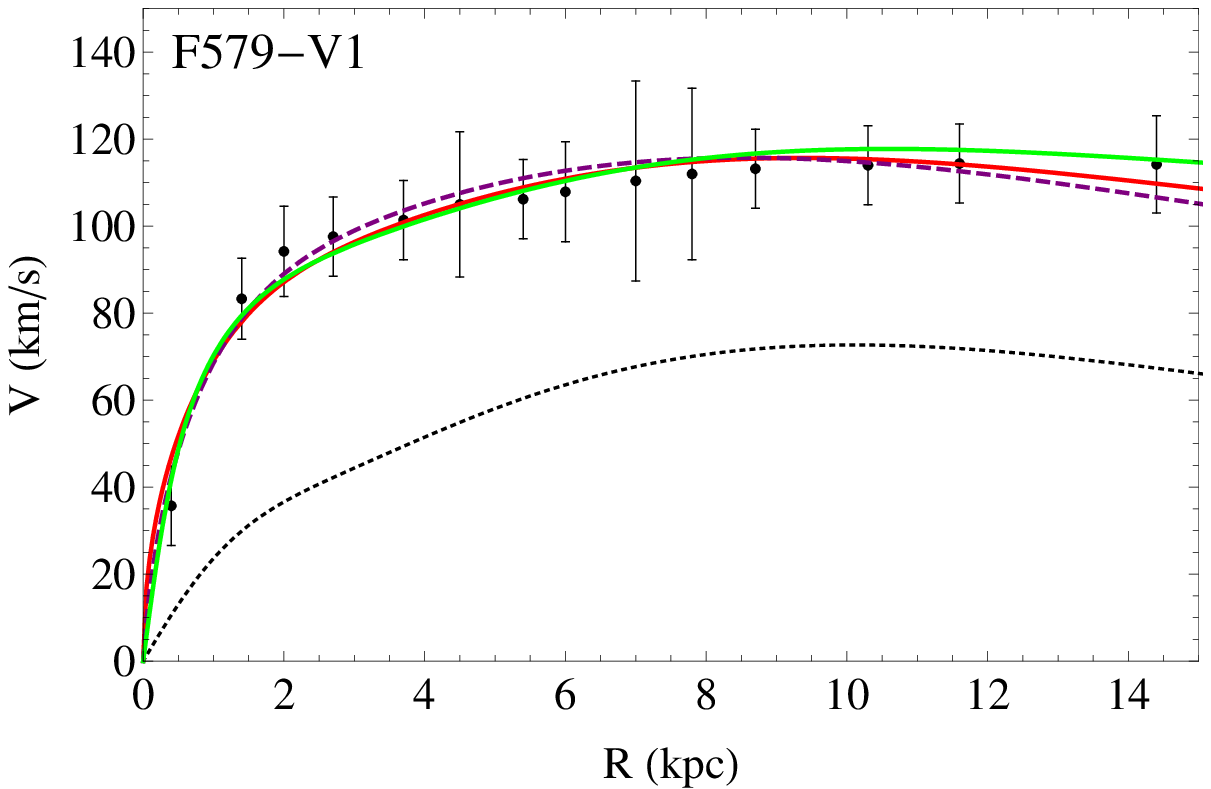}
\includegraphics[width=170pt,height=120pt]{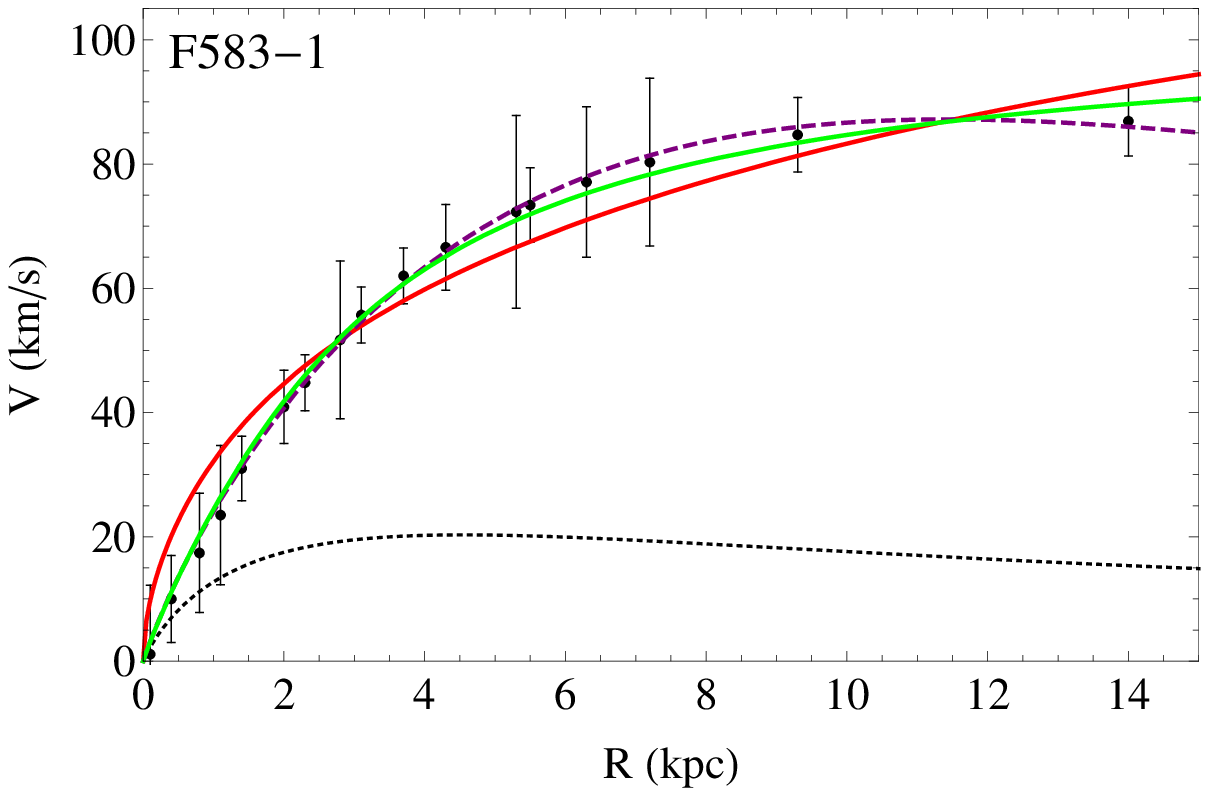}
\includegraphics[width=170pt,height=120pt]{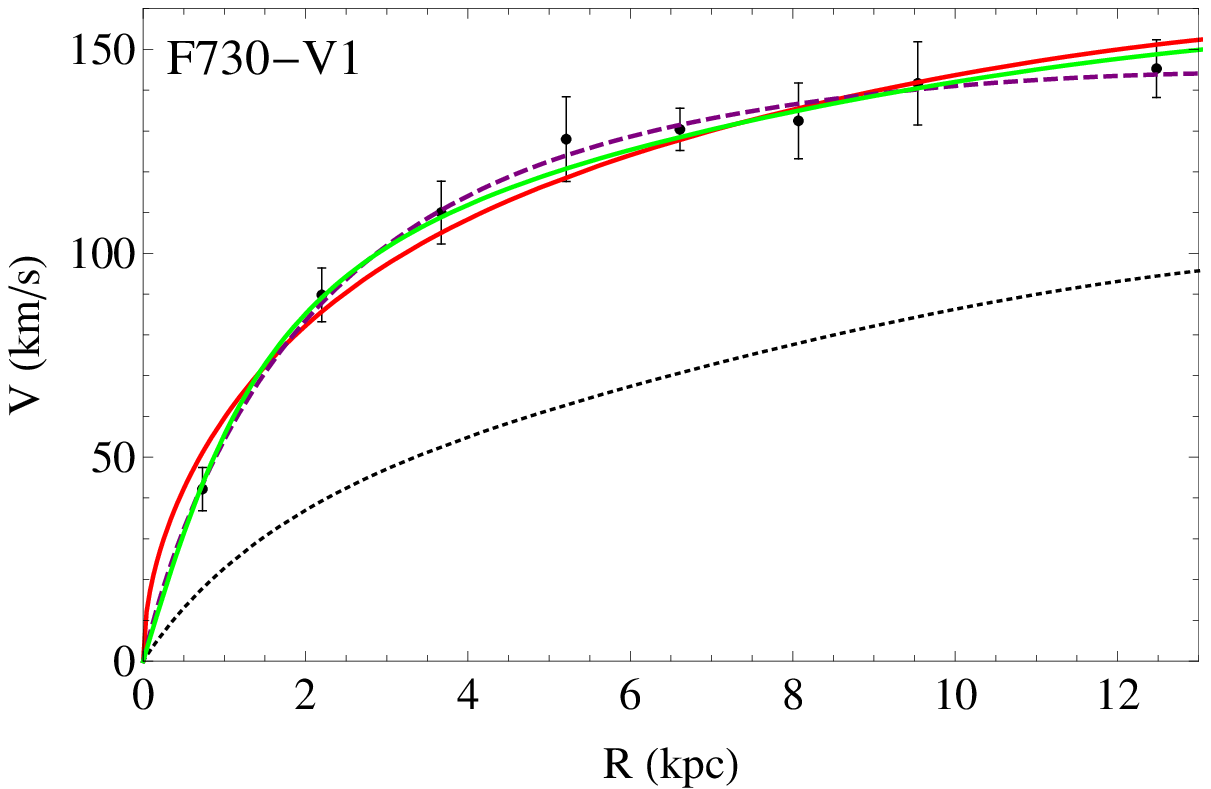}\newline
\includegraphics[width=170pt,height=120pt]{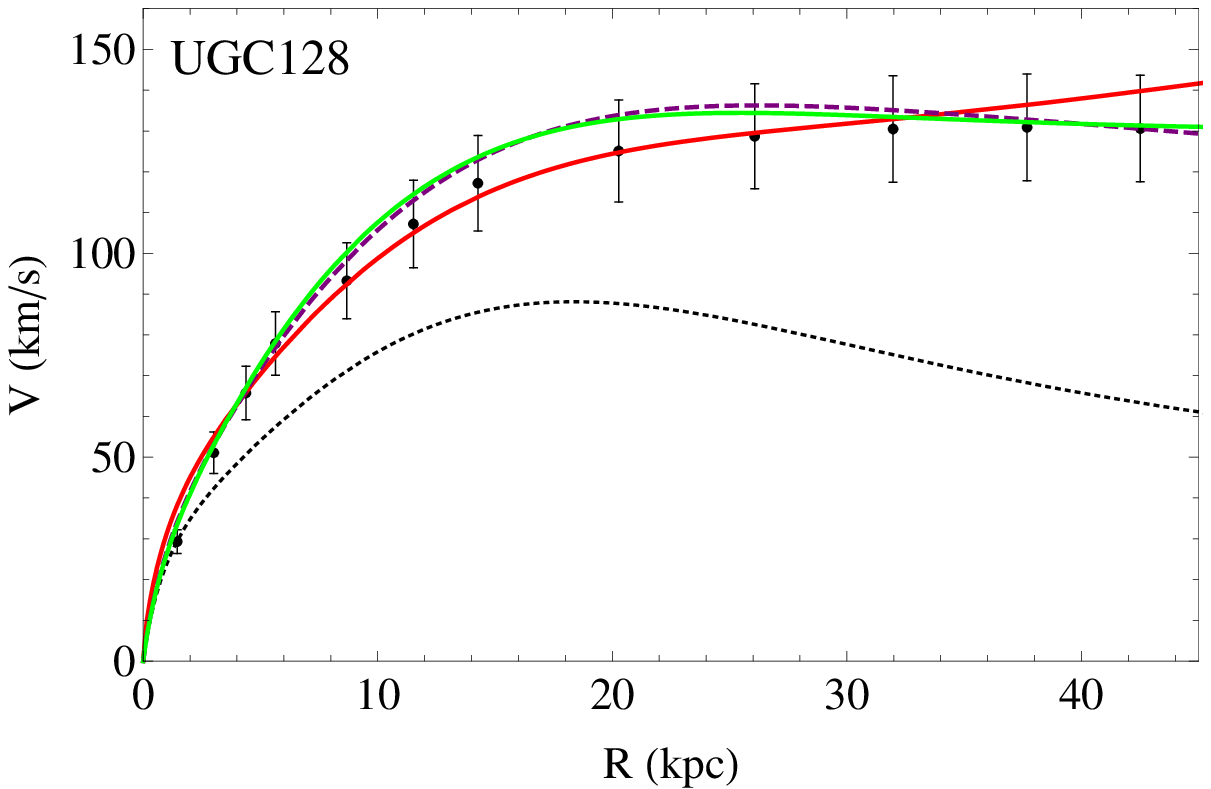}
\includegraphics[width=170pt,height=120pt]{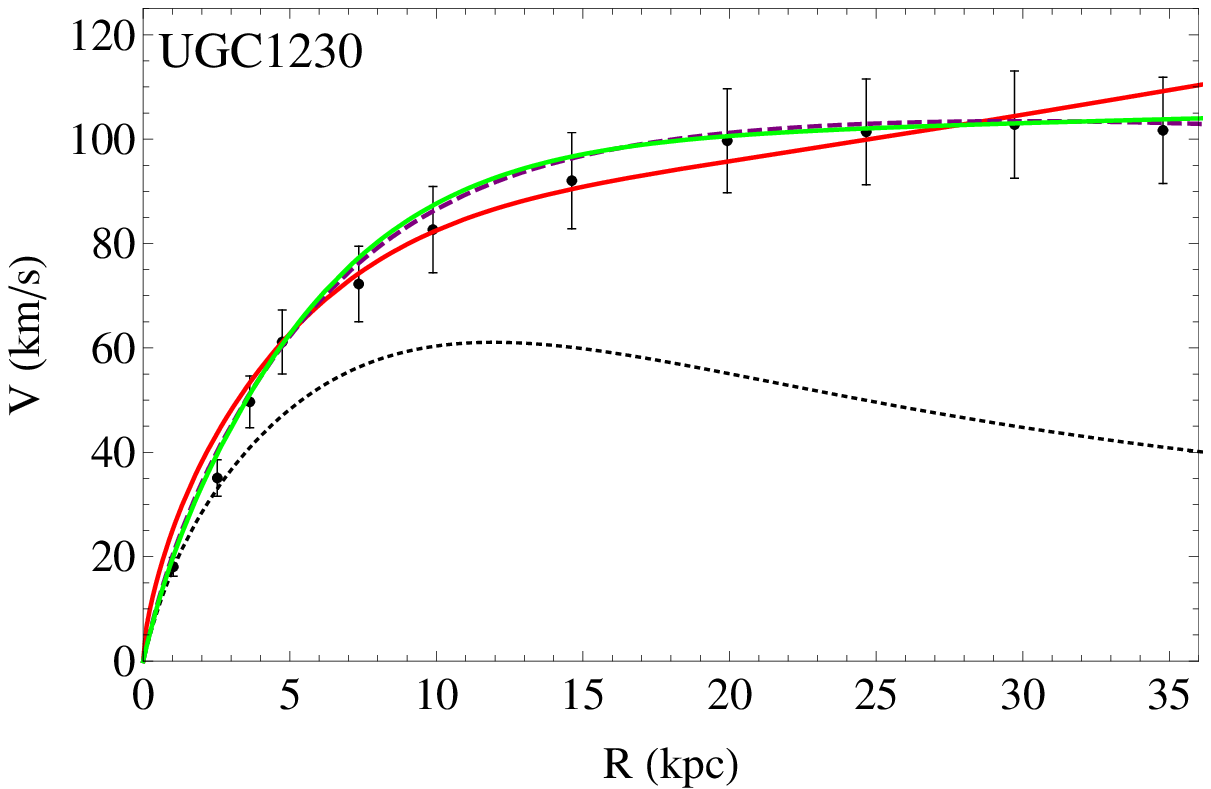}
\includegraphics[width=170pt,height=120pt]{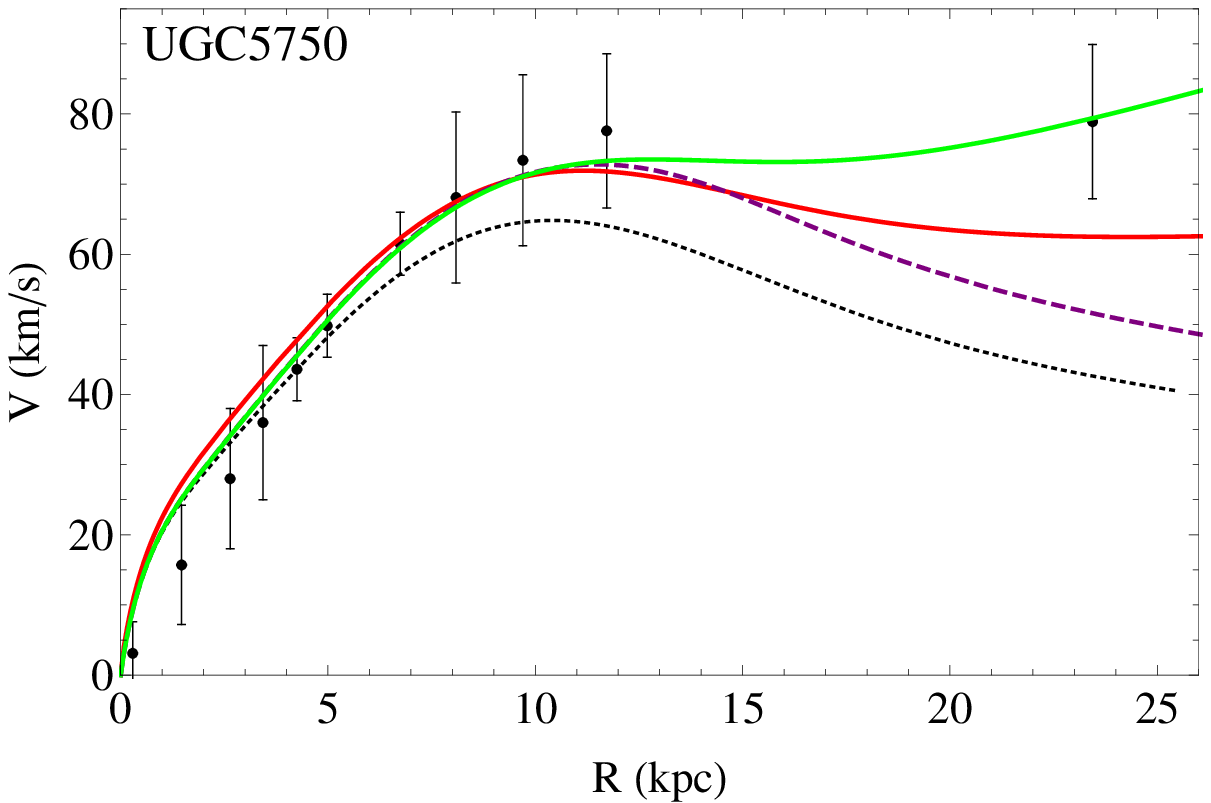}\newline
\includegraphics[width=170pt,height=120pt]{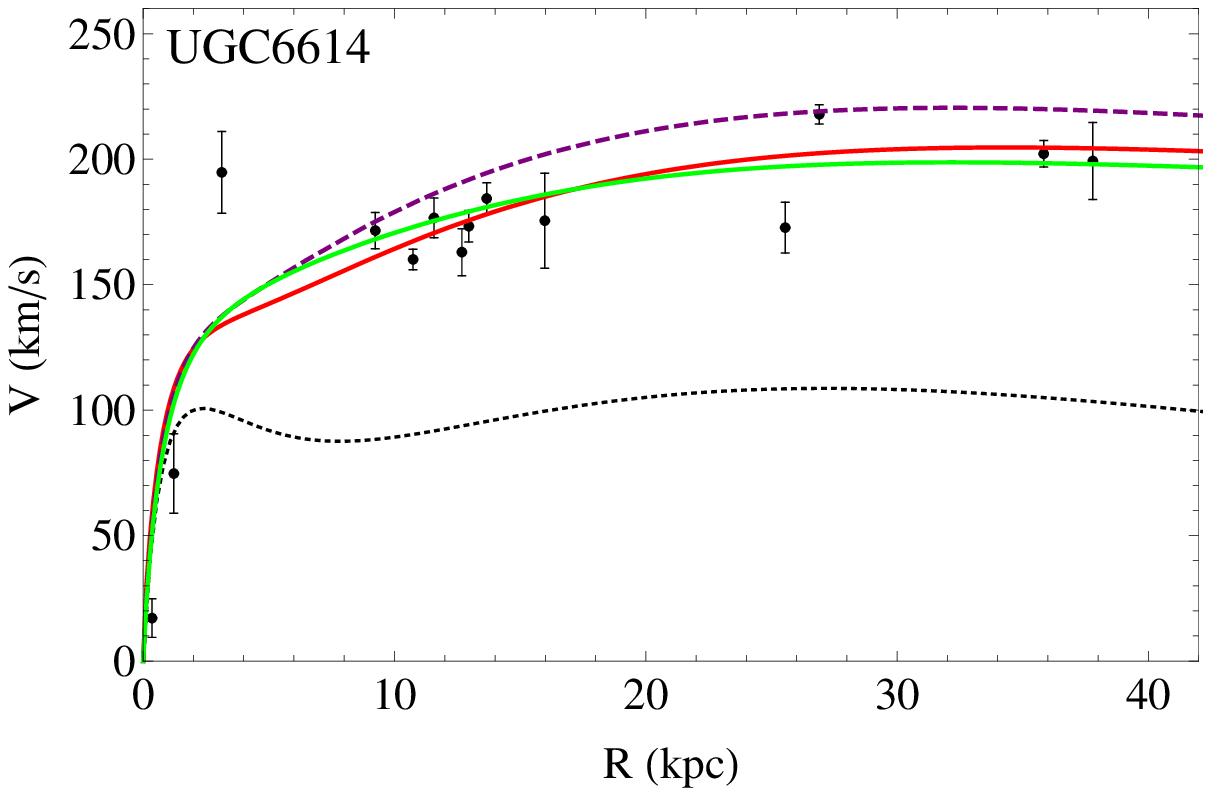}
\includegraphics[width=170pt,height=120pt]{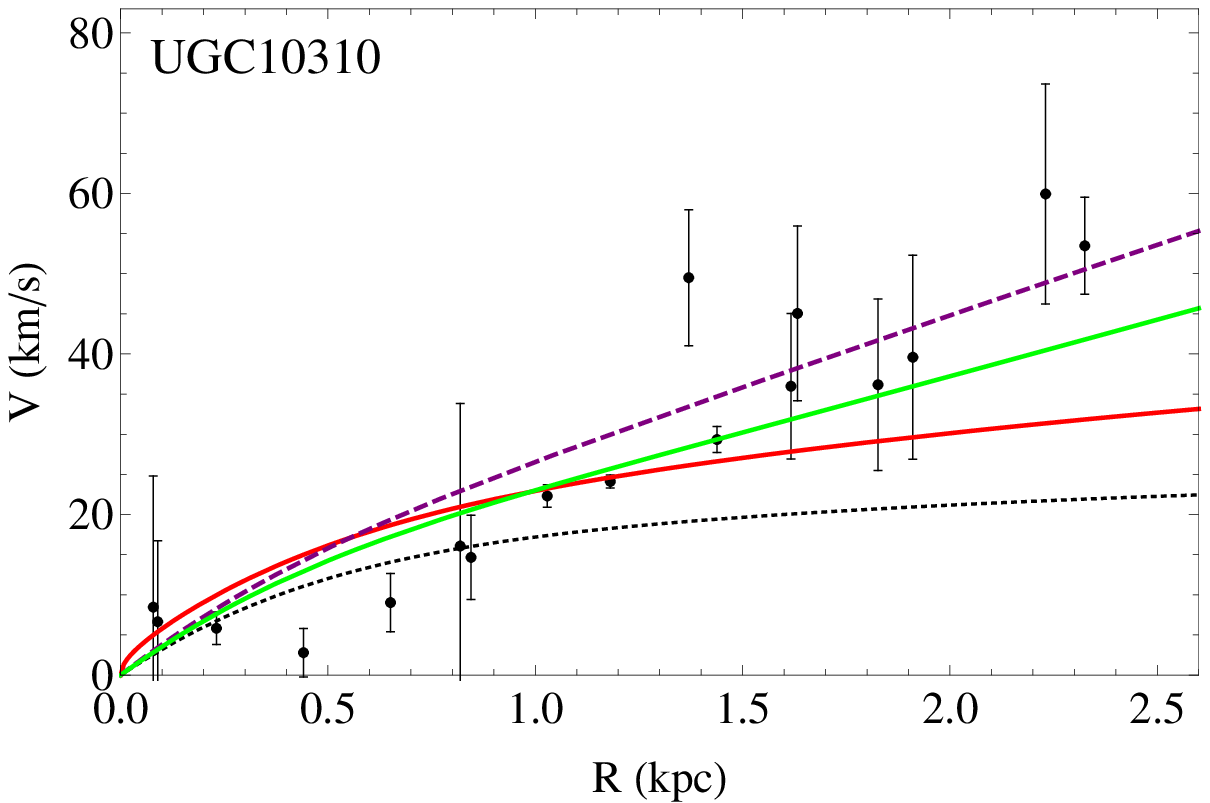}
\includegraphics[width=170pt,height=120pt]{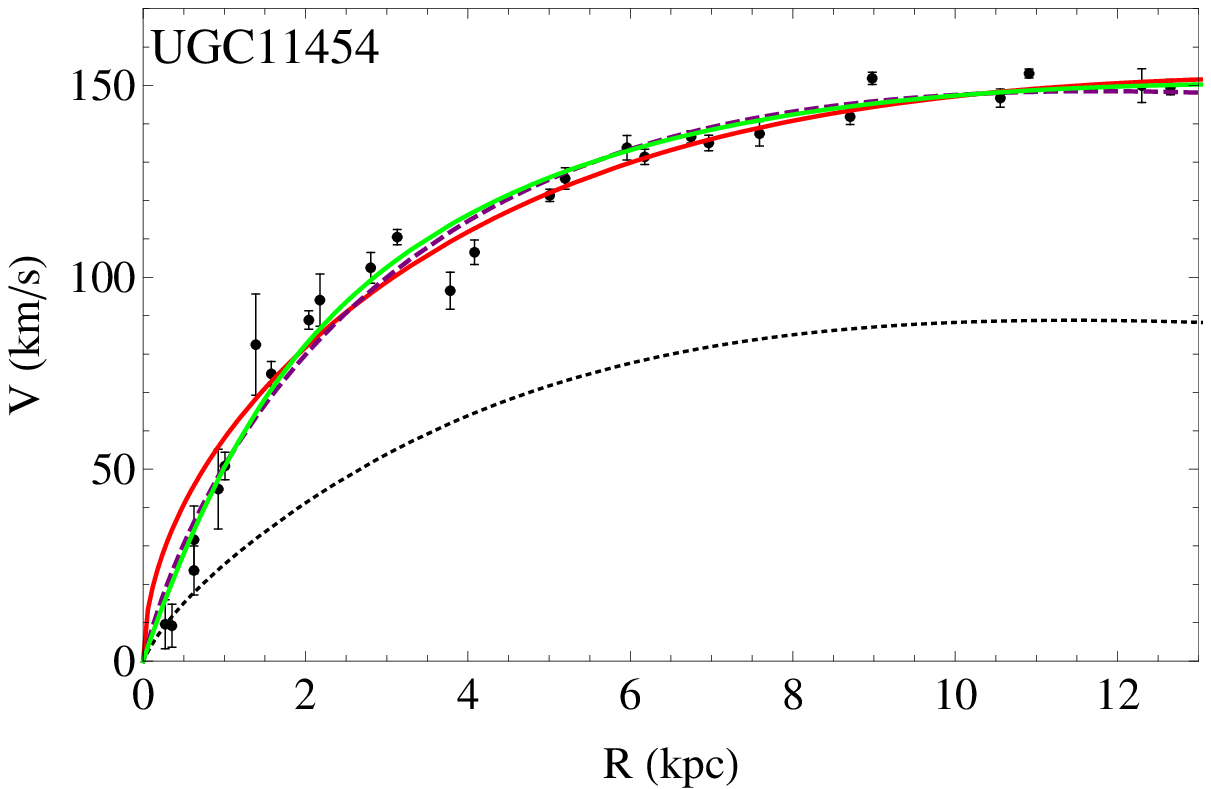}\newline
\includegraphics[width=170pt,height=120pt]{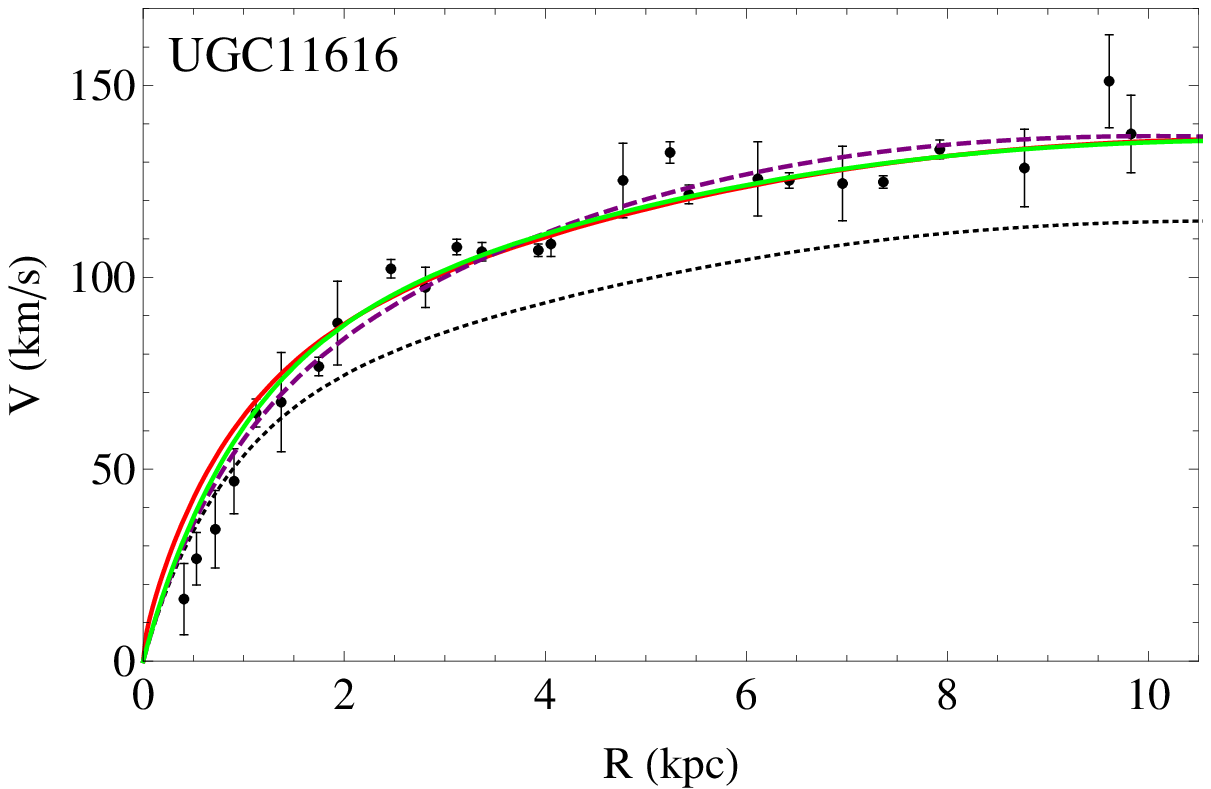}
\includegraphics[width=170pt,height=120pt]{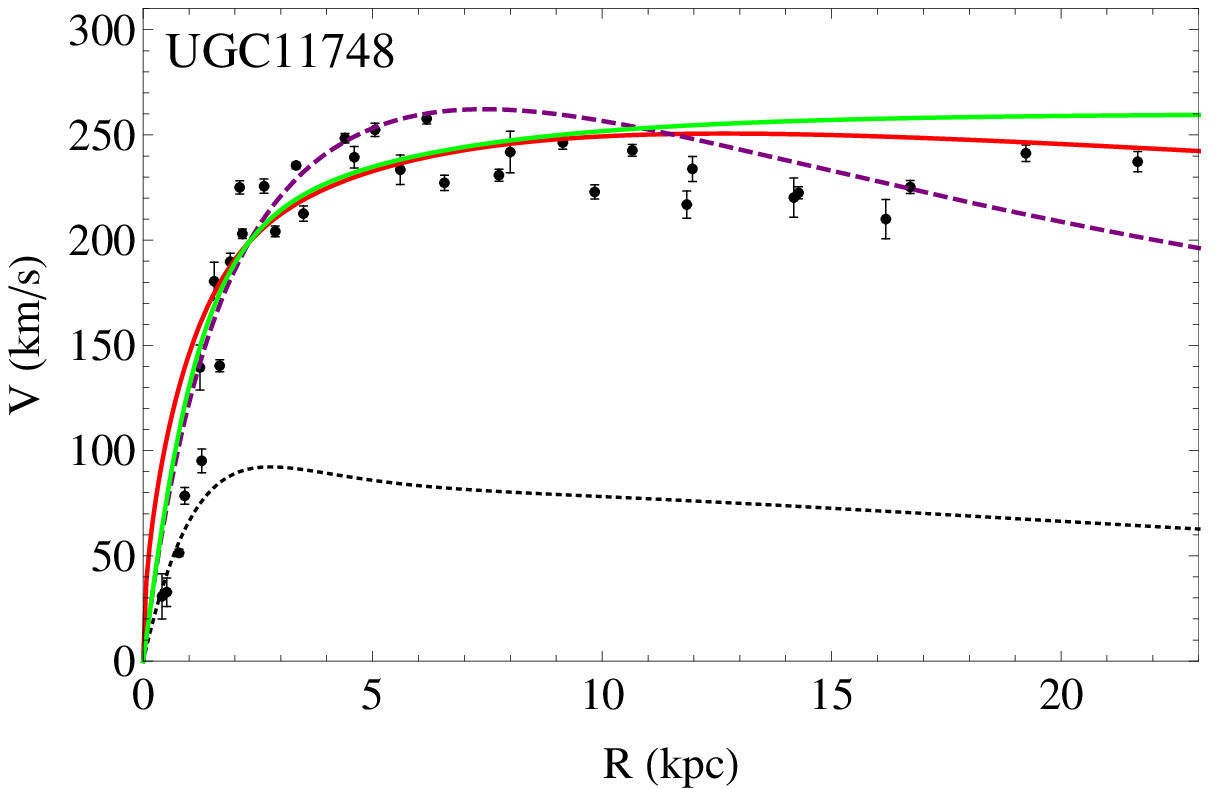}
\includegraphics[width=170pt,height=120pt]{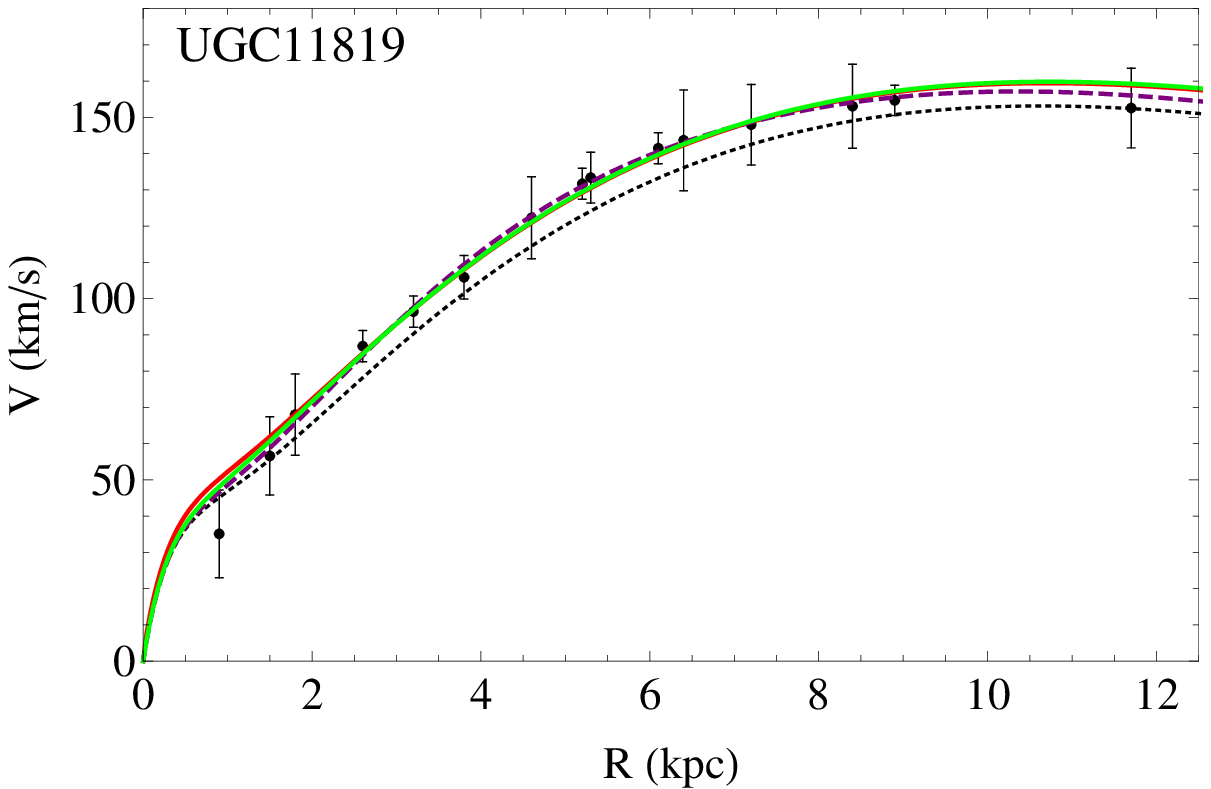}\newline
\caption{Best-fit rotational curves for the LSB galaxy sample. The dots with error bars denote archive rotational velocity curves derived from spectroscopic data. The fitted models are pure baryonic (black short-dashed curve),  baryonic + NFW (red continuous curve),  baryonic + Einasto (purple dashed curve), and  baryonic + PSE (green continuous curve).}
\label{fig:lsb_vrot}
\end{figure*}

We have explored a database of LSB galaxies taken from the literature as follows: the smoothed hybrid H$\alpha$-HI rotational velocity curves of the LSB galaxies F561-1, F563-1, F568-3, F583-1, F730-V1, UGC5750, UGC10310, UGC 11454, UGC11616, UGC11748, UGC11819, and
UGC6614 from \citet{deBlok2001b}, the HI rotational velocity curve of F579-V1 from \citet{deBlok1996}, the HI rotational velocity curves of UGC128 and UGC1230 from \citet{vanderHulst1993}. \citet{deBlok2001b} calculated the errors on the smoothed rotational curves as follows. The error bars consist of two components: (1) observational errors that are due to the measurement uncertainties in the individual raw data points, and (2) the differences between approaching and receding sides and noncircular motions. For the final error estimate, these two uncertainties were added quadratically. The original data for the approaching and receding sides are available in \citet{McGaugh2001}\footnote{http://astroweb.case.edu/ssm/data/RCHalpha.0701.dat}, showing only slight asymmetries. In the other four cases, no errors were published with the data. As the radial distribution of the velocities shows a quite regular pattern, we assumed for them $10$ percent errors. We have determined the best-fit parameters for each of the baryonic + NFW, baryonic + Einasto, baryonic + PSE models, and they are listed in Table \ref{table:lsb_vrot}. We present the rotational velocity curves of the best-fit models and the galaxy velocity curves in Fig. \ref{fig:lsb_vrot}.  

\begin{table*}
\caption{Parameters describing the best-fit pure baryonic, baryonic + NFW, baryonic + Einasto, and baryonic + PSE models of $15$ HSB galaxies. The rotational velocity data of the galaxies that were fit with the models are taken from \citet{Palunas2000}. When the best-fit is within the $1\sigma$ confidence level, the $\chi^2$-values are indicated in boldface.}
\label{table:hsb_vrot}
\resizebox{\textwidth}{!}{\begin{tabular}{lccccccccccccccc}
\hline
\hline
 & \multicolumn{3}{|c}{Baryonic} & \multicolumn{3}{|c}{NFW} & \multicolumn{4}{|c}{Einasto} & \multicolumn{3}{|c|}{PSE} & & \\
 \hline
ID & $\Upsilon_b$ & $\Upsilon_d$ & $\chi^2_\mathrm{B}$ & $\rho_s$ & $r_s$ & $\chi^2_\mathrm{NFW}$ & $\rho_e$ & $r_e$ & $n$ & $\chi^2_\mathrm{E}$ & $\rho_{0}$ & $r_c$ & $\chi^2_\mathrm{P}$ & $1\sigma_\mathrm{NFW,P}$& $1\sigma_\mathrm{E}$ \\
 &  &  &  & ($M_\odot \mathrm{kpc}^{-3}$) & ($\mathrm{kpc}$) & & ($M_\odot \mathrm{kpc}^{-3}$) & ($\mathrm{kpc}$) & & & ($M_\odot \mathrm{kpc}^{-3}$) & ($\mathrm{kpc}$) & & &\\
 \hline
ESO215G39 & 1.14 & 1.84 & 922.9 & 1.80E+07 & 10.74 & \textbf{5.76} & 8.04E+05 & 18.85 & 2.9 & \textbf{7.03} & 2.80E+08 & 1.26 & \textbf{6.59} & 36.3 & 35.24\\
ESO322G76 & 0.46 & 0.62 & 1064.8 & 3.74E+07 & 7.27 & \textbf{15.6} & 1.20E+04 & 120 & 6.1 & \textbf{13.62} & 5.80E+08 & 0.9 & \textbf{16.29} & 56.3 & 55.25\\
ESO322G77 & 0.93 & 1.96 & 179.15 & 1.89E+08 & 2.90 & \textbf{2.51} & 2.70E+04 & 76 & 6.5 & \textbf{3.09} & 2.15E+09 & 0.42 & \textbf{3.23} & 14.84 & 13.74\\
ESO322G82 & 1.28 & 1.69 & 663.215 & 1.37E+08 & 3.19 & 76.1 & 7.16E+05 & 15.97 & 4.01 & 88.1 & 2.01E+09 & 0.38 & 77.9 & 39.48 & 38.42\\
ESO323G25 & 1.47 & 2.05 & 324.71 & 2.80E+08 & 2.2 & \textbf{31.28} & 3.60E+03 & 120 & 9.6 & \textbf{29.01} & 7.40E+09 & 0.19 & \textbf{26.01} & 68.83 & 67.79\\
ESO374G02 & 0.84 & 1.15 & 1928.88 & 9.19E+07 & 5.35 & \textbf{37.4} & 7.83E+04 & 55.7 & 5.6 & \textbf{40.9} & 7.67E+08 & 0.90 & \textbf{37.6} & 88.58 & 87.54\\
ESO375G12 & 1.25 & 1.58 & 1052 & 2.02E+08 & 3.32 & \textbf{8.4} & 1.66E+06 & 13.98 & 3.58 & \textbf{14.0} & 1.54E+10 & 0.17 & \textbf{7.95} & 53.15 & 52.11\\
ESO376G02 & 0.93 & 1.48 & 2164.92 & 4.31E+07 & 6.09 & 200.2 & 5.33E+06 & 8.28 & 1.74 & 162.1 & 3.43E+08 & 1.09 & 158.8 & 63.61 & 62.57\\
ESO383G02 & 1.08 & 1.56 & 564.68 & 8.48E+07 & 4.5 & \textbf{5.93} & 1.62E+06 & 13.36 & 2.92 & \textbf{5.84} & 1.41E+09 & 0.54 & \textbf{6.69} & 49.64 & 41.59\\
ESO383G88 & 1.43 & 2.32 & 616.92 & 7.54E+07 & 2.94 & 57.2 & 6.01E+06 & 5.32 & 1.5 & 67.2 & 9.06E+08 & 0.40 & \textbf{53.0} & 53.15 & 52.11\\
ESO445G19 & 0.92 & 1.47 & 363.27 & 2.35E+07 & 8.23 & \textbf{3.56} & 4.60E+04 & 60 & 4.9 & \textbf{0.33} & 2.78E+08 & 1.16 & \textbf{4.26} & 40.53 & 39.48\\
ESO446G01 & 0.78 & 1.05 & 735.53 & 3.56E+07 & 8.38 & \textbf{11.88} & 8.70E+03 & 150 & 6.5 & \textbf{10.53} & 4.72E+08 & 1.11 & \textbf{10.31} & 46.85 & 45.80\\
ESO502G02 & 1.72 & 2.64 & 33.0 & 1.26E+08 &0.42 & \textbf{32.9} & 4.70E+07 & 1.2 & 0.08 & \textbf{31.8} & 8.43E+06 & 1.49 & \textbf{30.1} & 43.7 & 42.64\\
ESO509G80 & 1.37 & 1.94 & 1124.27 & 1.83E+07 & 14.41 & \textbf{8.30} & 1.30E+06 & 22 & 2.4 & \textbf{7.66} & 1.85E+08 & 2.24 & \textbf{7.42} & 39.48 & 38.42\\
ESO569G17 & 0.74 & 1.15 & 126.62 & 1.10E+07 & 13.03 & \textbf{4.64} & 1.30E+06 & 11 & 2.9 & \textbf{7.82} & 2.29E+08 & 1.21 & \textbf{5.06} & 21.36 & 20.28\\
\hline
\end{tabular}}
\end{table*}

\begin{table*}
\caption{Parameters describing the best-fit pure baryonic, baryonic + NFW, baryonic + Einasto, and baryonic + PSE models of $15$ LSB galaxies. The rotational velocity data of the galaxies are taken from: $^{1}$\citet{deBlok2001b}, $^{2}$\citet{deBlok1996}, $^3$\citet{vanderHulst1993}. The superscript $^\star$ singles out galaxies, for which a bulge component fully describes the surface brightness density. Whenever the best-fit is within the $1\sigma$ confidence level, the $\chi^2$-values are indicated in boldface.}
\label{table:lsb_vrot}
\resizebox{\textwidth}{!}{\begin{tabular}{lccccccccccccccc}
\hline
\hline
 & \multicolumn{3}{|c}{Baryonic} & \multicolumn{3}{|c}{NFW} & \multicolumn{4}{|c}{Einasto} & \multicolumn{3}{|c|}{PSE} & & \\
 \hline
ID & $\Upsilon_b$ & $\Upsilon_d$ & $\chi^2_\mathrm{B}$ & $\rho_s$ & $r_s$ & $\chi^2_\mathrm{NFW}$ & $\rho_e$ & $r_e$ & $n$ & $\chi^2_\mathrm{E}$ & $\rho_{0}$ & $r_c$ & $\chi^2_\mathrm{P}$ & $1\sigma_\mathrm{NFW,P}$& $1\sigma_\mathrm{E}$\\
 &  &  &  & ($M_\odot \mathrm{kpc}^{-3}$) & ($\mathrm{kpc}$) & & ($M_\odot \mathrm{kpc}^{-3}$) & ($\mathrm{kpc}$) & & & ($M_\odot \mathrm{kpc}^{-3}$) & ($\mathrm{kpc}$) & & &\\
 \hline
F561-1$^1$ & 0.85 & 1.57 & 249.32 & 1.50E+06 & 9.04 & 7.06 & 3.53E+06 & 3.57 & 0.39 & \textbf{1.13} & 1.60E+07 & 1.40 & \textbf{4.10} & 4.72 & 3.53\\
F563-1$^{1 \star}$ & 1.55 & x & 57.49 & 9.90E+04 & 160 & \textbf{1.39} & 1.80E+06 & 11.54 & 0.14 & \textbf{0.51} & 3.36E+06 & 9.73 & \textbf{0.84} & 8.18 & 7.04\\
F568-3$^1$ & 1.21 & 2.68 & 103.32 & 2.98E+05 & 80.24 & 14.03 & 1.00E+07 & 4.70 & 0.13 & \textbf{1.71} & 1.94E+07 & 2.92 & \textbf{6.52} & 9.3 & 8.18\\
F579-V1$^2$ & 2.03 & 3.08 & 278.90 & 7.24E+07 & 3.13 & \textbf{2.88} & 1.87E+06 & 8.13 & 2.90 & \textbf{2.22} & 8.05E+08 & 0.46 & \textbf{1.75} & 12.64 & 11.54\\
F583-1$^{2 \star}$ & 0.81 & x & 683.03 & 8.67E+05 & 38.57 & \textbf{10.82} & 2.03E+06 & 9.22 & 0.99 & \textbf{0.43} & 2.60E+07 & 2.76 & \textbf{0.89} & 15.94 & 14.84\\
F730-V1$^1$ & 1.20 & 2.80 & 423.92 & 1.30E+07 & 9.83 & \textbf{5.48} & 3.76E+06 & 8.03 & 1.67 & \textbf{0.59} & 1.98E+08 & 1.21 & \textbf{1.00} & 5.89 & 4.72\\
UGC128$^3$  & 1.37 & 4.89 & 123.49 & 3.70E+04 & 420 & \textbf{8.15} & 1.88E+05 & 37.49 & 1.48 & \textbf{1.58} & 6.95E+06 & 6.74 & \textbf{0.95} & 10.42 & 9.3\\
UGC1230$^3$  & 1.12 & 5.60 & 146 & 4.20E+04 & 290 & \textbf{4.49} & 1.61E+05 & 34.17 & 1.38 & \textbf{0.70} & 4.53E+06 & 7.19 & \textbf{0.69} & 9.3 & 8.18\\
UGC5750$^1$ & 1.16 & 3.51 & 15.39 & 1.20E+03 & 2000 & \textbf{7.79} & 4.70E+05 & 19 & 0.15 & \textbf{3.98} & 5.39E+05 & 40.97 & \textbf{4.02}& 9.3 & 8.18\\
UGC6614$^1$ & 2.05 & 3.73 & 2229.43 & 6.4E+06 & 24.0 & 81.3 & 1.14E+05 & 59.89 & 3.09 & 83.42 & 9.36E+07 & 2.50 & 75.06 & 12.64 & 11.54\\
UGC10310$^1$ & 0.83 & 3.22 & 173 & 4.1E+04 & 210 & 62.02 & 1.7E+06 & 38 & 0.99 & 108.48 & 1.3E+07 & 1.8E+04 & 35.53 & 15.94 & 14.84\\
UGC11454$^1$ & 1.1 & 0.82 & 13330 & 8.6E+06 & 13 & 125.16 & 1.64E+06 & 12.49 & 1.91 & 124.43 & 1.3E+08 & 1.6 & 98.37 & 26.73 & 25.66  \\
UGC11616$^1$ & 0.69 & 3.99 & 846 & 5.3E+06 & 9.7 & 80.86 & 3.1E+06 & 6.7 & 0.89 & 86.31 & 6.1E+07 & 1.4 & 68.54 & 24.59 & 23.51\\
UGC11748$^1$ & 0.74 & 3.62 & 57534 & 1.17E+08 & 6.47 & 2775 & 3.05E+07 & 6.57 & 1.33 & 2752 & 1.01E+09 & 1.12 & 2627 & 34.18 & 33.12\\
UGC11819$^1$ & 1.43 & 2.01 & 19.23 & 3.18E+06 & 7.39 & \textbf{3.54} & 4.41E+06 & 3.82 & 0.35 & \textbf{2.05} & 2.34E+07 & 1.43 & \textbf{3.09} & 13.74 & 12.64\\
\hline
\end{tabular}}
\end{table*}

\section{Relevance of dark matter models}
\label{stat_ranking}

\subsection{Comparison of dark matter models}

Since the dark matter models are not submodels of each other, comparing them by exact statistical tests, such as the likelihood ratio test, is not conclusive. Instead we applied the Akaike information criterion $\textrm{AIC}=2N+\chi^2$ \citep{Akaike1974} that is usually employed in the literature \citep[e.g.,][]{Chemin2011}.
The AIC is a measure of the fit of a given model to the statistical data based on both the residual sum of squares $\chi^2$ and the number of parameters $N$. A lower AIC value represents a better model performance. In Table \ref{table:gx_vrot_stat} the columns $\textrm{AIC}_{\textrm{NFW}}$, $\textrm{AIC}_{\textrm{E}}$, and $\textrm{AIC}_{\textrm{P}}$ are the calculated $\textrm{AIC}$ values for the baryonic + NFW, the baryonic + Einasto, and the baryonic + PSE model, respectively, for those galaxies for which at least one model fits the dataset within the $1\sigma$ confidence level. The lowest values are marked in boldface.

A model is worse than the best-fit model when the difference $\Delta$ of the corresponding $\textrm{AIC}$ values is higher. These differences are presented in the last three columns of Table \ref{table:gx_vrot_stat}. We establish the following thresholds: $\Delta\leq 2$ refers to approximately equal performances, $4\leq\Delta\leq 7$ represents a measurable difference in the fit of the two models, while $\Delta>10$ clearly favors one of the models over the other. We stress, however, that the imposition of these limits is somewhat arbitrary. For related considerations see Chapter $2$ of \citet{Burnham2003}.

For $7$ galaxies (2 HSB, and 5 LSB) neither the pure baryonic nor any of the dark matter models fit the dataset within the $1\sigma$ confidence level. In order to identify the performances of the best-fit models in the case of the other $23$ galaxies, the $\Delta$-values are listed in the last three columns of Table \ref{table:gx_vrot_stat}, and their interpretation is summarized in Table \ref{table:gx_vrot_disruletable}. The best-fit PSE model is favored in $\Sigma_{++}+\Sigma_{+}=23$ cases (14 HSB, and 9 LSB) and cannot be ruled out in either case. The best-fit NFW is favored in $\Sigma_{++}+\Sigma_{+}=13$ cases (10 HSB, and
3 LSB), and is ruled out in $\Sigma_{--}=2$ cases ($1$ HSB, and
$1$ LSB). The best-fit Einasto model is favored in $\Sigma_{++}+\Sigma_{+}=10$ cases (3 HSB, and 7 LSB), and is ruled out in $\Sigma_{--}=1$ cases ($1$ HSB).

\subsection{Pure baryonic model with M/L ratio fit}

The rotational curves cannot be explained by the pure baryonic matter with the $M/L$ ratios derived in Section \ref{estimated_ml}. However, the estimation of $M/L$ depends on the applied stellar population model and the IMF. Previously, we used the self-consistent model of \citet{Bell2003}. 

Now we compare the pure baryonic model with fitted $M/L$ to the models that contain baryonic matter with fixed $M/L$ and dark matter. For this purpose we employ lower and upper limits on the $M/L$ ratios based on the nine different models of \citet{Bell2001} and the corresponding CMLRs, tabulated for different color indices, and for cosmological redshifts $z=0.008$ and $z=0.02$. We used the CMLRs evaluated at $z=0.02$, as this redshift applies for our galaxy sample. In all cases the CMLR based on the Bruzal \& Charlot population synthesis model with a modified Salpeter IMF gave the lower limits, and the CMLR based on the PEGASE model with $x=-1.85$ IMF gave the upper limits to the $M/L$ ratios. These $M/L$ ratios (separately for the bulge and the disk) are summarized in Table \ref{table:comp_ml_ratios}, from columns 5 to 8. The best-fit values of the $M/L$ ratios are presented in columns $9$ and $10$ of the same table. In Table \ref{table:comp_ml_ratios} we also present the $M/L$ ratios of the baryonic+dark matter models (in columns $3$ and $4$) in order to compare the best-fit $M/L$ ratios of the pure baryonic model to them.
 
\begin{figure*}
\centering
\includegraphics[width=170pt,height=120pt]{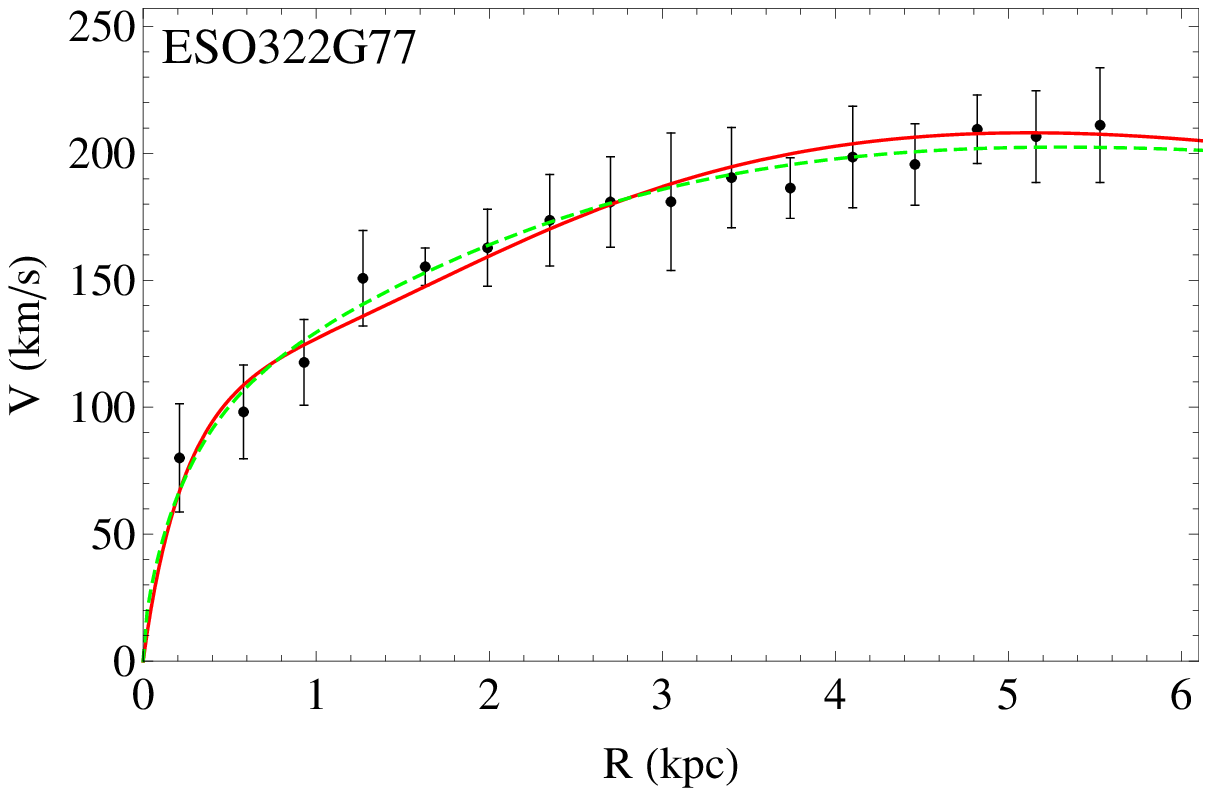}
\includegraphics[width=170pt,height=120pt]{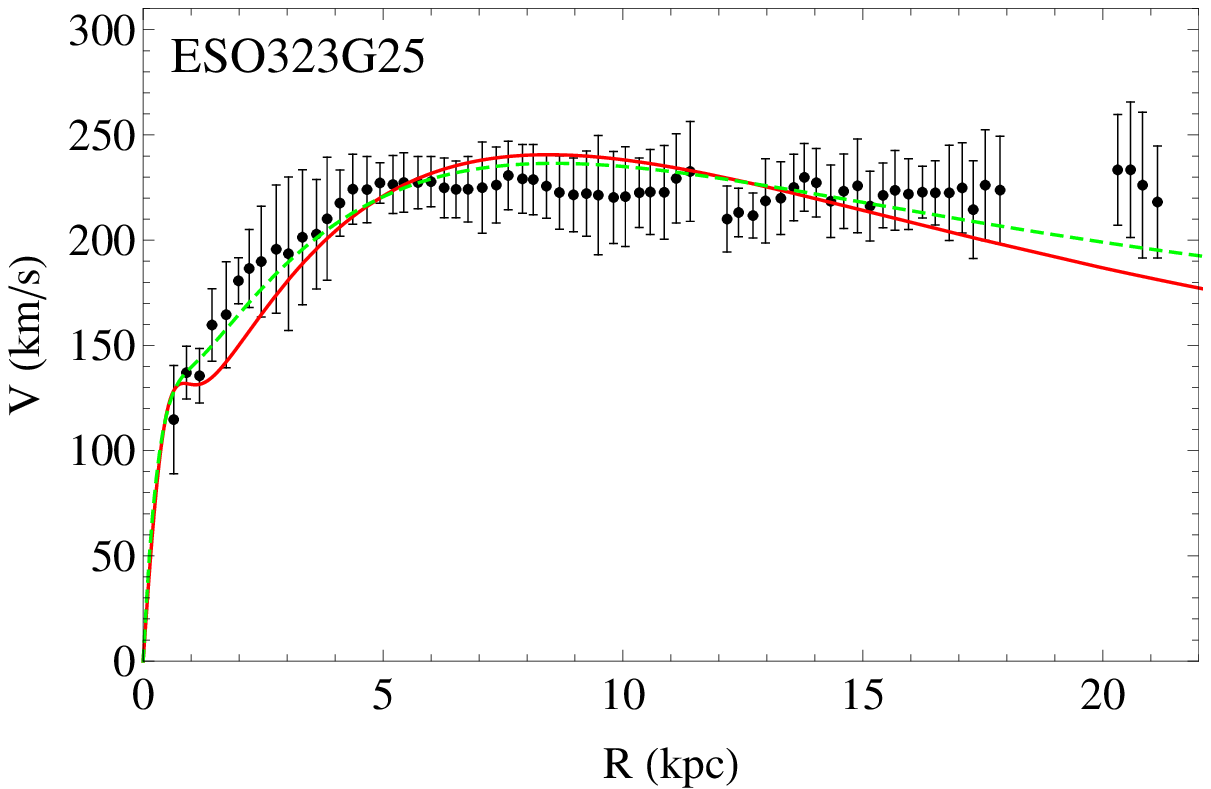}
\includegraphics[width=170pt,height=120pt]{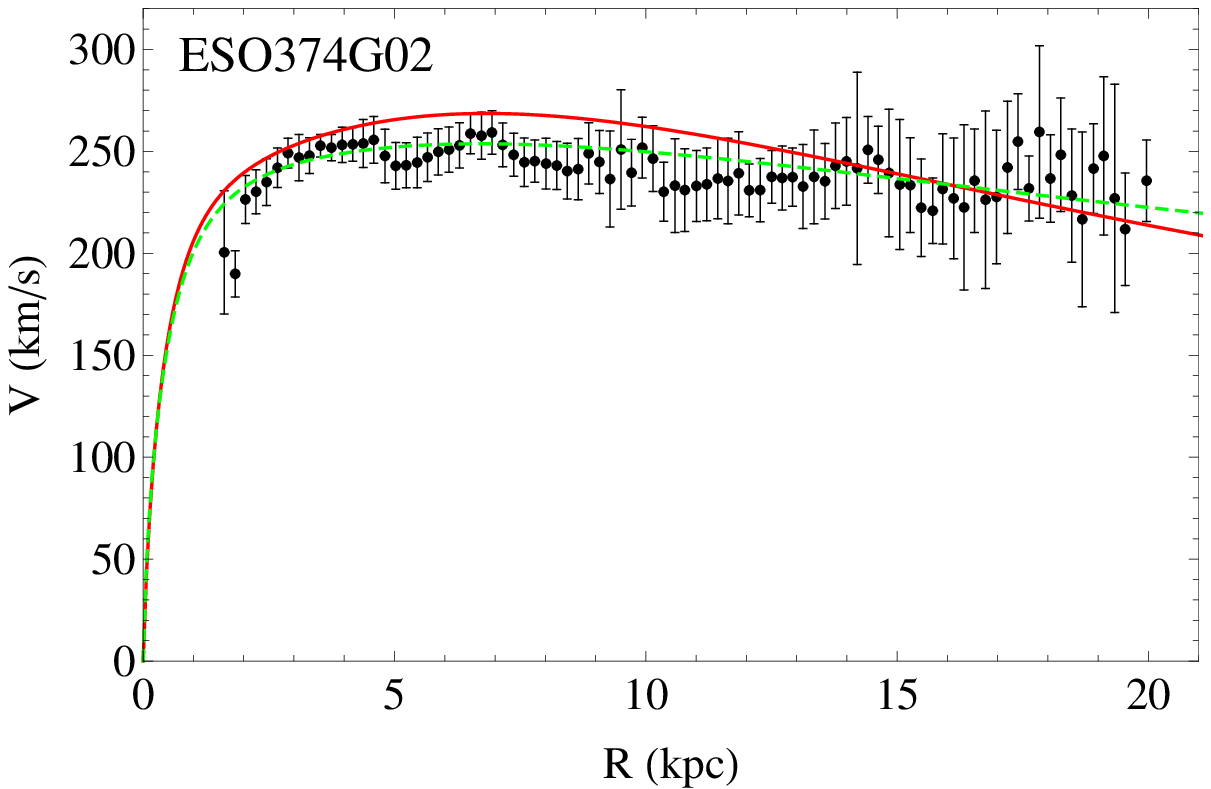}\newline
\includegraphics[width=170pt,height=120pt]{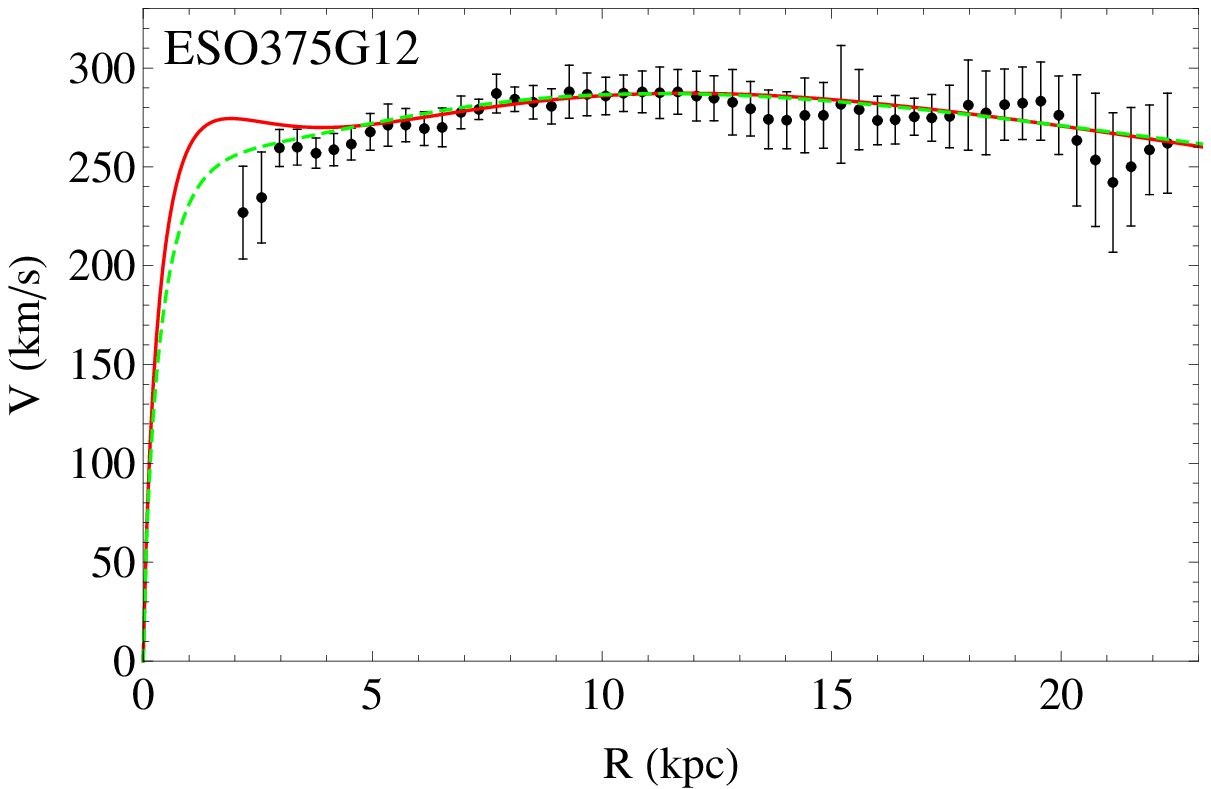}
\includegraphics[width=170pt,height=120pt]{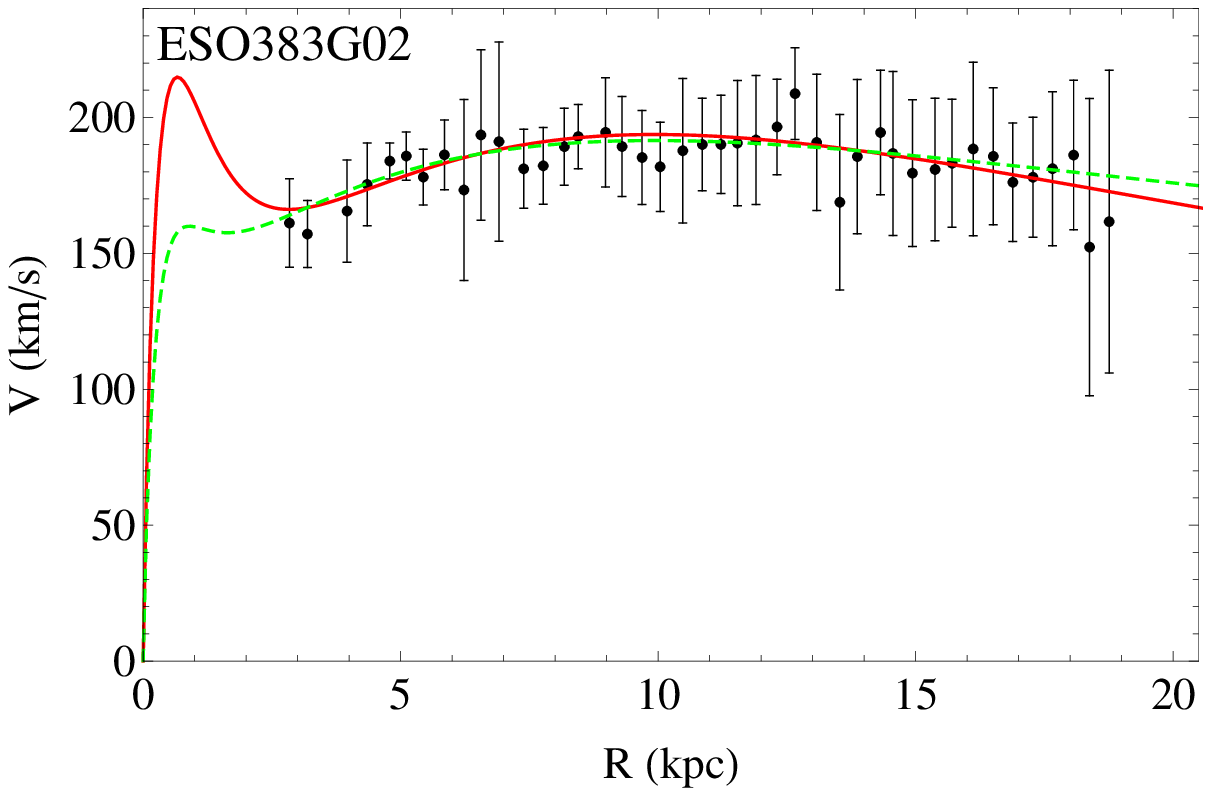}
\includegraphics[width=170pt,height=120pt]{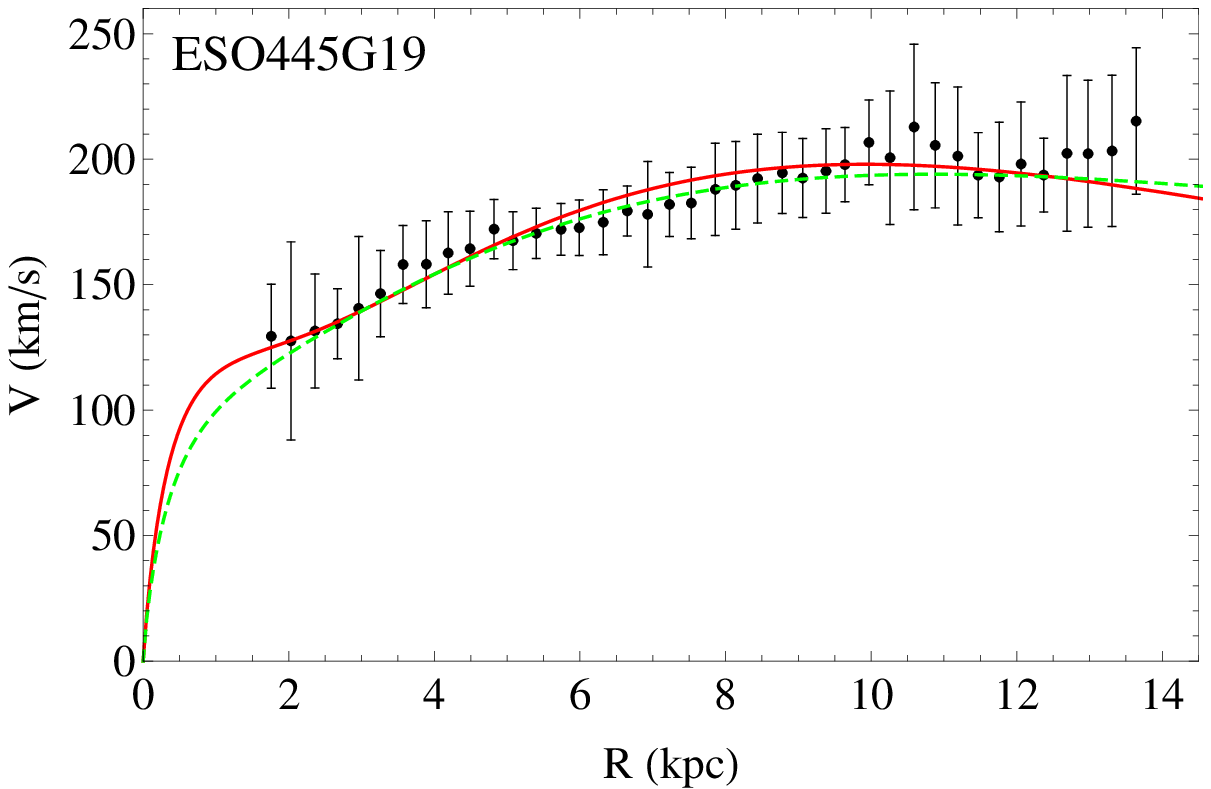}\newline
\includegraphics[width=170pt,height=120pt]{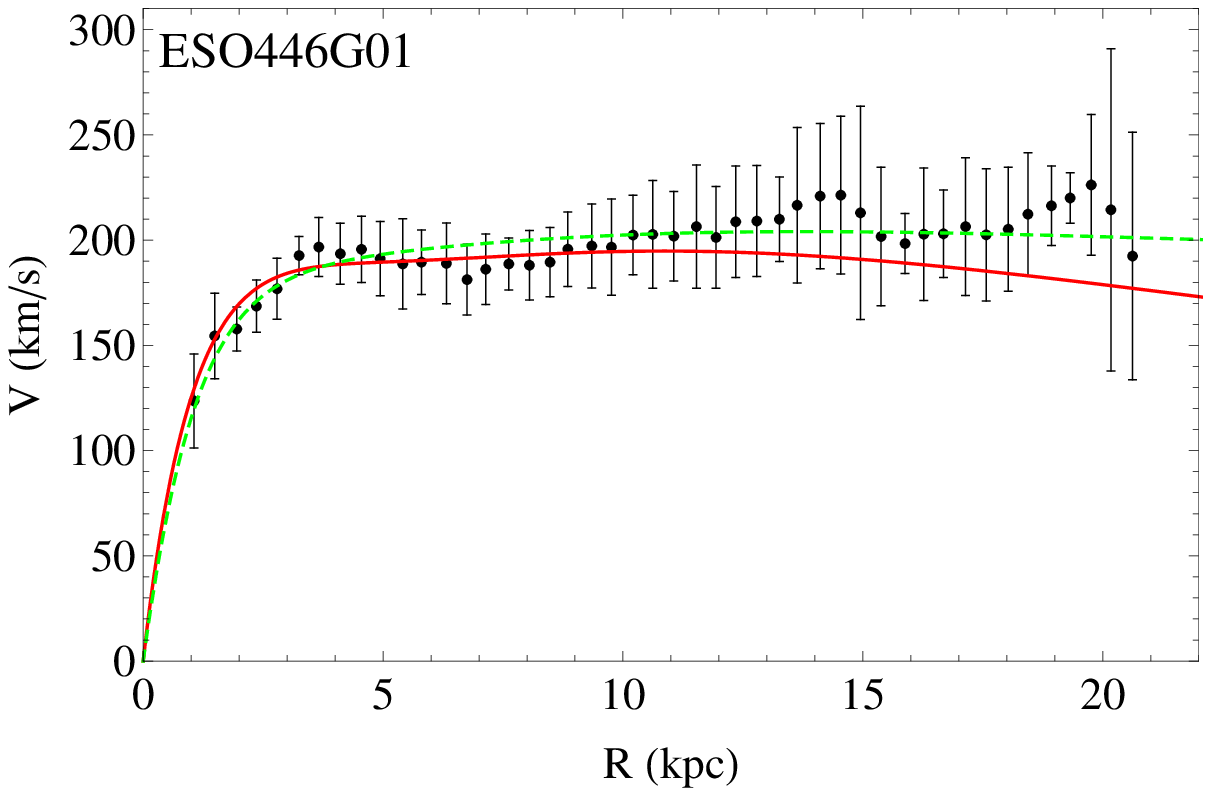}
\includegraphics[width=170pt,height=120pt]{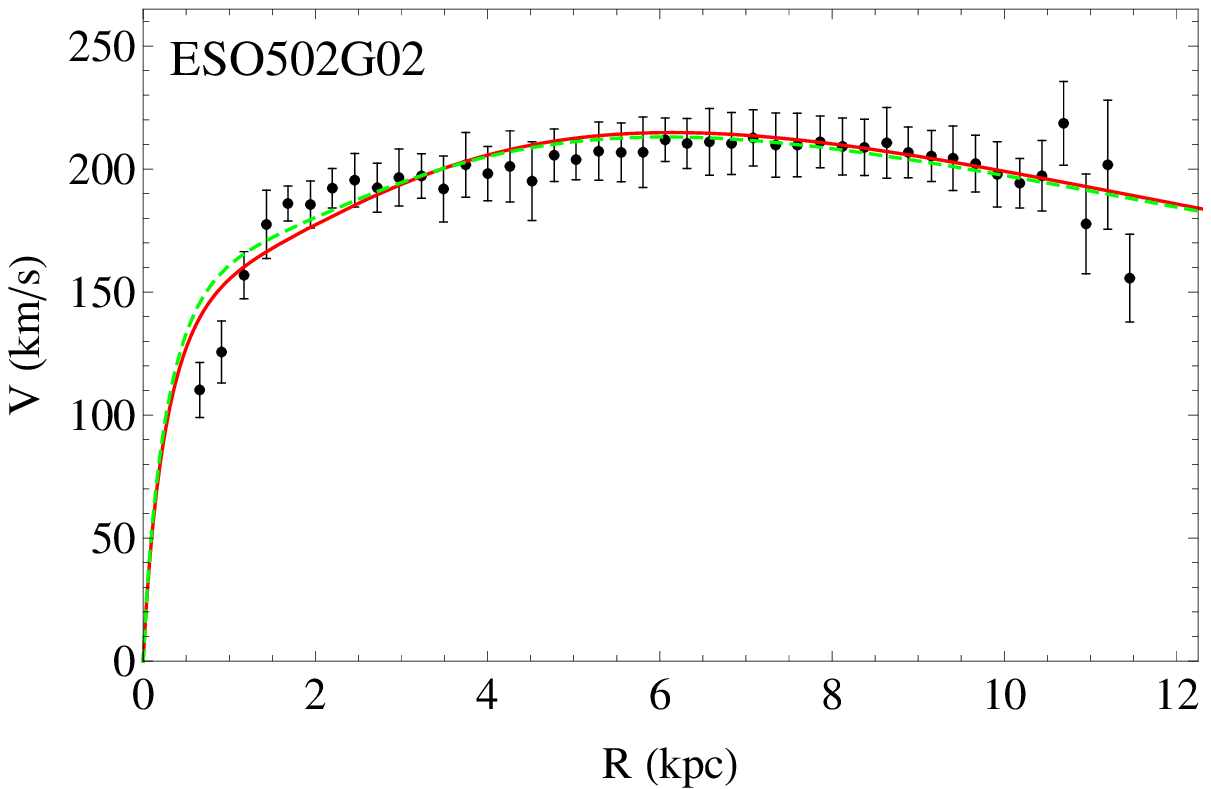}
\includegraphics[width=170pt,height=120pt]{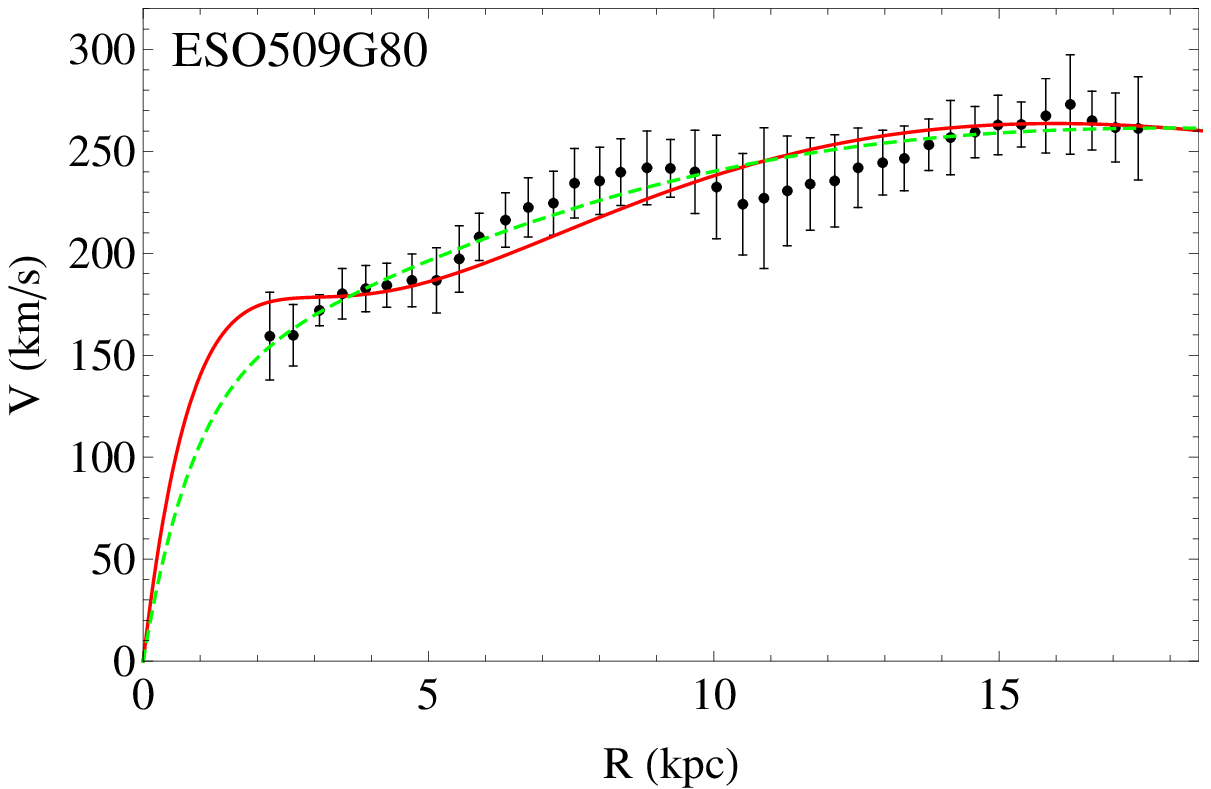}\newline
\includegraphics[width=170pt,height=120pt]{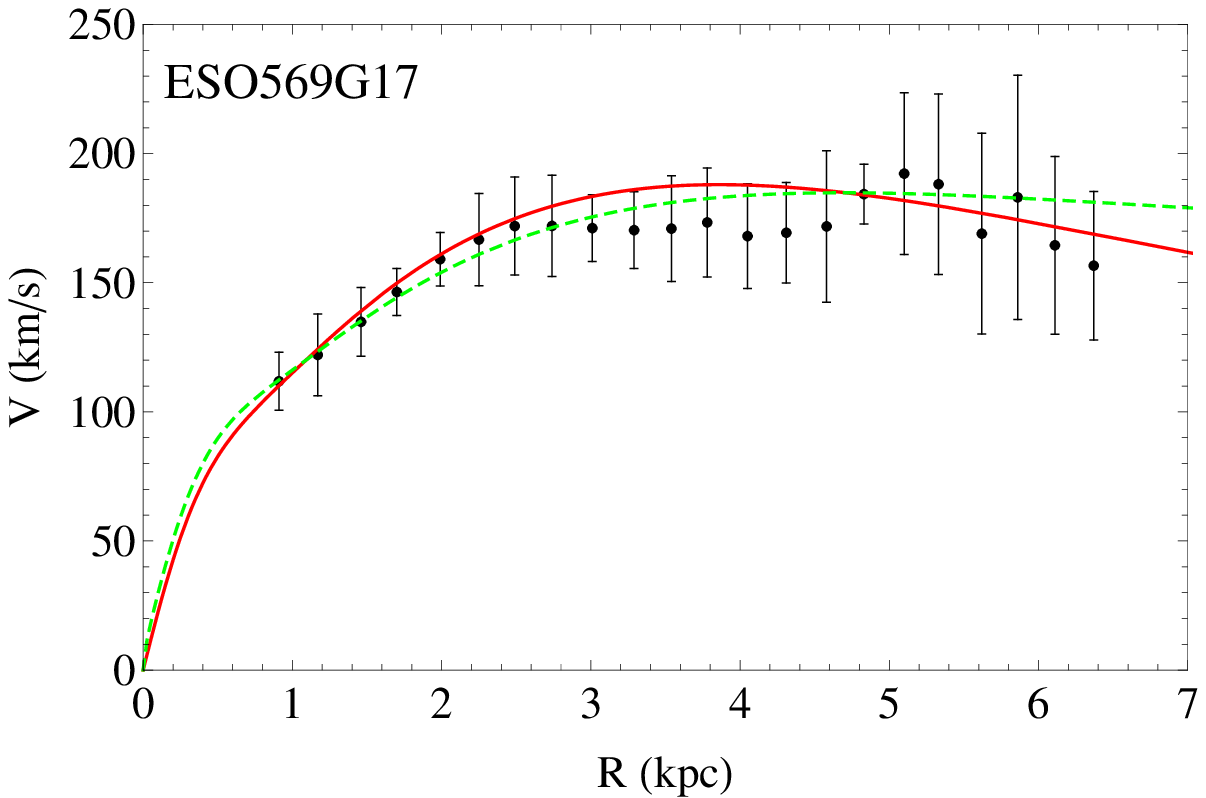}
\includegraphics[width=170pt,height=120pt]{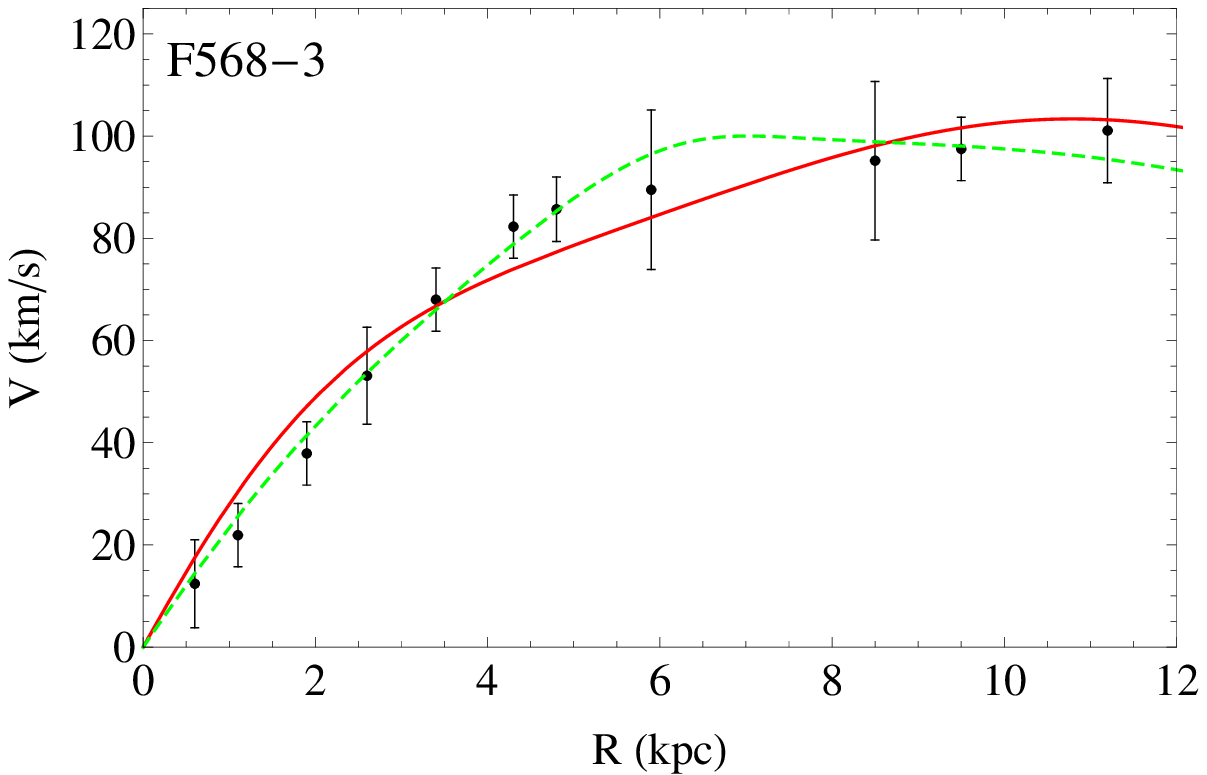}
\includegraphics[width=170pt,height=120pt]{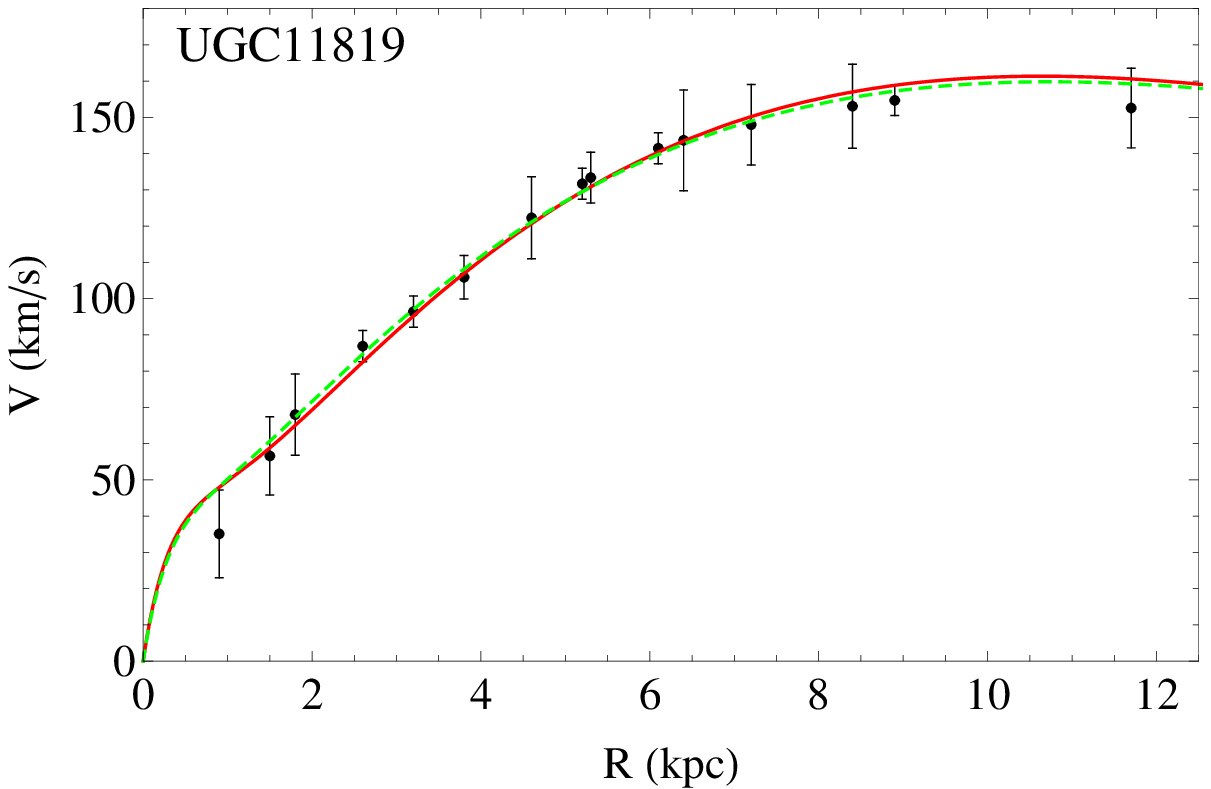}\newline
\caption{Best-fit rotational curves for the HSB and LSB galaxies for which the $\chi^2$ values with the fit $M/L$ ratios are within the $1\sigma$ confidence level of the fitting. The best-fit pure baryonic models are indicated by solid red curves, and the best-fit models composed of baryonic matter with fixed $M/L$ and dark matter are indicated by dashed green curves.}
\label{fig:hsblsb_vrotfitml}
\end{figure*}

Columns $9$ and $10$ of Table \ref{table:comp_ml_ratios} show the best-fit $M/L$ ratios for the bulge and the disk, respectively. In Fig. \ref{fig:hsblsb_vrotfitml} we represent the pure baryonic model fits with rotational curve data within the $1\sigma$ confidence level. We found goods fits like this for ten HSB and two LSB galaxies. Although the fitting is within $1\sigma$ for the HSB galaxies ESO323G25, ESO374G02, ESO445G19, and ESO446G01,  the pure baryonic model still does not reproduce the plateau as well
as the best-fit dark matter models, which are also indicated in Fig. \ref{fig:hsblsb_vrotfitml}. This appears most clearly for ESO446G01.

Columns $2$ and $6$ of Table \ref{table:gx_vrot_stat} show the AIC$_B$ and $\Delta$-value (i.e., the difference of AIC$_B$ to
the best-fit AIC), respectively. The second column of Table \ref{table:gx_vrot_disruletable} indicates the relevance (based on the AIC) of the pure baryonic model that is favored in $\Sigma_{++}+\Sigma_{+}=2$ cases (1 HSB, and 1 LSB), which is ruled out in $\Sigma_{--}=13$ cases ($7$ HSB, and $6$ LSB).

\begin{table*}
\centering
\caption{Best-fit (BF) dark matter models in column 2, where N/E/P marks hold for the Navarro-Frank-White/Einasto/pseudo-isothermal sphere models, and the estimated $M/L$ ratios of the baryonic+dark matter models ($\Upsilon_\mathrm{b}$ and $\Upsilon_\mathrm{d}$ for the bulge and disk, respectively) in columns $3$ and $4$. Columns $5$ to $8$ show the lower ($\Upsilon_\mathrm{b,min}$, $\Upsilon_\mathrm{d,min}$) and the upper limits ($\Upsilon_\mathrm{b,max}$, $\Upsilon_\mathrm{d,max}$) of the $M/L$ ratios for the bulge and disk, respectively, derived from the CMLR relations given in \citet{Bell2001}. The best-fit values of the bulge and disk $M/L$ ratios ($\Upsilon_\mathrm{b,bf}$, $\Upsilon_\mathrm{d,bf}$) are given in columns $9$ and $10$, respectively, obtained from fitting the pure baryonic model to the rotational curves. The $\chi^2$ and the $1\sigma$ confidence intervals are given in columns $11$ and $12$. When the fit is within the $1\sigma$ confidence level, the $\chi^2$-values are indicated in boldface.}
\label{table:comp_ml_ratios}
\begin{tabular}{lccc|cccccccc}
\hline
\hline
ID & \multicolumn{3}{c|}{Dark matter} & \multicolumn{8}{c}{Baryonic}\\
\hline
& BF & $\Upsilon_\mathrm{b}$ & $\Upsilon_\mathrm{d}$ & $\Upsilon_\mathrm{b,min}$ & $\Upsilon_\mathrm{b,max}$ & $\Upsilon_\mathrm{d,min}$& $\Upsilon_\mathrm{d,max}$& $\Upsilon_\mathrm{b,bf}$ & $\Upsilon_\mathrm{d,bf}$ & $\chi^2_\mathrm{B}$ & $1\sigma$ \\
\hline
ESO215G39 & E & 1.14 & 1.84 & 0.88 & 3.35 & 1.43 & 5.42 & 3.35 & 5.42 & 203.39 & 36.3\\
ESO322G76 & E & 0.46 & 0.62 & 0.32 & 1.43 & 0.43 & 1.95 & 1.43 & 1.95 & 57.9 & 56.3\\
ESO322G77 & E & 0.93 & 1.96 & 0.66 & 2.61 & 1.39 & 5.49 & 2.6 & 3.7 & \textbf{4.72} & 14.84\\
ESO322G82 & B & 1.28 & 1.69 & 0.92 & 3.43 & 1.22 & 4.54 & 2.6 & 2.4 & 75.47 & 39.48\\
ESO323G25 & P & 1.47 & 2.05 & 1.06 & 3.85 & 1.48 & 5.38 & 3.7 & 2.8 & \textbf{52.8} & 68.83\\
ESO374G02 & N & 0.84 & 1.15 & 0.59 & 2.4 & 0.81 & 3.28 & 1.1 & 2.9 & \textbf{49.84} & 88.58\\
ESO375G12 & P & 1.25 & 1.58 & 0.99 & 3.7 & 1.25 & 4.65 & 2.1 & 2.2 & \textbf{12.75} & 53.15\\
ESO376G02 & P & 0.93 & 1.48 & 0.69 & 2.74 & 1.1 & 4.34 & 1.1 & 2.9 & 190.42 & 63.61\\
ESO383G02 & N & 1.08 & 1.56 & 0.77 & 2.96 & 1.12 & 4.3 & 2.6 & 3.5 & \textbf{7.06} & 49.64\\
ESO383G88 & P & 1.43 & 2.32 & 1.03 & 3.77 & 1.68 & 6.11 & 3.7 & 3 & 87.33 & 53.15\\
ESO445G19 & E & 0.92 & 1.47 & 0.65 & 2.58 & 1.04 & 4.13 & 2.4 & 2.7 & \textbf{5.63} & 40.53\\
ESO446G01 & P & 0.78 & 1.05 & 0.55 & 2.25 & 0.74 & 3.04 & 1.6 & 3 & \textbf{25.77} & 46.85\\
ESO502G02 & P & 1.72 & 2.64 & 1.25 & 4.4 & 1.92 & 6.76 & 1.55 & 2.78 & \textbf{32.14} & 43.7\\
ESO509G80 & P & 1.37 & 1.94 & 0.98 & 3.62 & 1.4 & 5.14 & 3.2 & 4.9 & \textbf{19.16} & 39.48\\
ESO569G17 & N & 0.74 & 1.15 & 0.52 & 2.16 & 0.81 & 3.34 & 0.64 & 1.8 & \textbf{8.84} & 21.36\\
\hline
F561-1 & E & 0.85 & 1.57 & 0.77 & 3.19 & 1.43 & 5.91 & 1.1 & 3.2 & 75.53 & 4.72\\
F563-1 & P & 1.55 & - & 1.25 & 4.94 & - & - & 2.7 & - & 10.77 & 8.18\\
F568-3 & E & 1.21 & 2.68 & 1.02 & 4.12 & 2.27 & 9.15 & 2.8 & 5.3 & \textbf{8.93} & 9.3\\
F579-V1 & P & 2.03 & 3.08 & 1.54 & 5.99 & 2.34 & 9.08 & 5.99 & 9.07 & 39.87 & 12.64\\
F583-1 & P & 0.81 & - & 0.74 & 3.08 & - & - & 3.08 & - & 291.53 & 15.94\\
F730-V1 & P & 1.2 & 2.8 & 0.31 & 4.27 & 0.71 & 9.96 & 4.27 & 7.4 & 22.3 & 5.89\\
UGC128 & P & 1.37 & 4.89 & 1.13 & 4.51 & 4.03 & 16.1 & 1.13 & 11.64 & 17.6 & 10.42\\
UGC1230 & P & 1.12 & 5.6 & 0.96 & 3.9 & 4.8 & 19.49 & 0.96 & 11.46 & 62.57 & 9.3\\
UGC5750 & P & 1.16 & 3.51 & 0.99 & 4 & 2.99 & 12.12 & 0.99 & 3.59 & 14 & 9.3\\
UGC6614 & P & 2.05 & 3.73 & 0.31 & 6.33 & 0.56 & 11.5 & 5.31 & 11.5 & 182.6 & 12.64\\
UGC10310 & P & 0.83 & 3.22 & 0.3 & 3.26 & 1.18 & 12.64 & 0.3 & 1.4 & 57.69 & 15.94\\
UGC11454 & P & 1.1 & 0.82 & 0.31 & 8.85 & 0.35 & 9.94 & 1.1 & 0.82 & 142.59 & 26.73\\
UGC11616 & P & 0.69 & 3.99 & 0.31 & 5.54 & 0.52 & 9.33 & 0.88 & 0.63 & 90.15 & 24.59\\
UGC11748 & P & 0.74 & 3.62 & 0.3 & 3 & 1.48 & 14.63 & 3 & 14.63 & 12934 & 34.18\\
UGC11819 & P & 1.43 & 2.01 & 0.31 & 4.86 & 0.43 & 6.84 & 1.63 & 2.22 & \textbf{4.52} & 13.74\\
\hline
\end{tabular}
\end{table*}

\begin{table*}
\centering
\caption{Akaike information criterion of the pure baryonic model with the $M/L$ ratio fit (AIC$_\mathrm{B}$), baryonic model with estimated $M/L$ ratio combined with the NFW (AIC$_\mathrm{NFW}$), the Einasto (AIC$_\mathrm{E}$), and the PSE dark matter models (AIC$_\mathrm{P}$). The number of the fitted parameters are $2$, $2$, $3$, and $2$, respectively. We indicate the smallest AIC number for each galaxy in boldface, and give the $\Delta$ values.}
\label{table:gx_vrot_stat}
\begin{tabular}{lcccccccc}
\hline
\hline
ID  & AIC$_\mathrm{B}$ & AIC$_\mathrm{NFW}$ & AIC$_\mathrm{E}$ & AIC$_\mathrm{P}$ & $\Delta_\mathrm{B}$ & $\Delta_\mathrm{NFW}$ & $\Delta_\mathrm{E}$ & $\Delta_\mathrm{P}$ \\
\hline
ESO215G39 & 207.39 & $\mathbf{9.76}$ & 13.03 & 10.59 & 197.63 & 0 & 3.27 & 0.83\\
ESO322G76 & 61.90  &$\mathbf{19.60}$ & 19.62 & 20.29 & 42.3 & 0 & 0.02 & 0.69\\
ESO322G77 & 8.72 & $\mathbf{6.51}$ & 9.09 & 7.23 & 2.21 & 0 & 2.58 & 0.72\\
ESO323G25 & 56.80 & 35.28 & 35.01 & $\mathbf{30.01}$ & 26.79 & 5.27 & 5 & 0\\
ESO374G02 & 53.84 & $\mathbf{41.4}$ & 46.9 & 41.6 & 12.44 & 0 & 5.5 & 0.2\\
ESO375G12 & 16.75 & 12.4 & 20 & $\mathbf{11.95}$ & 4.8 & 0.45 & 8.05 & 0.0\\
ESO383G02 & 11.06 & $\mathbf{9.63}$ & 11.84 & 10.69 & 1.13 & 0 & 1.91 & 0.76\\
ESO383G88 & 91.33 & 61.2 & 73.2 & $\mathbf{57}$ & 34.33 & 4.2 & 16.2 & 0\\
ESO445G19 & 9.63 & 7.56 & $\mathbf{6.33}$ & 8.26 & 3.3 & 1.23 & 0 & 1.93\\
ESO446G01 & 29.77 & 15.88 & 16.53 & $\mathbf{14.31}$ & 15.46 & 1.57 & 2.22 & 0\\
ESO502G02 & 36.14 & 36.9 & 37.8 & $\mathbf{34.1}$ & 2.04 & 2.8 & 3.7 & 0.0\\
ESO509G80 & 23.16 & 12.30 & 13.66 & $\mathbf{11.42}$ & 11.74 & 0.88 & 2.24 & 0\\
ESO569G17 & 12.84 & $\mathbf{8.64}$ & 13.82 & 9.06 & 4.2 & 0 & 5.18 & 0.42\\
\hline
F561-1 & 79.53 & 11.06 & $\mathbf{7.13}$ & 8.1 & 72.4 & 3.93 & 0 & 0.97\\
F563-1 & 14.77 & 5.39 & 6.51 & $\mathbf{4.84}$ & 9.93 & 0.55 & 1.67 & 0\\
F568-3 & 12.93 & 18.03 & $\mathbf{7.71}$ & 10.52 & 5.22 & 10.32 & 0 & 2.81\\
F579-V1 & 43.87 & 6.88 & 8.22 & $\mathbf{5.75}$ & 38.12 & 1.13 & 2.47 & 0\\
F583-1 & 295.53 & 14.82 & 6.43 & $\mathbf{4.89}$ & 290.64 & 9.93 & 1.54 & 0\\
F730-V1 & 26.30 & 9.48 & 6.59 & $\mathbf{5}$ & 21.3 & 4.48 & 1.59 & 0\\
UGC128 & 21.60 & 12.15 & 7.58 & $\mathbf{4.95}$ & 16.65 & 7.2 & 2.63 & 0\\
UGC1230 & 66.57 & 8.49 & 6.7 & $\mathbf{4.69}$ & 61.88 & 3.8 & 2.01 & 0\\
UGC5750 & 18 & 11.79 & 9.98 & $\mathbf{8.02}$ & 9.98 & 3.77 & 1.96 & 0\\
UGC11819 & 8.52 & 7.54 & 8.05 & $\mathbf{7.09}$ & 1.43 & 0.45 & 0.96 & 0\\
\hline
\end{tabular}
\end{table*}

\begin{table*}[h!]
\centering
\caption{Model performances based the $\Delta$-values. The models marked $\text{with a plus}$ (with $0 < \Delta \leq 2$) refer to comparably good performances, $\text{double pluses}$ represent the best fit (with $\Delta=0$). The analysis of the fitness of the models marked as $0$ (with $2 < \Delta < 4$) is inconclusive, models marked $\text{with a minus}$ (with $4\leq \Delta \leq 10$) are disfavored, while the models marked $\text{with
a double minus}$ (with $10<\Delta$) are clearly ruled out. The number of cases where the given model is favored ($\Sigma_{++}+\Sigma_{+}$) or ruled out ($\Sigma_{--}+\Sigma_{-}$) is also represented.}
\label{table:gx_vrot_disruletable}
\begin{tabular}{lcccc|lcccc}
\hline
\multicolumn{5}{c|}{HSB galaxies} & \multicolumn{5}{c}{LSB galaxies}\\
\hline
ID &  $\Delta_\mathrm{B}$ & $\Delta_\mathrm{NFW}$ & $\Delta_\mathrm{E}$ & $\Delta_\mathrm{P}$ & ID &  $\Delta_\mathrm{B}$ & $\Delta_\mathrm{NFW}$ & $\Delta_\mathrm{E}$ & $\Delta_\mathrm{P}$\\
\hline
ESO215G39 &  $--$ & $++$ & $0$ & $+$  &  F561-1 &  $--$ & $0$ & $++$ & $+$ \\
ESO322G76 &  $--$ & $++$ & $+$ & $+$  &  F563-1 &  $-$ & $+$ & $+$ & $++$ \\
ESO322G77 &  $0$ & $++$ & $0$ & $+$   &  F568-3 &  $-$ & $--$ & $++$ & $0$ \\
ESO323G25 &  $--$ & $-$ & $-$ & $++$  &  F579-V1 &  $--$ & $+$ & $0$ & $++$ \\
ESO374G02 &  $--$ & $++$ & $-$ & $+$ &   F583-1 &  $--$ & $-$ & $+$ & $++$ \\
ESO375G12 &  $-$ & $+$ & $-$ & $++$  &   F730-V1 &  $--$ & $-$ & $+$ & $++$ \\
ESO383G02 &  $+$ & $++$ & $+$ & $+$    & UGC128 &  $--$ & $-$ & $0$ & $++$ \\
ESO383G88 &  $--$ & $--$ & $--$ & $++$ & UGC1230 &  $--$ & $0$ & $0$ & $++$ \\
 ESO445G19 &  $0$ & $+$ & $++$ & $+$ & UGC5750 &  $-$ & $0$ & $+$ & $++$ \\ 
ESO446G01 &  $--$ & $+$ & $0$ & $++$ &UGC11819 &  $+$ & $+$ & $+$ & $++$  \\
ESO502G02 &  $0$ & $0$ & $0$ & $++$ &  \\
ESO509G80 &  $--$ & $+$ & $0$ & $++$ & \\
ESO569G17 &  $-$ & $++$ & $-$ & $+$  & \\
\hline
\hline
  $\Sigma_{++}$  & 0 & 6 & 1 & 6 &   $\Sigma_{++}$& 0 & 0 & 2 & 8\\
  $\Sigma_{+}$   & 1 & 4 & 2 & 6 &   $\Sigma_{+}$ & 1 & 3 & 5 & 1\\
  $\Sigma_{-}$   & 2 & 1 & 4 & 0 &   $\Sigma_{-}$ & 3 & 3 & 0 & 0\\
  $\Sigma_{--}$  & 7 & 1 & 1 & 0 &   $\Sigma_{--}$& 6 & 1 & 0 & 0\\
\hline
\end{tabular}
\end{table*}

\section{Summary and concluding remarks}
\label{summary}

In this paper we have assembled a database consisting of $15$ HSB and $15$ LSB galaxies that are representative for the various possible galaxy morphologies of both types. In particular, the HSB galaxy set contained spiral galaxies of various brightness profiles, while the LSB galaxy set contained both disk and dwarf galaxies. For the selected galaxies, both surface brightness density data and spectroscopic rotation curve data were available in the literature.

We explored this dataset for a comparative testing of frequently applied and well-established dark matter models (NFW, Einasto, and PSE). We investigated the compatibility of the pure baryonic model and baryonic plus various dark matter models with observations on the galaxy database. The mass distribution of the baryonic component of the galaxies was derived from the spatial luminosity distribution by estimating the $M/L$ ratios through color--to-mass-to-light relations and gas mass fractions. For our analysis we constrained the axial ratio of the galaxies based on SDSS results as $0.4<q_b<1$, and $0<q_d<0.3$.

We calculated the Akaike information criterion to characterize the goodness of the best-fit models. In case of the pure baryonic model, the $M/L$ ratios were varied between reasonable limits in the fitting to the rotation curves, while in case of baryonic + dark matter combined models, the baryonic component was inferred using $M/L$ ratios derived based on the CMLR of \citet{Bell2003} and \citet{McGaugh2014}. In case of 7 galaxies (2 HSB, and 5 LSB), neither model fits the dataset within the $1\sigma$ confidence level. According to the Akaike information criterion, the pseudo-isothermal sphere emerges as most favored in $14$ cases out of the remaining 23 galaxies, followed by the Navarro-Frenk-White ($6$ cases)
and the Einasto ($3$ cases) dark matter models.

The pure baryonic model with an $M/L$ ratio fit did not provide the best model performance in any of the cases, and it was ruled
out in case of 7 HSB and 6 LSB galaxies based on the AIC. On the other hand, the pure baryonic model fit the dataset within
the $1\sigma$ confidence level in case of 10 HSB and 2 LSB galaxies. From these 12 galaxies, the pure baryonic model provided a poor fit compared to the best-fit DM model in 5 cases, giving $\Delta>10$. This clearly disfavors the baryonic model.

By cross-correlating the results of the fits by the two methods, we found that the following seven galaxies could not be described with any of the considered models: ESO322G82, ESO376G02, UGC6614, UGC10310, UGC11454, UGC11616, and UGC11748. The proper modeling of these galaxies needs more sophisticated descriptions. Employing massive datasets of 2D velocity fields from integral field unit surveys, such as SAMI \citep{Allen2015} and MaNGA in SDSS-IV \citep{Bundy2015}, will help to further develop the method we presented and test it for non-axisymmetry on high-quality velocity fields with well-defined errors.

The remaining 23 galaxies from the dataset, which require a dark matter component to be modeled, might be useful in the comparative testing of spherically symmetric dark matter substitutes emerging in either alternative gravity theories or from the study of non-standard dark matter candidates, like Bose-Einstein condensates.


\begin{acknowledgements}
The authors are grateful to the anonymous referee for the valuable suggestions provided throughout the refereeing process, which have contributed to significant improvement of the paper. EK, ZK, and L\'{A}G acknowledge the support of the Hungarian National Research, Development and Innovation Office (NKFI) in the form of the grant 123996. ZK and L\'{A}G were further supported by COST Action CA15117 “Cosmology and Astrophysics Network for Theoretical Advances and Training Actions” (CANTATA).
\end{acknowledgements}

\newpage
\section*{Appendix}

\begin{figure*}
\centering
\includegraphics[width=150pt,height=130pt]{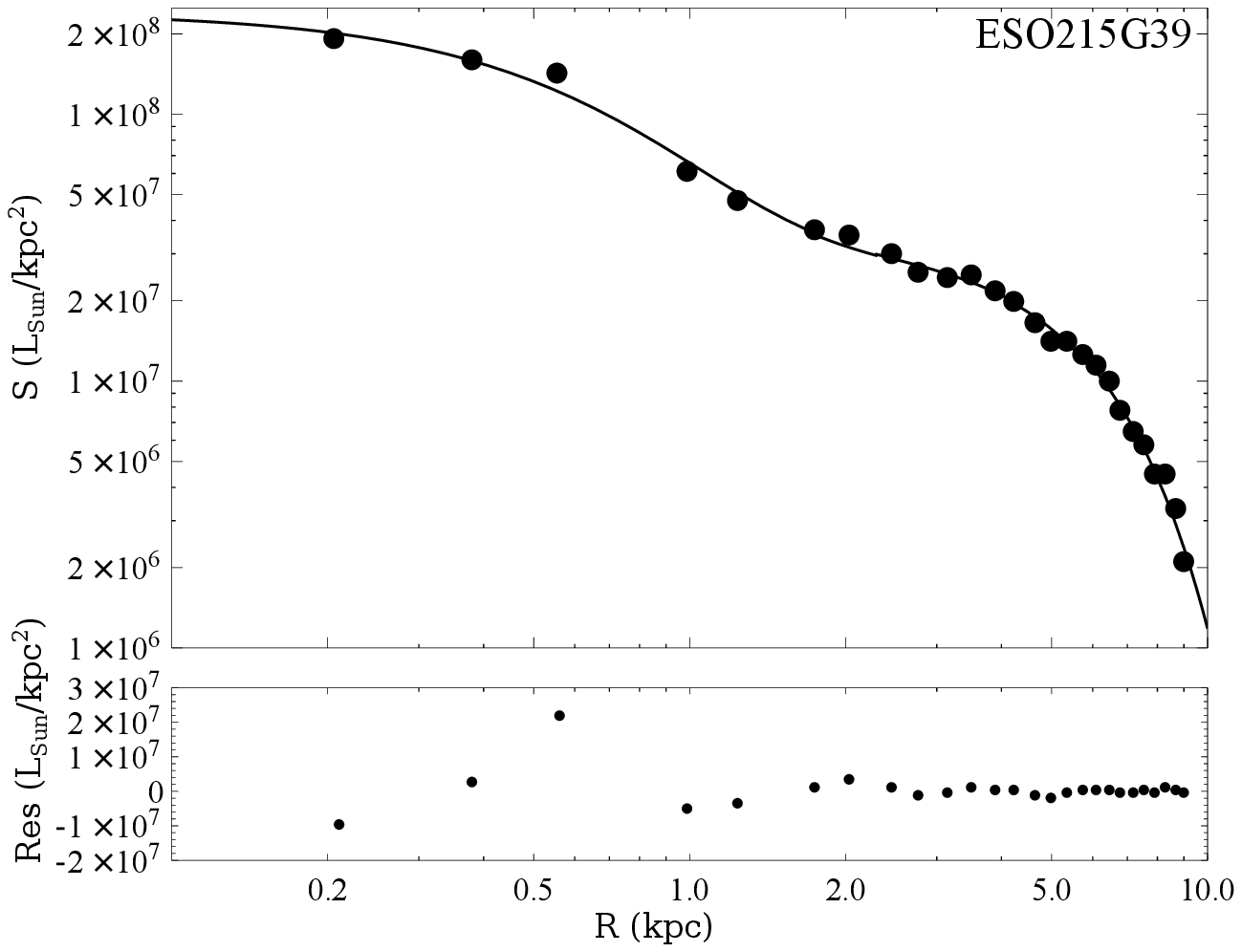}
\includegraphics[width=150pt,height=130pt]{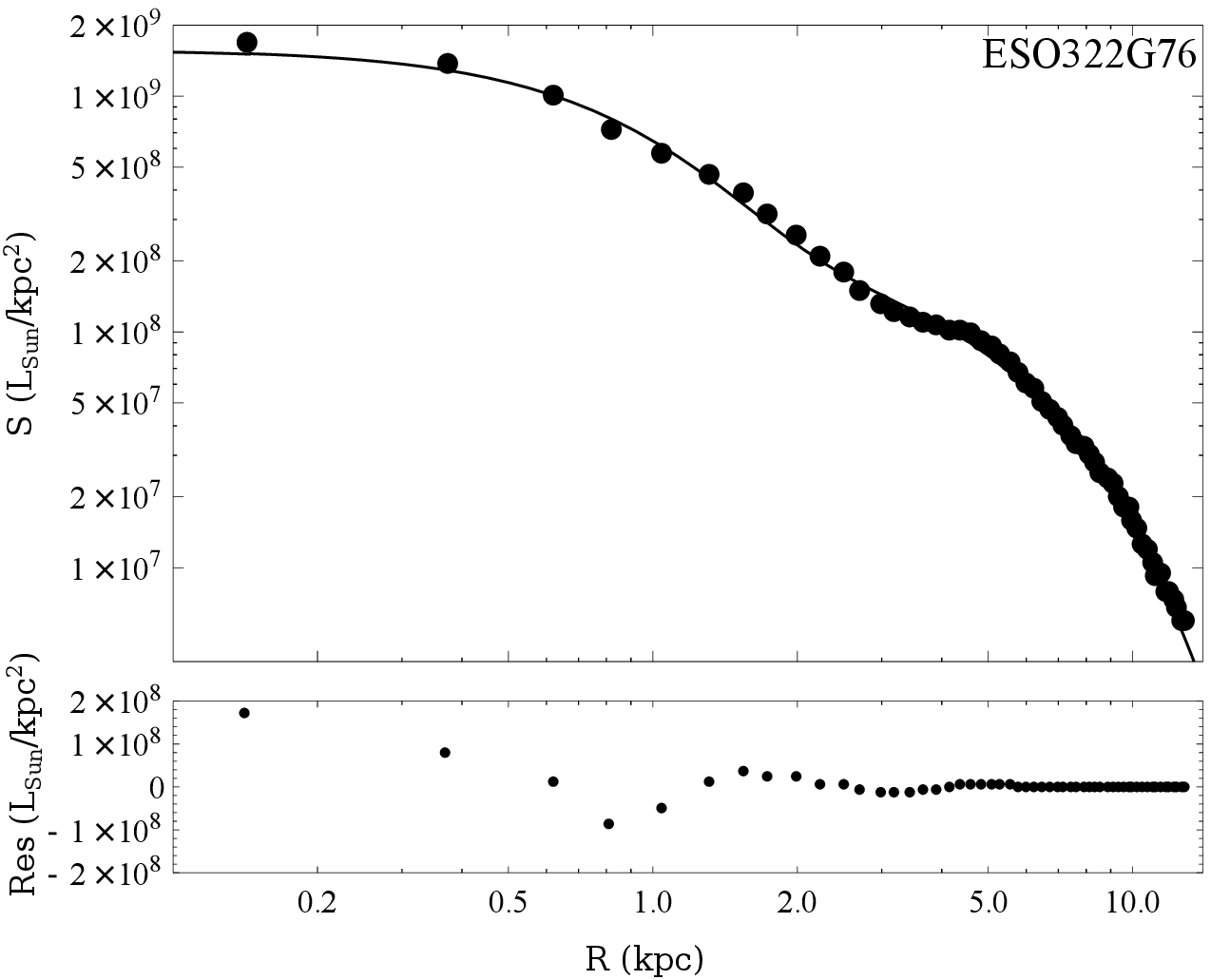}
\includegraphics[width=150pt,height=130pt]{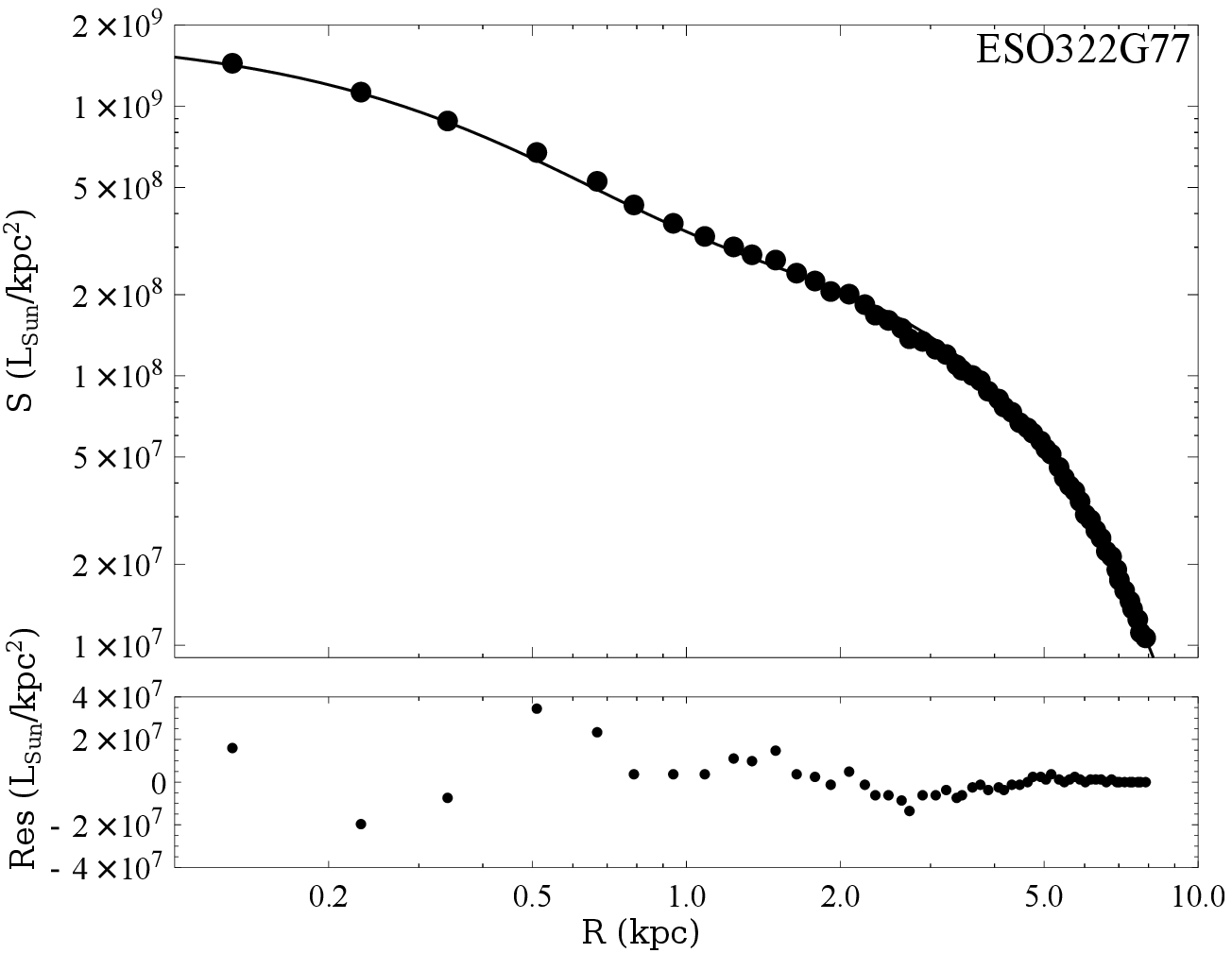}\newline
\includegraphics[width=150pt,height=130pt]{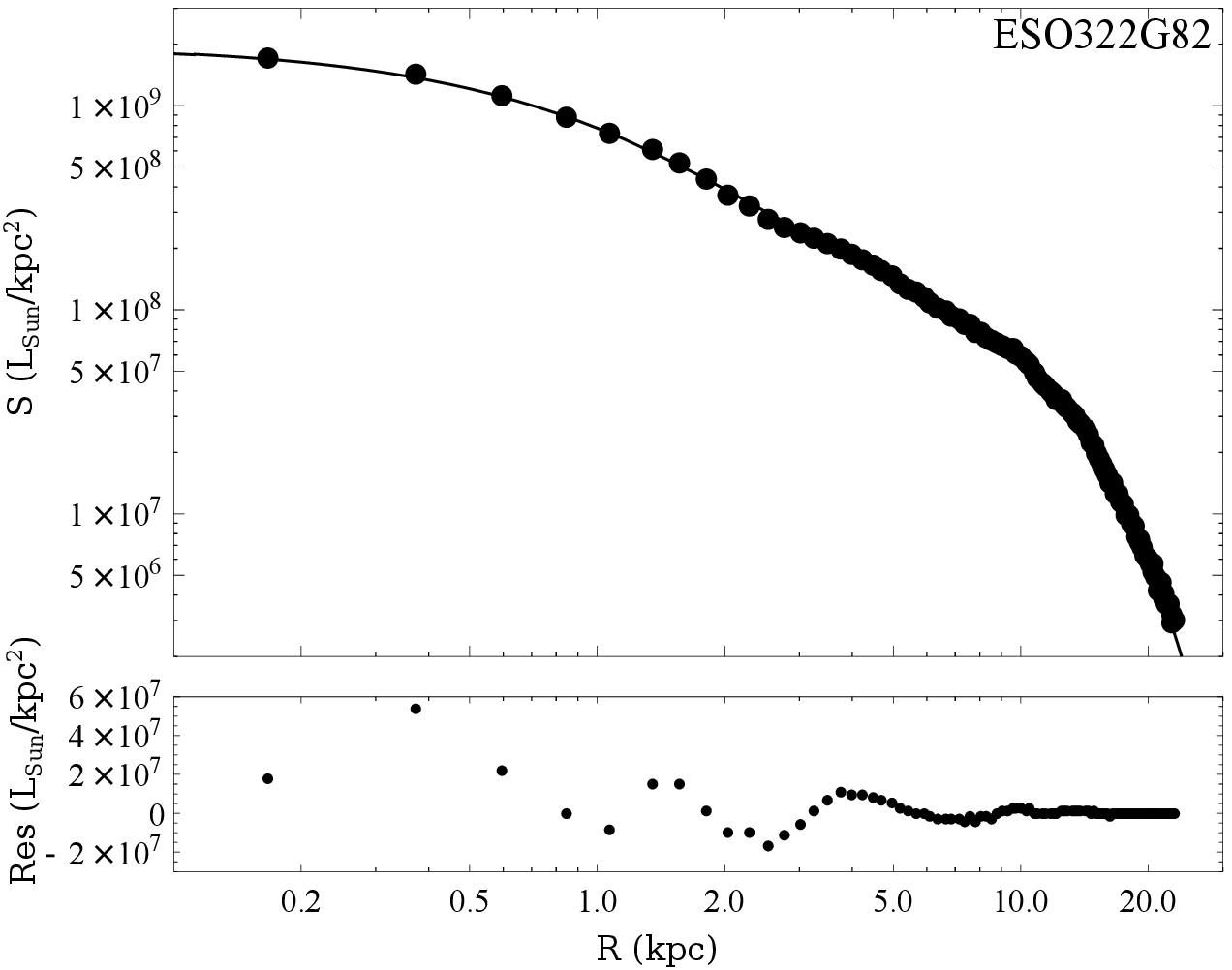}
\includegraphics[width=150pt,height=130pt]{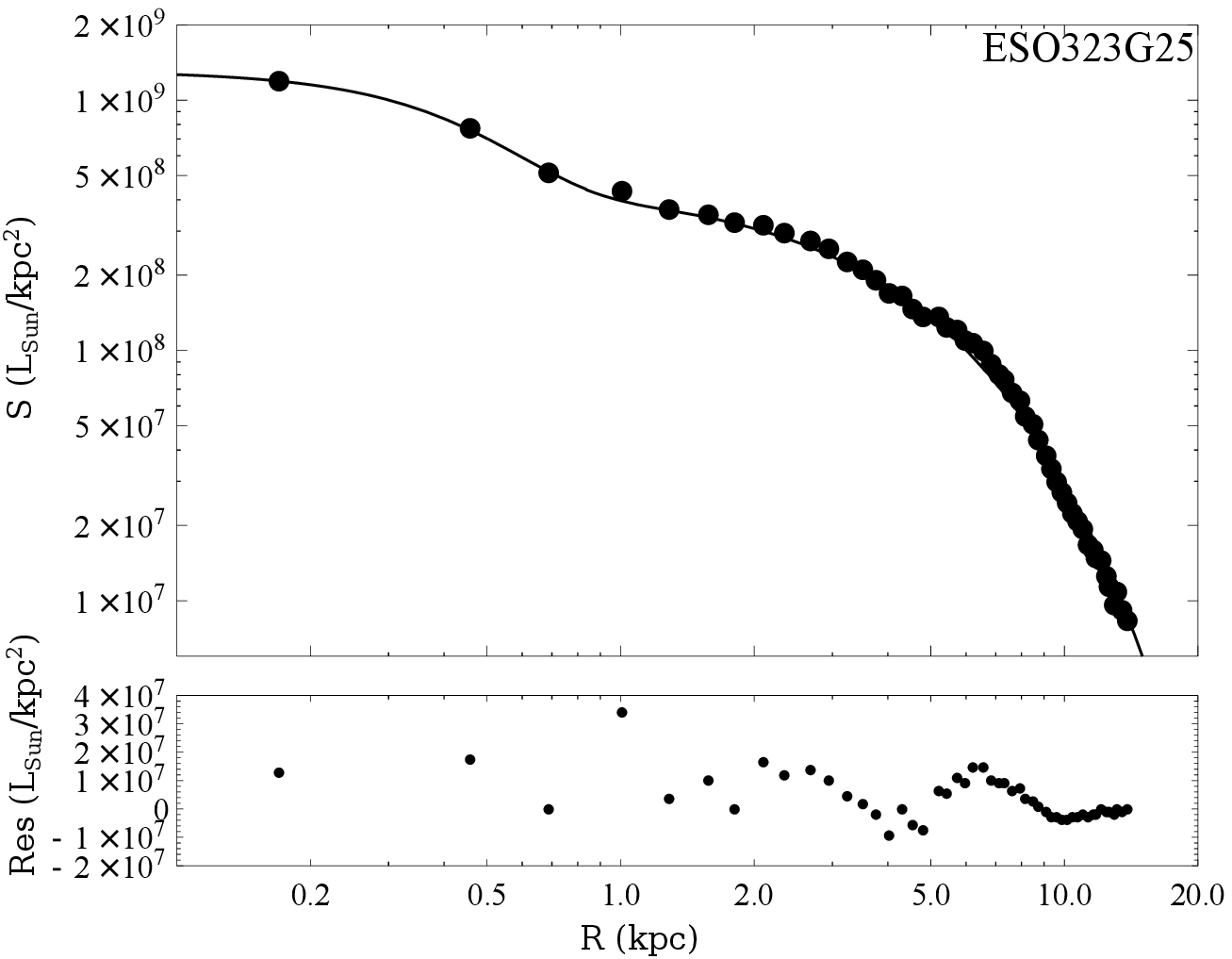}
\includegraphics[width=150pt,height=130pt]{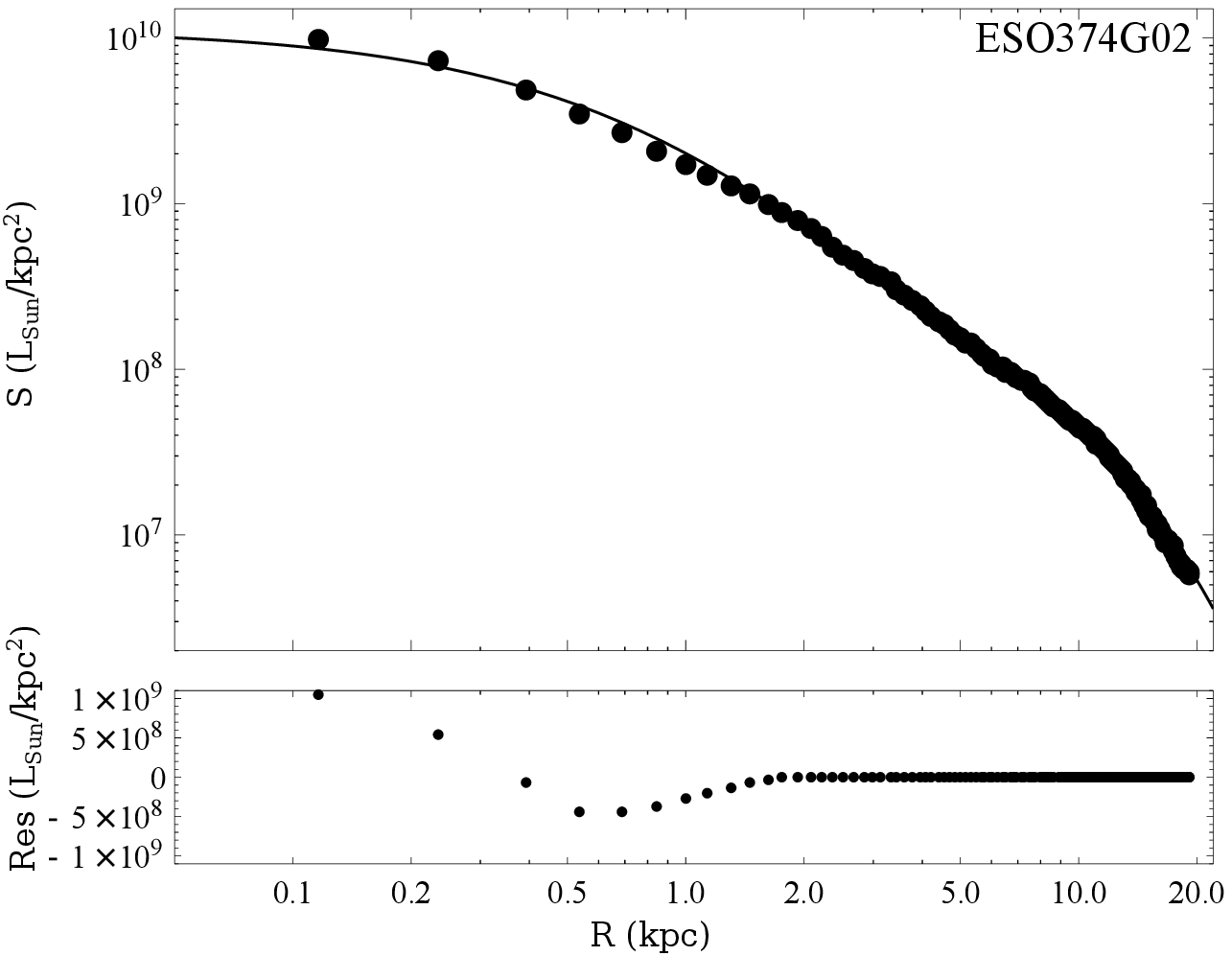}\newline
\includegraphics[width=150pt,height=130pt]{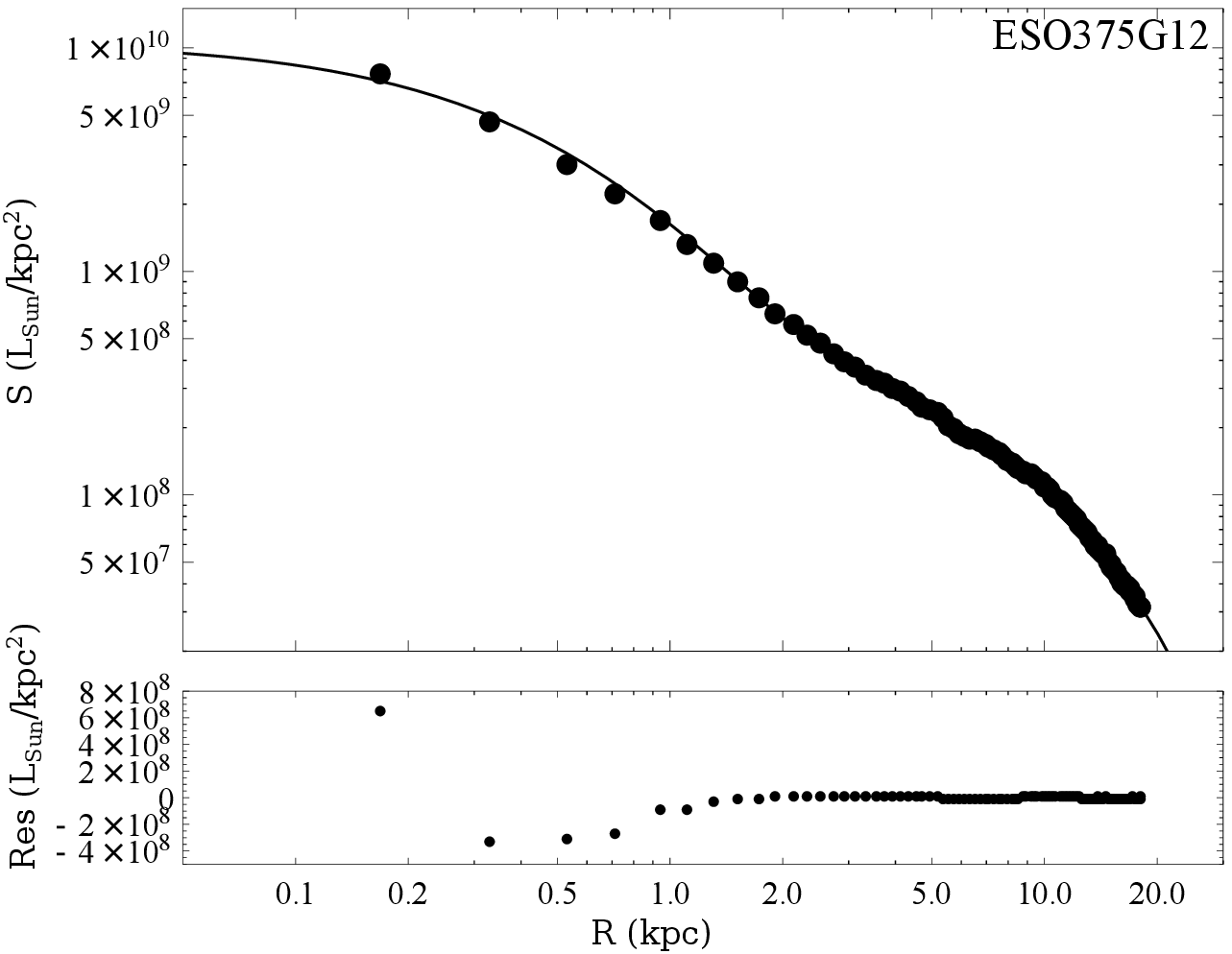}
\includegraphics[width=150pt,height=130pt]{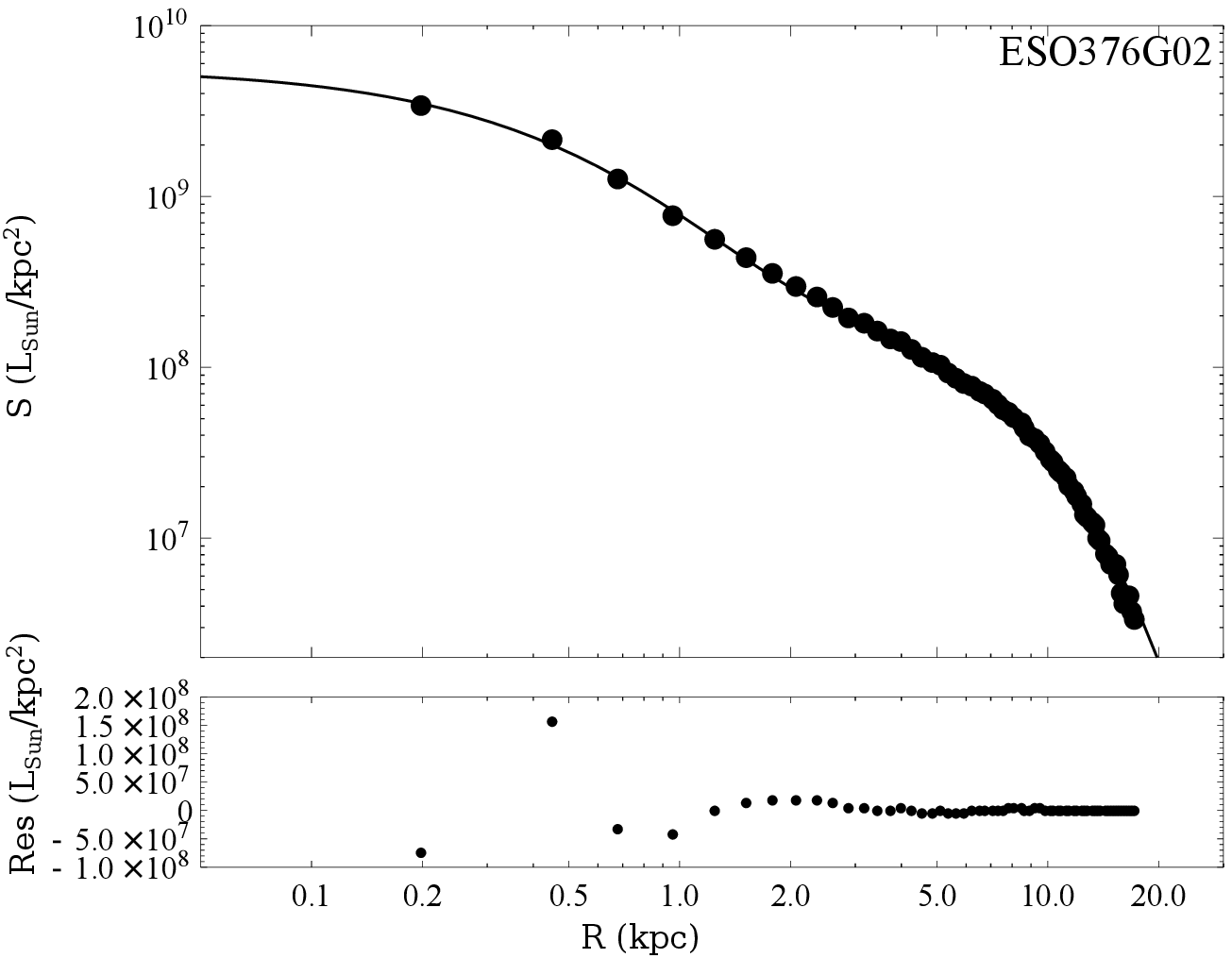}
\includegraphics[width=150pt,height=130pt]{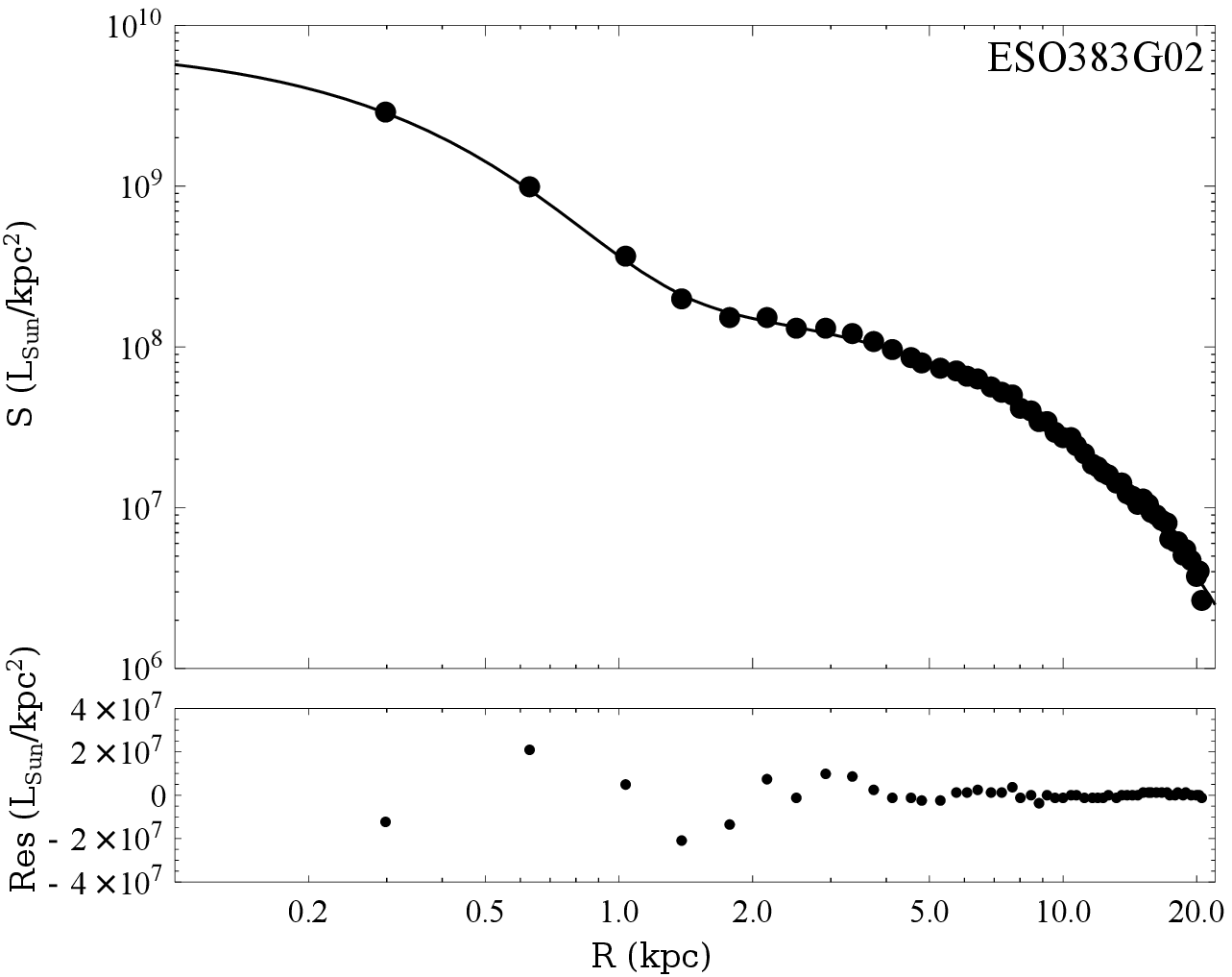}\newline
\includegraphics[width=150pt,height=130pt]{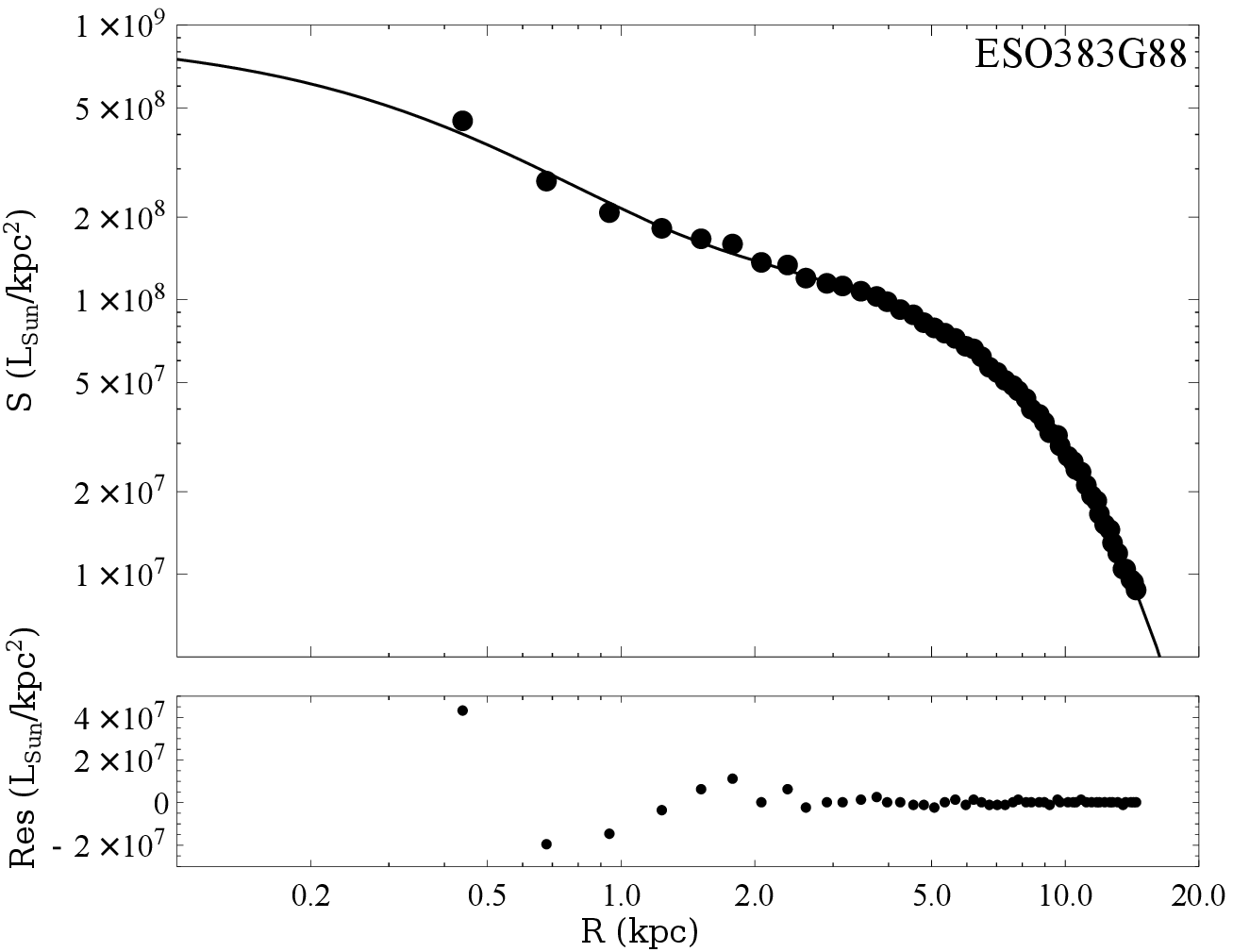}
\includegraphics[width=150pt,height=130pt]{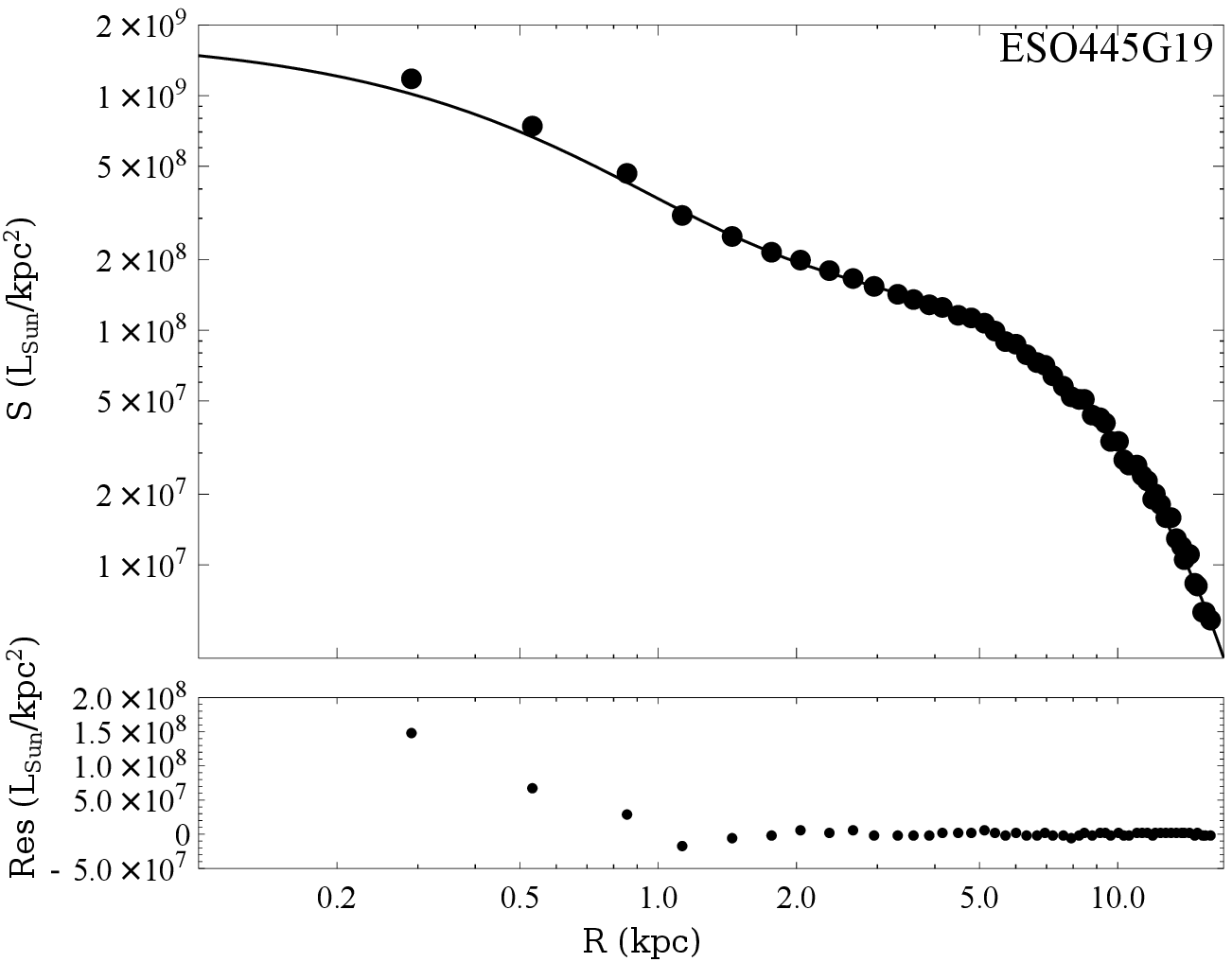}
\includegraphics[width=150pt,height=130pt]{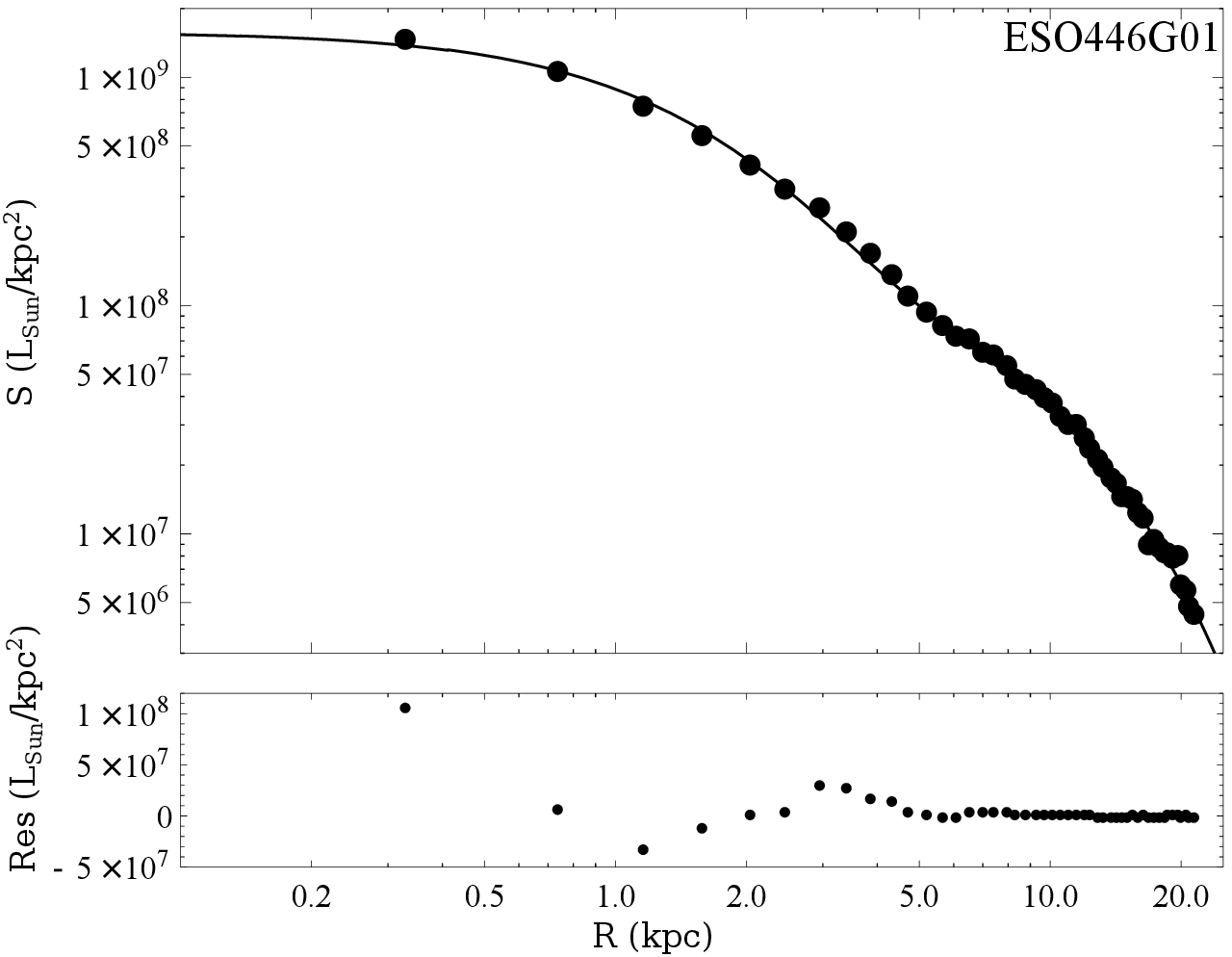}\newline
\includegraphics[width=150pt,height=130pt]{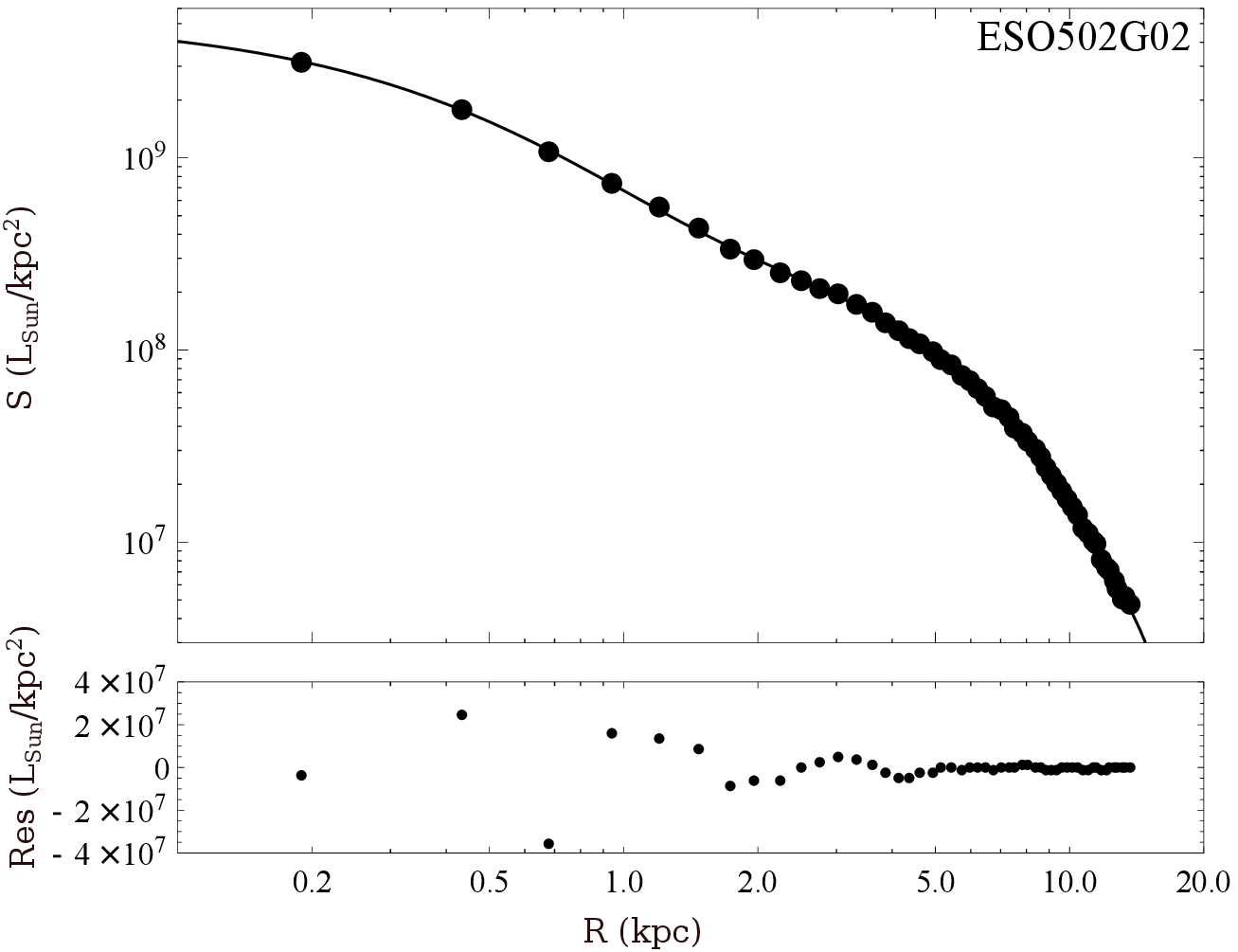}
\includegraphics[width=150pt,height=130pt]{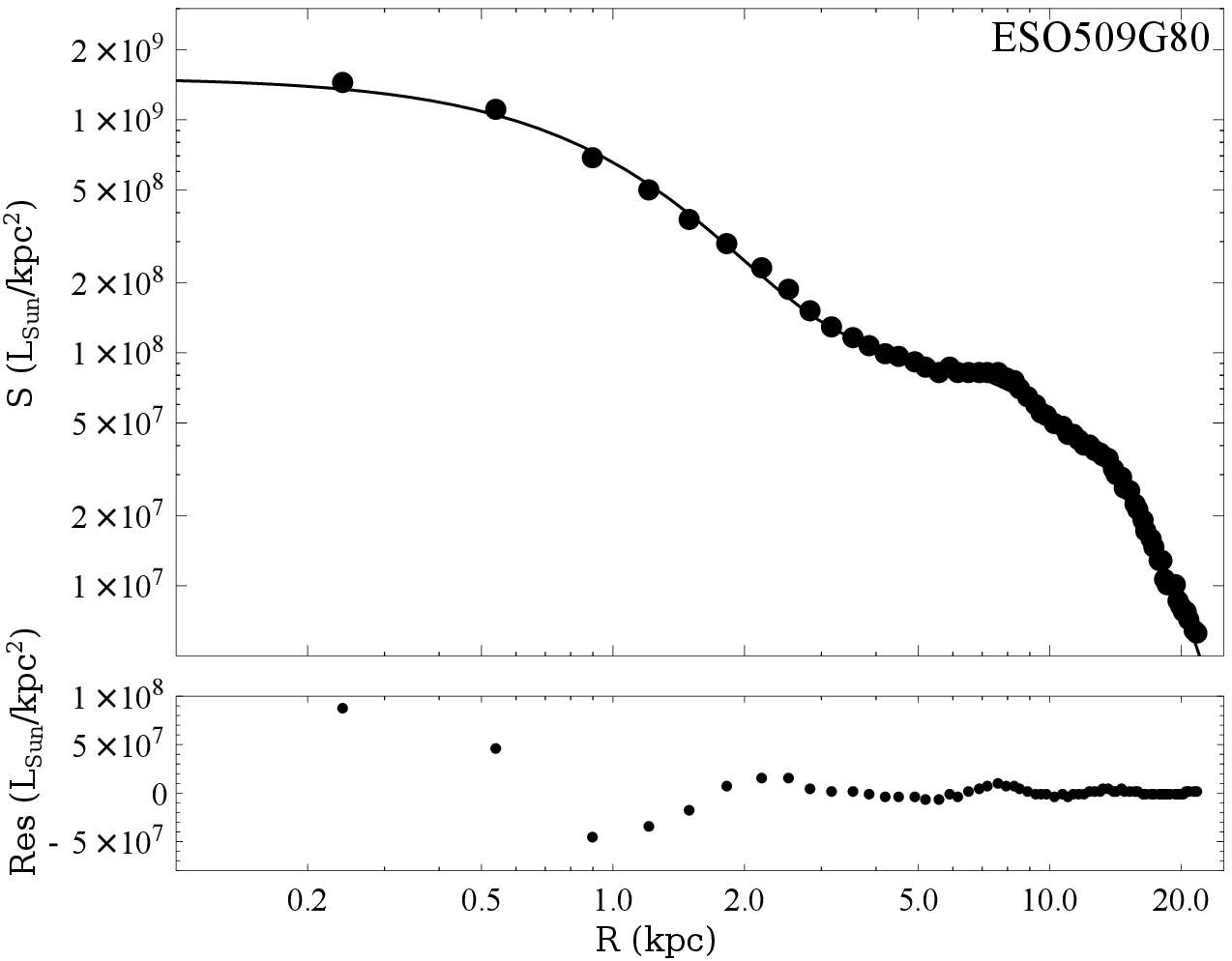}
\includegraphics[width=150pt,height=130pt]{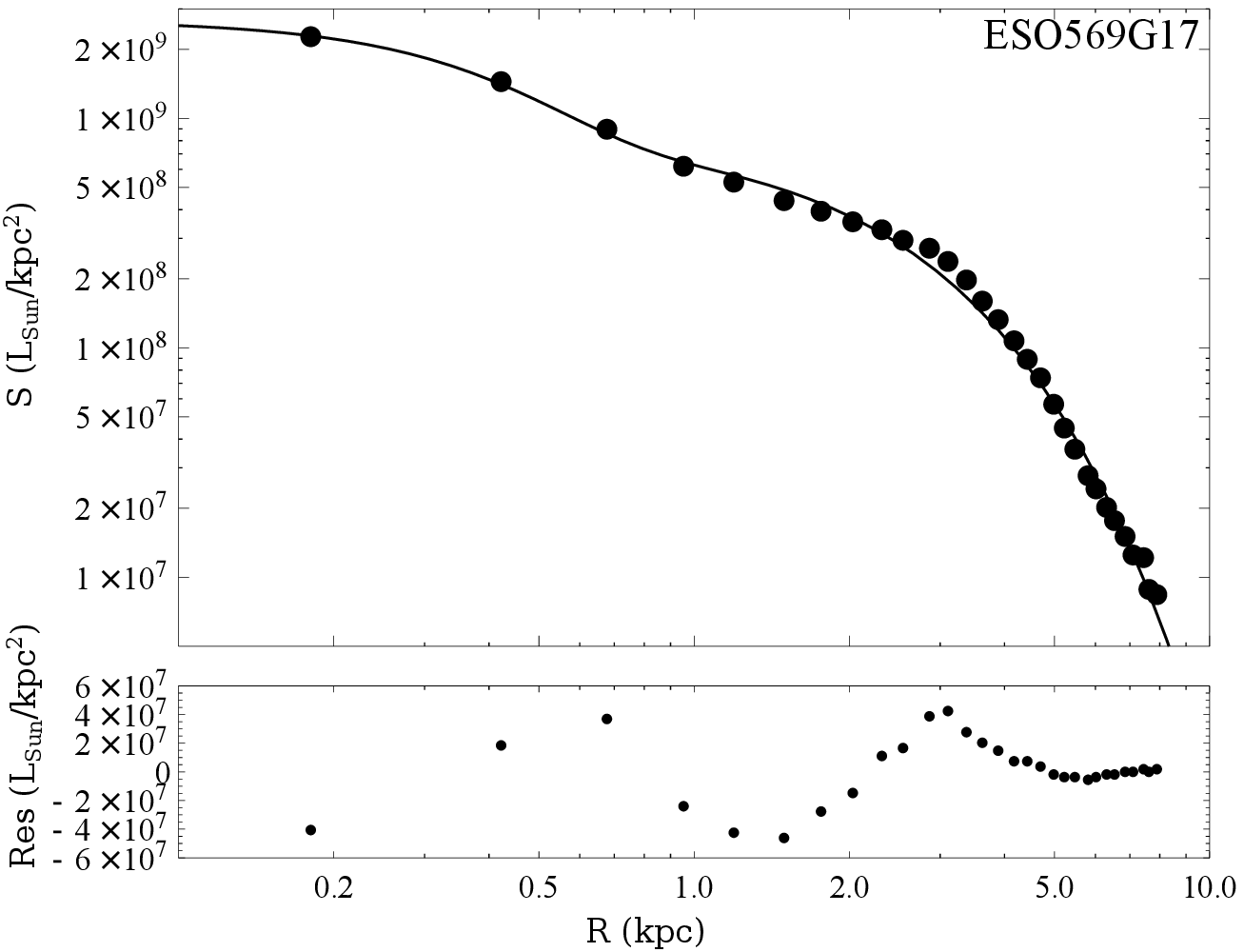}\newline
\caption{Best-fit 1D surface brightness models of the 15 HSB galaxies (upper panels) and their residuals (lower panels). The measurements are represented as larger black dots, the residuals by points, while the model predictions are shown as black continuous lines.}
\label{fig:bright_fit_plots1}
\end{figure*}

\begin{figure*}
\centering
\includegraphics[width=150pt,height=130pt]{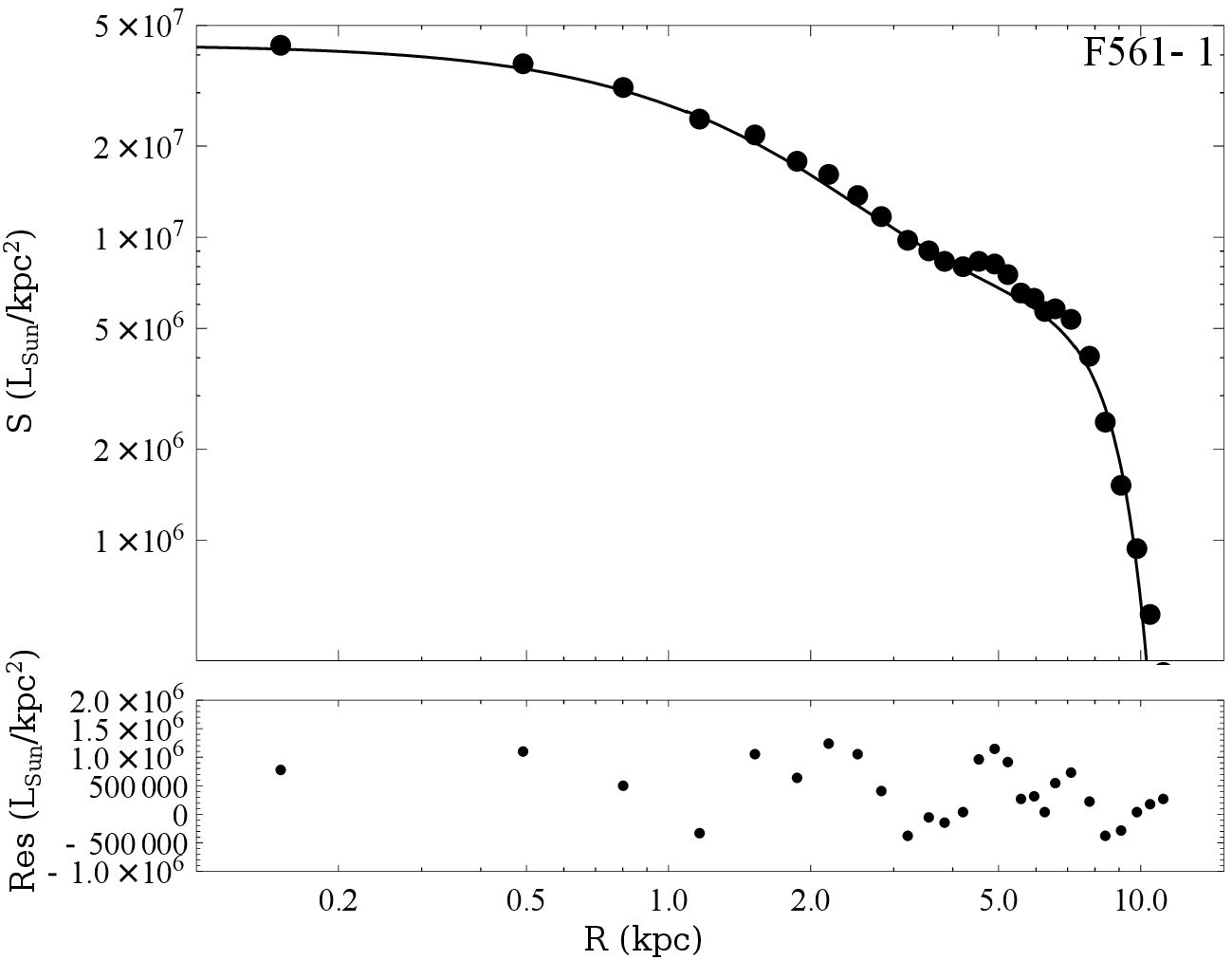}
\includegraphics[width=150pt,height=130pt]{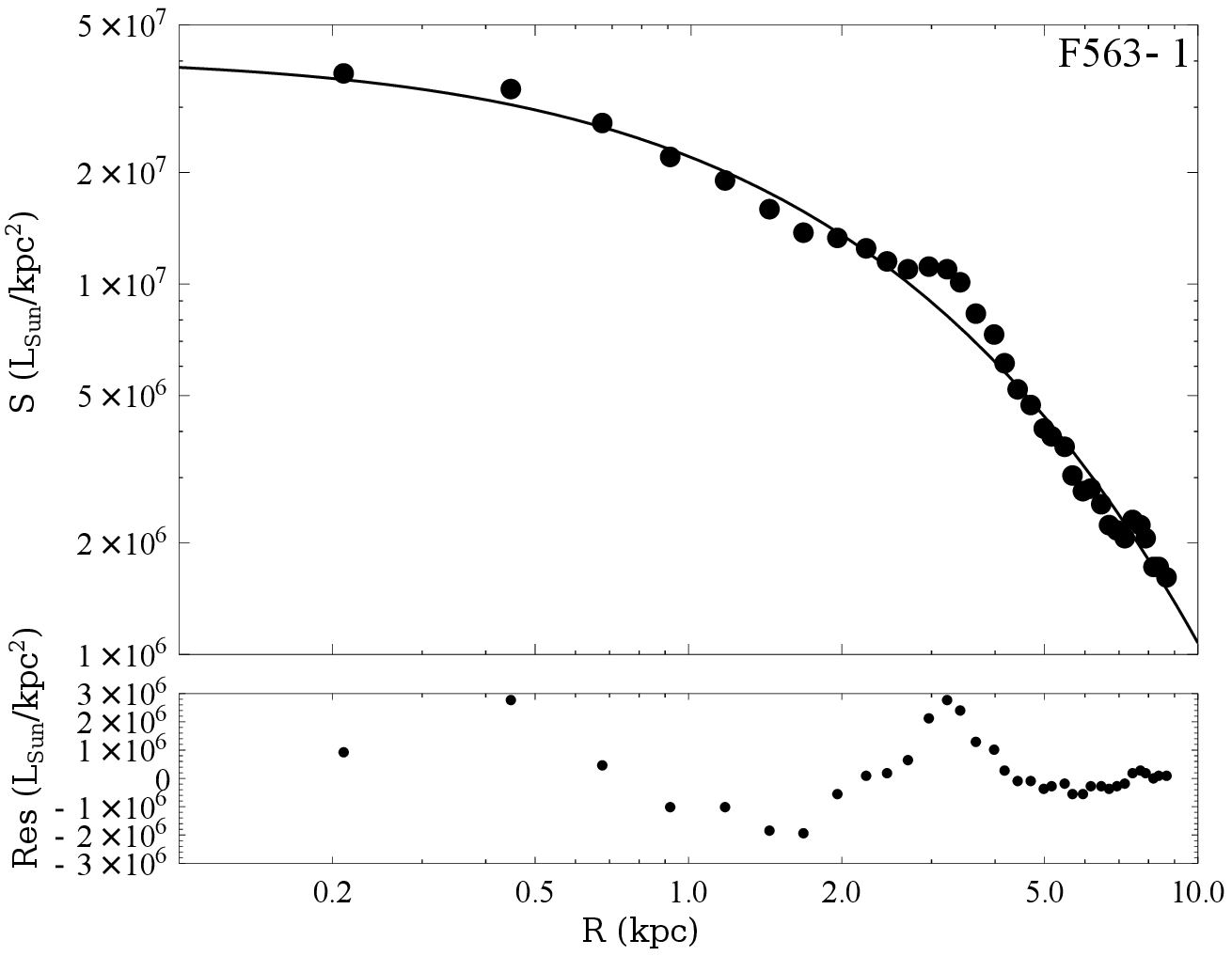}
\includegraphics[width=150pt,height=130pt]{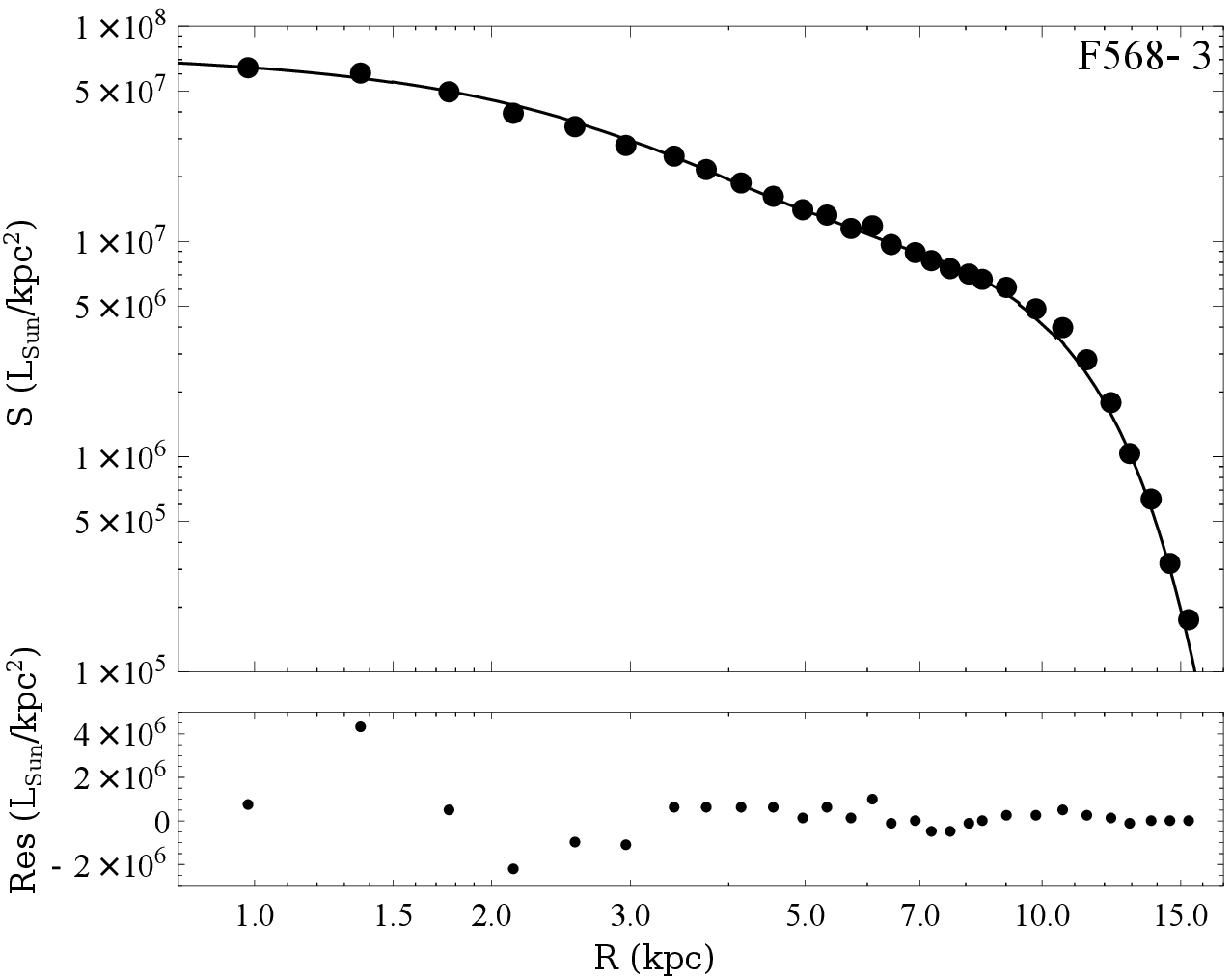}\newline
\includegraphics[width=150pt,height=130pt]{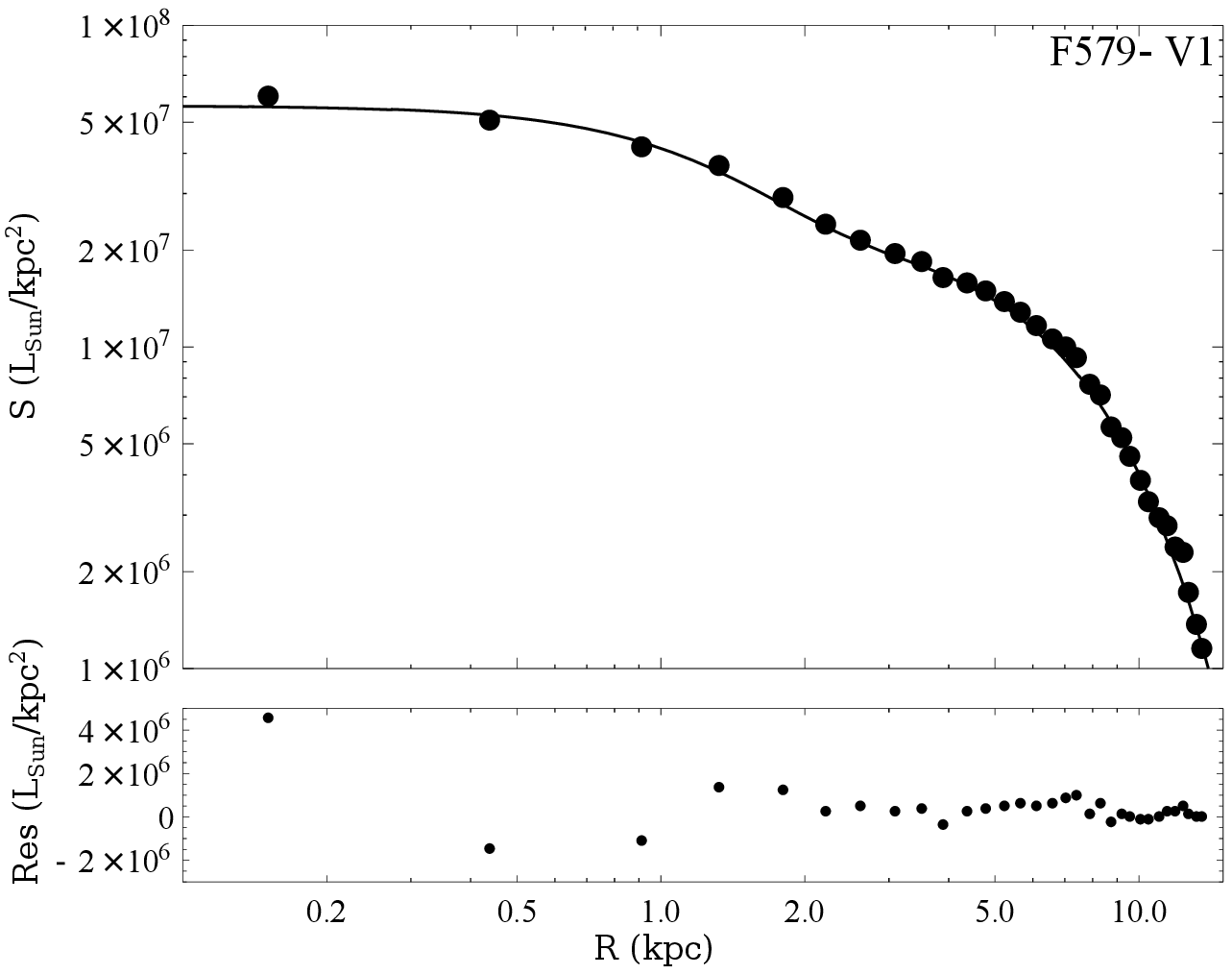}
\includegraphics[width=150pt,height=130pt]{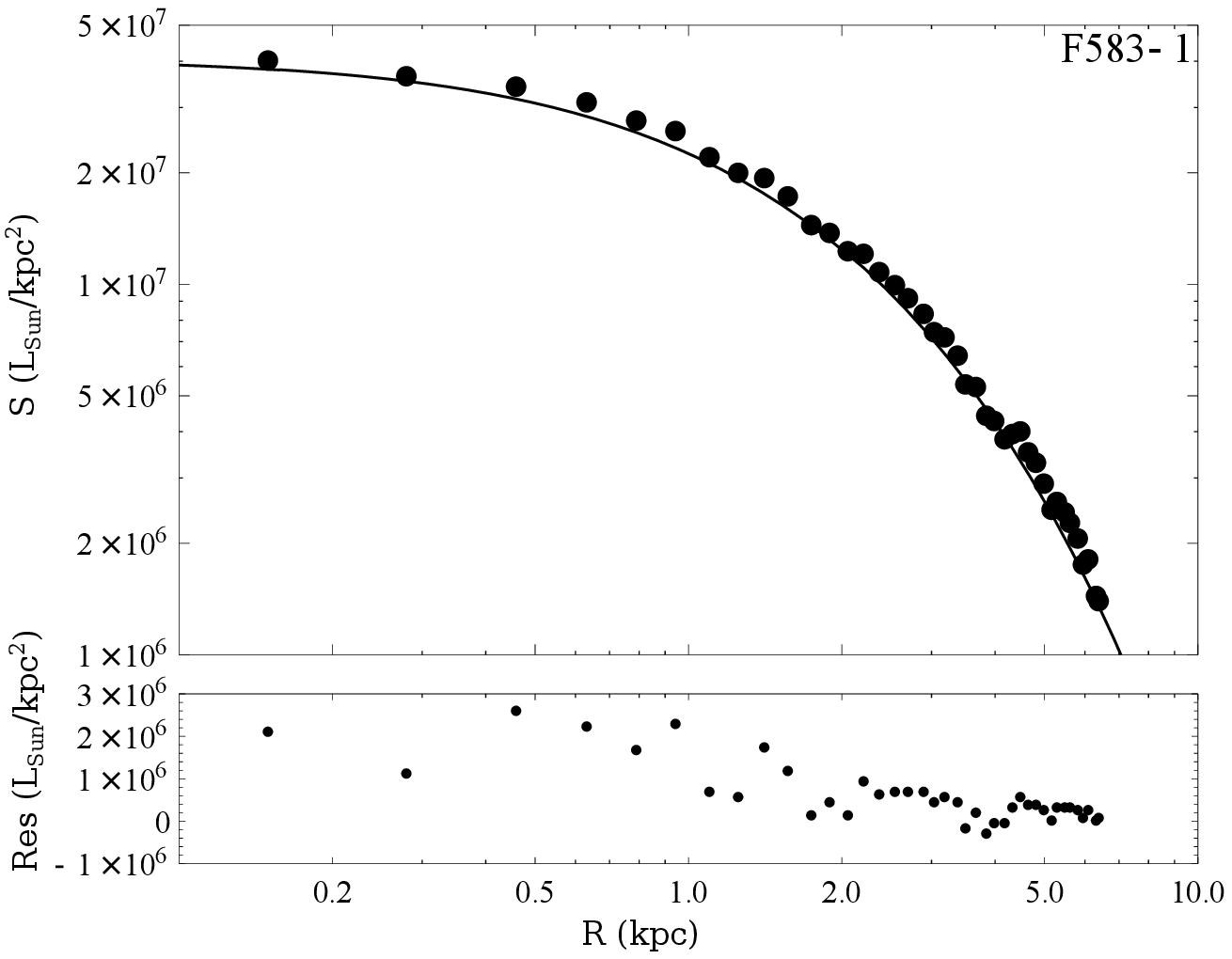}
\includegraphics[width=150pt,height=130pt]{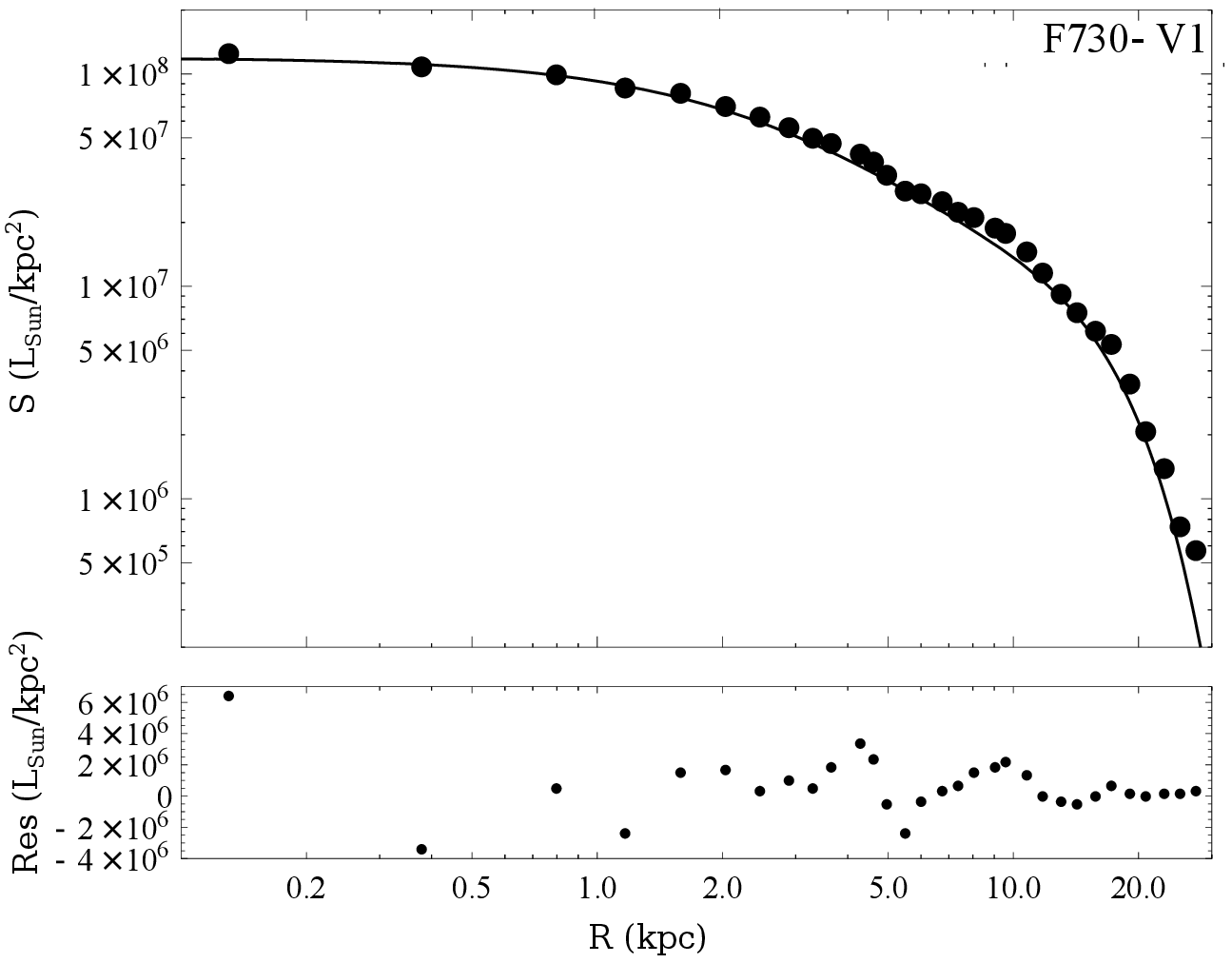}\newline
\includegraphics[width=150pt,height=130pt]{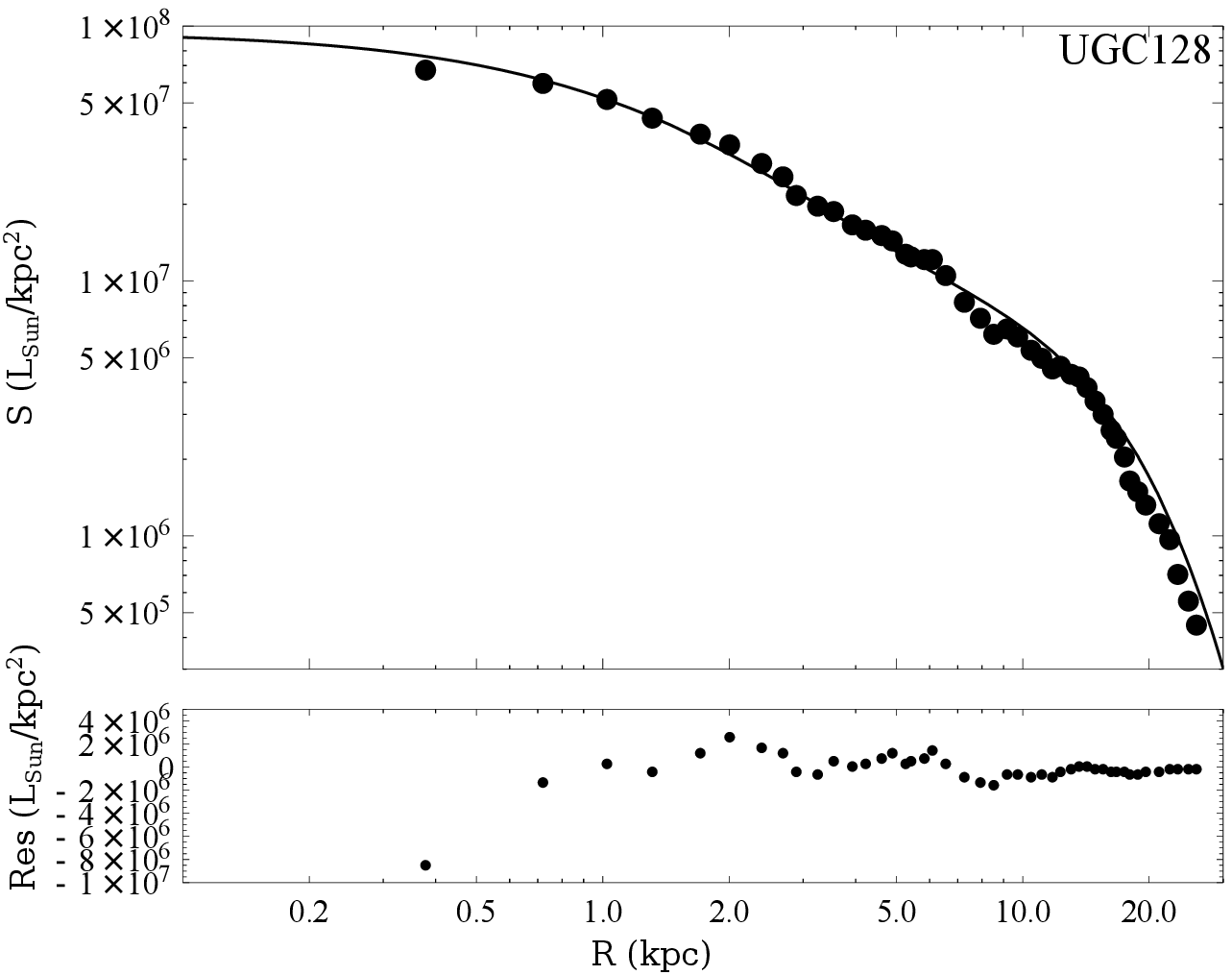}
\includegraphics[width=150pt,height=130pt]{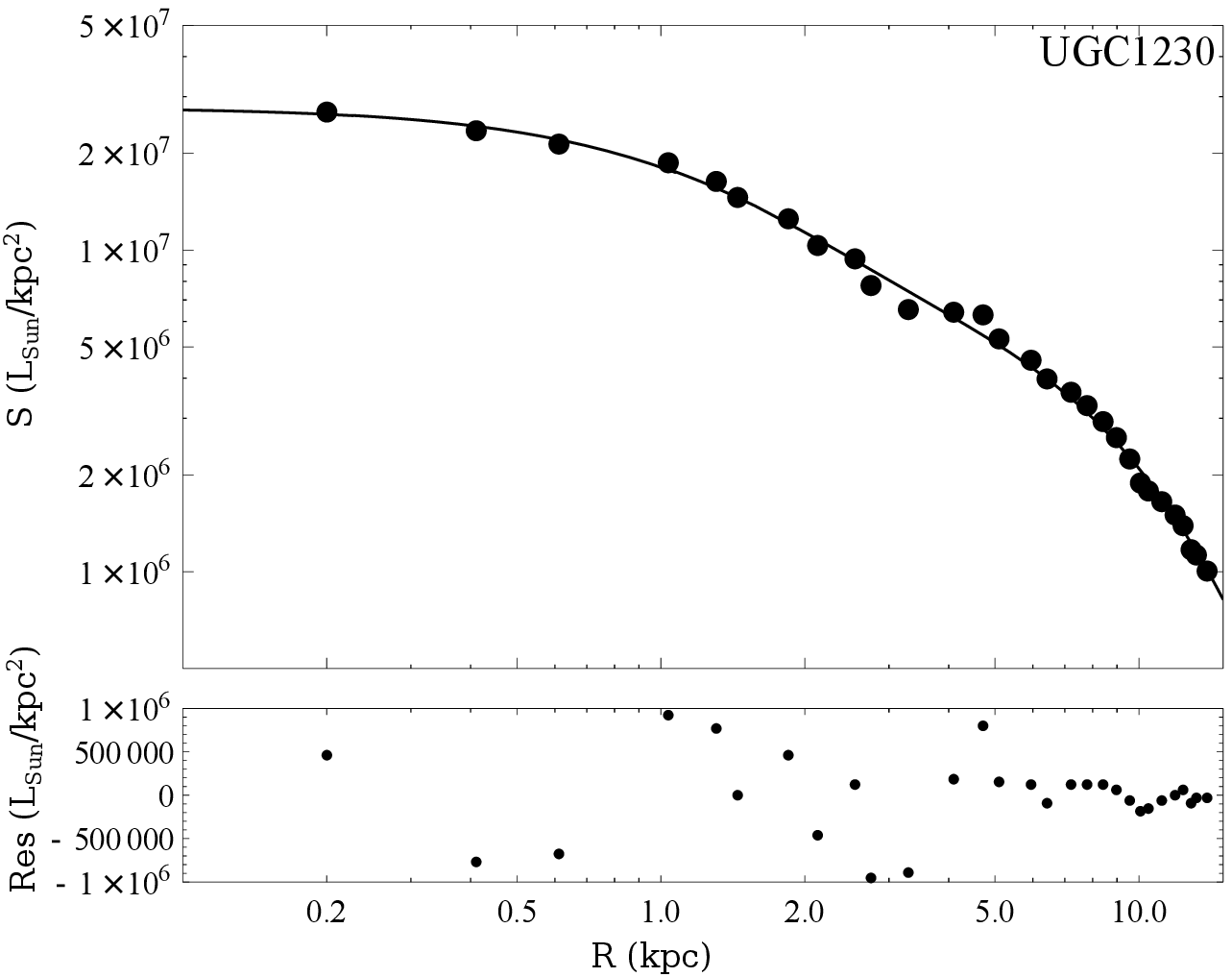}
\includegraphics[width=150pt,height=130pt]{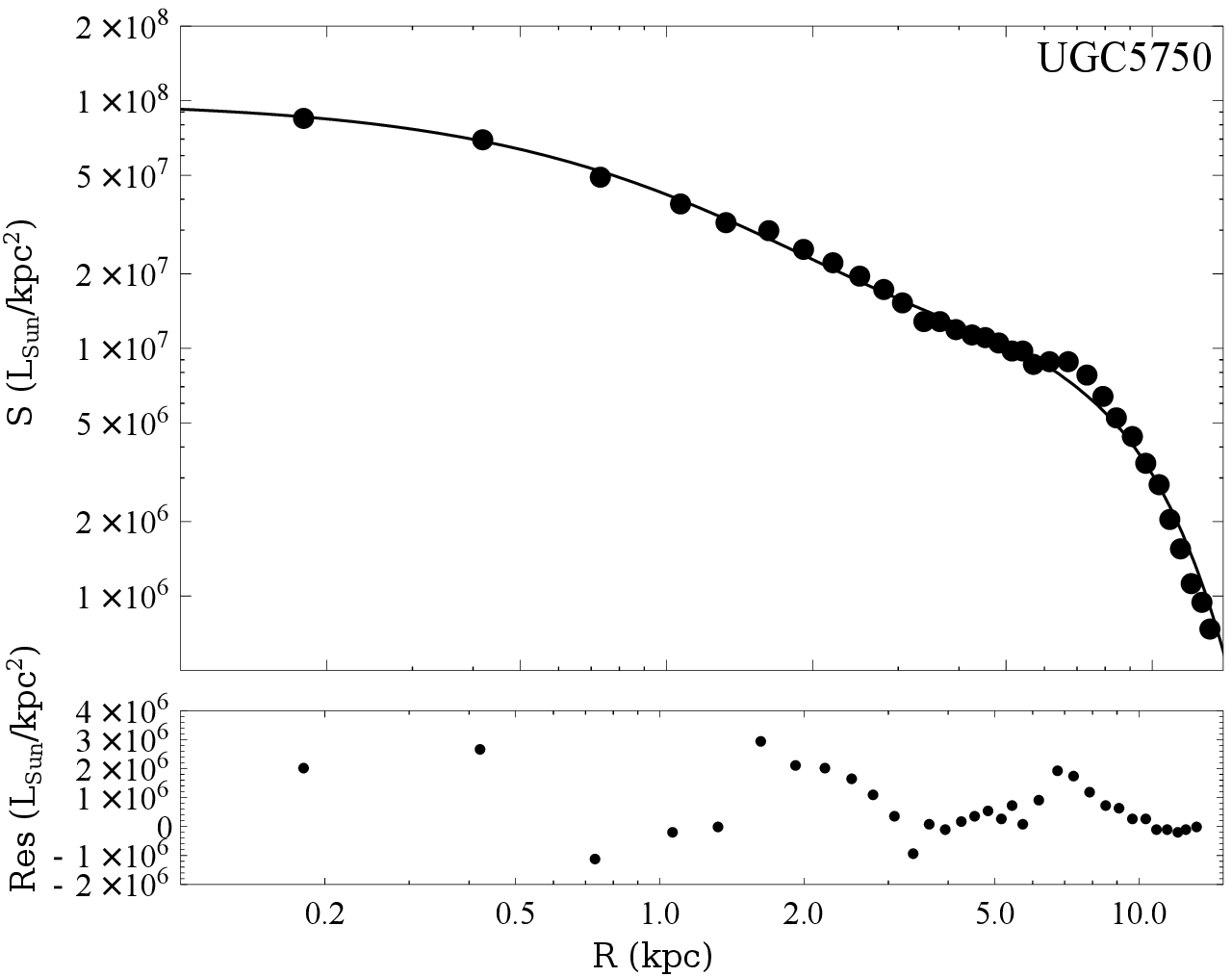}\newline
\includegraphics[width=150pt,height=130pt]{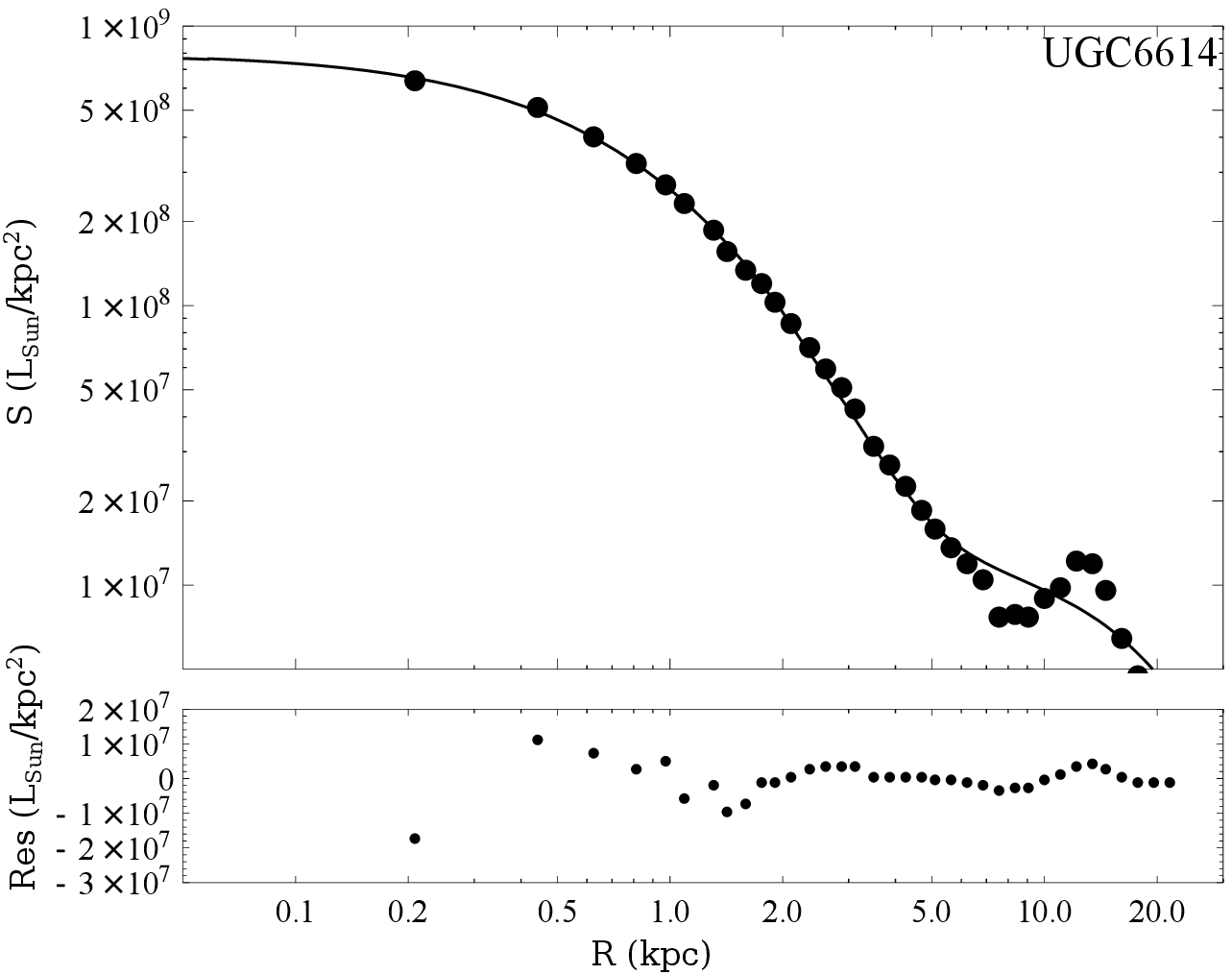}
\includegraphics[width=150pt,height=130pt]{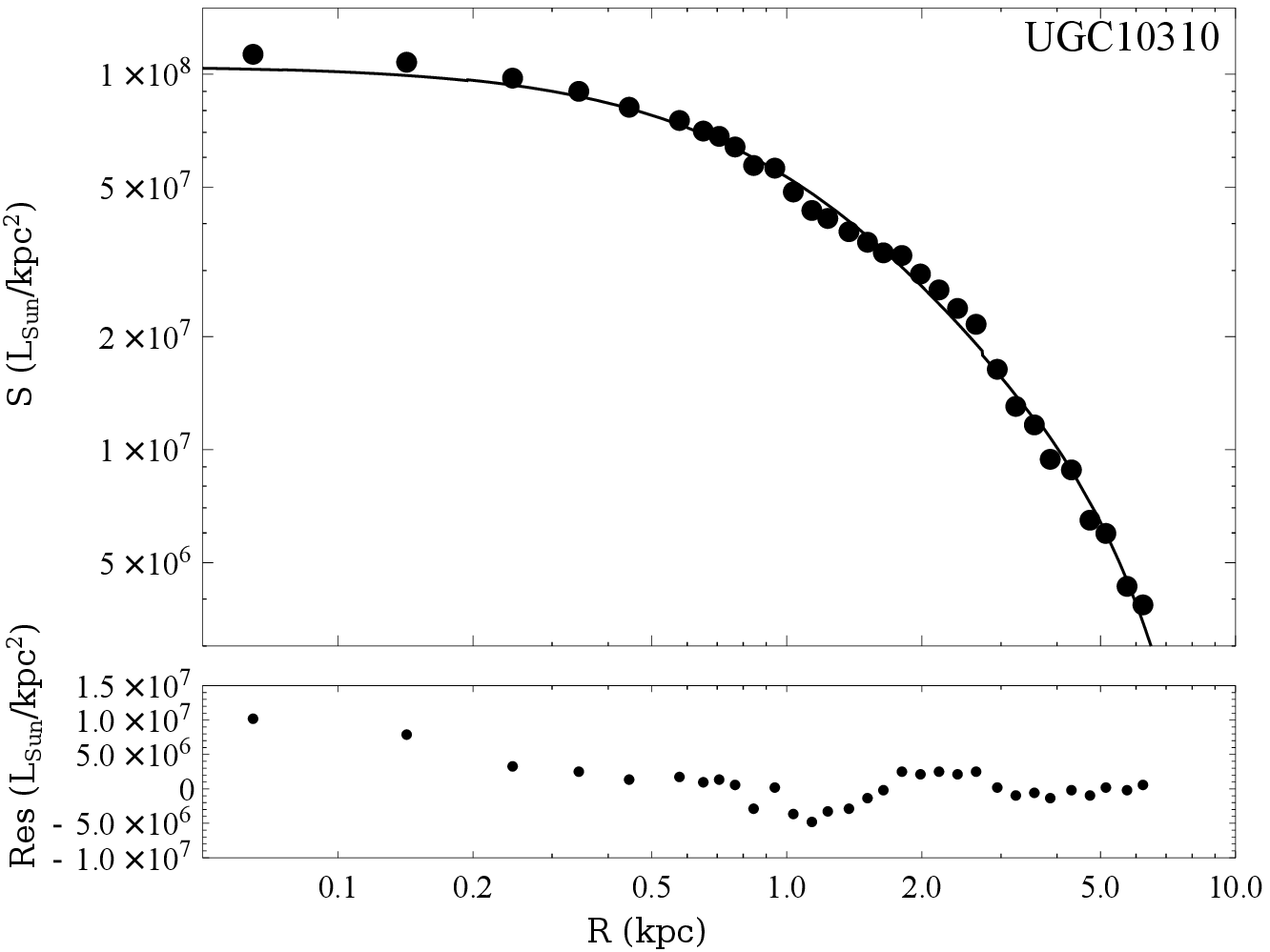}
\includegraphics[width=150pt,height=130pt]{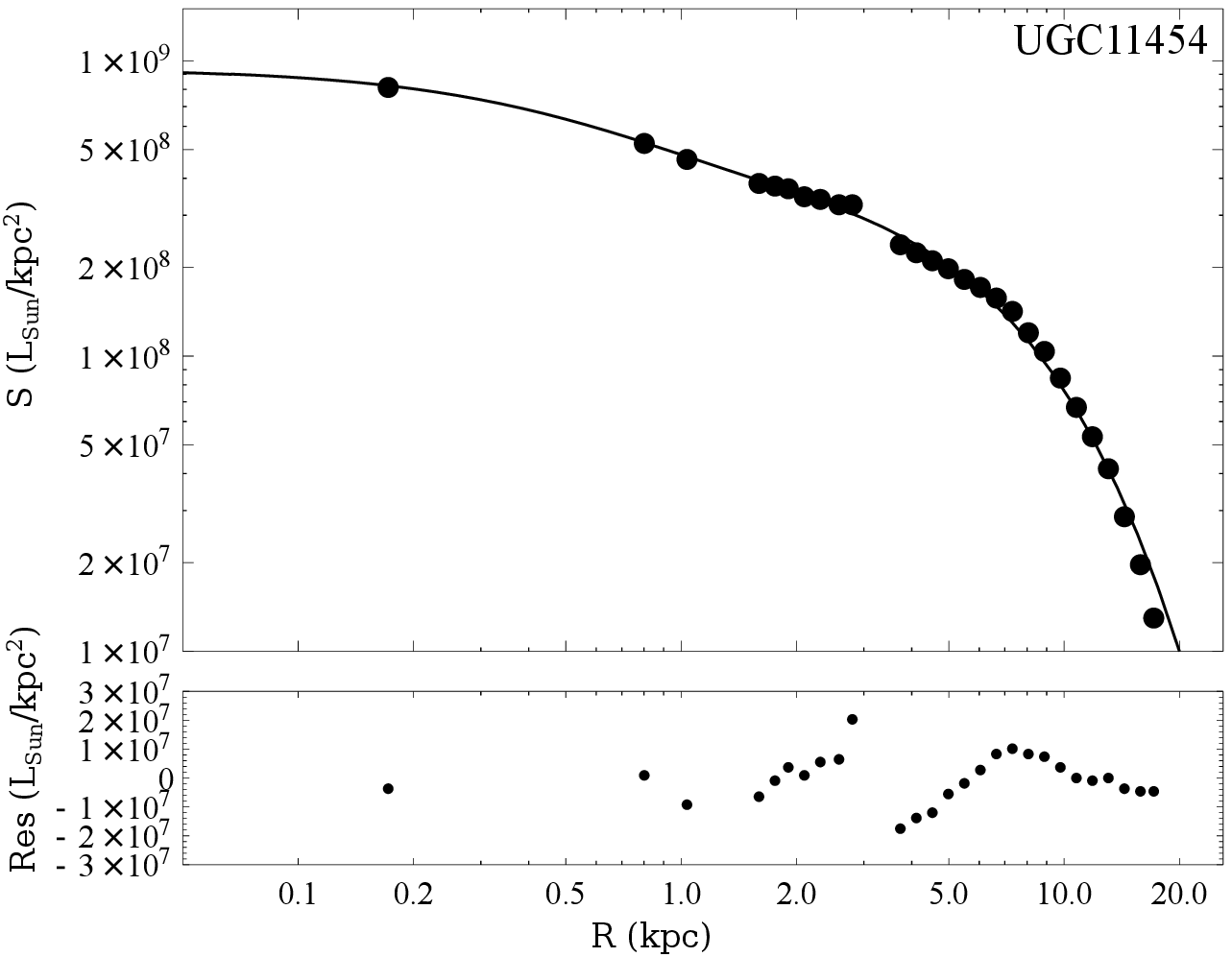}\newline
\includegraphics[width=150pt,height=130pt]{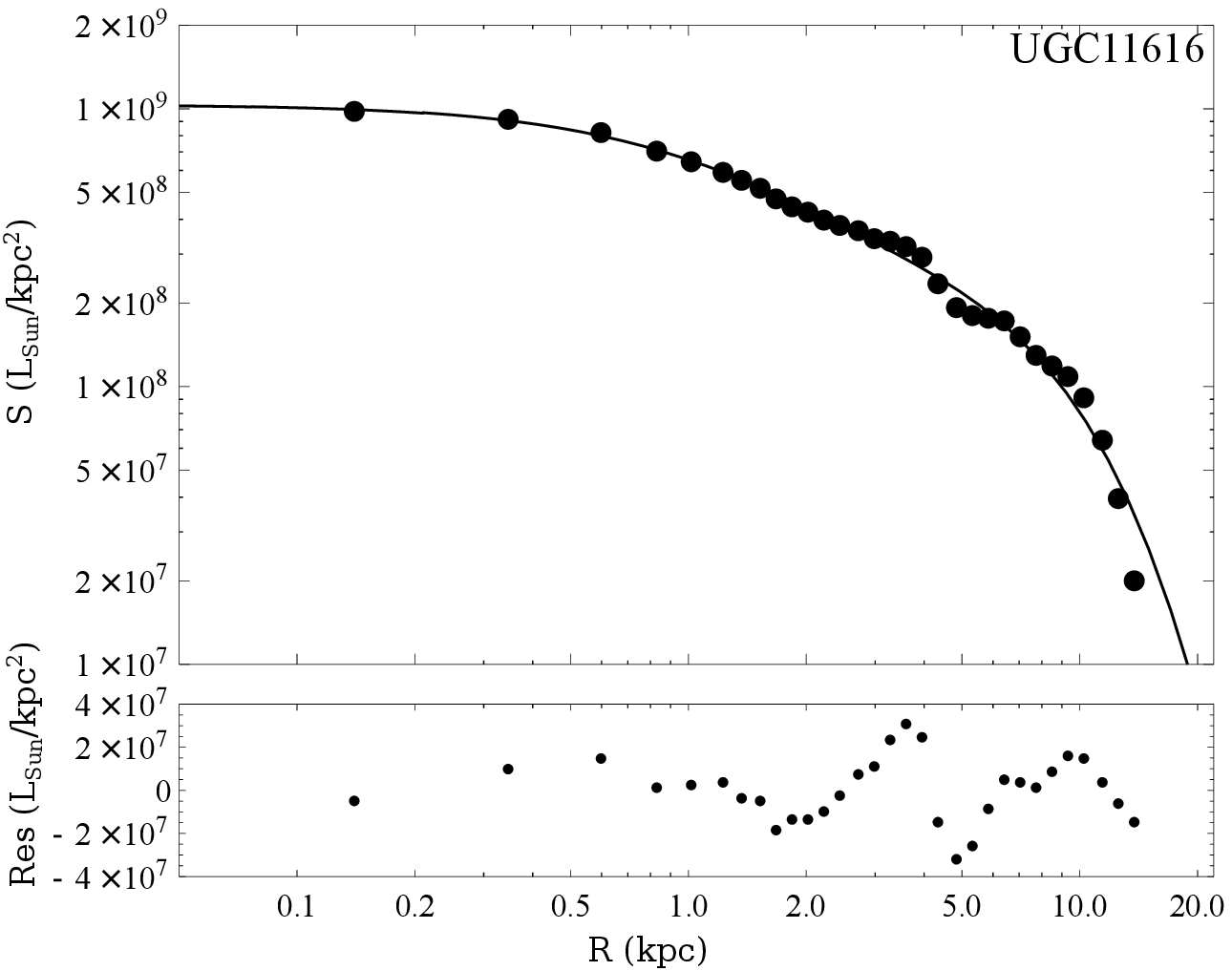}
\includegraphics[width=150pt,height=130pt]{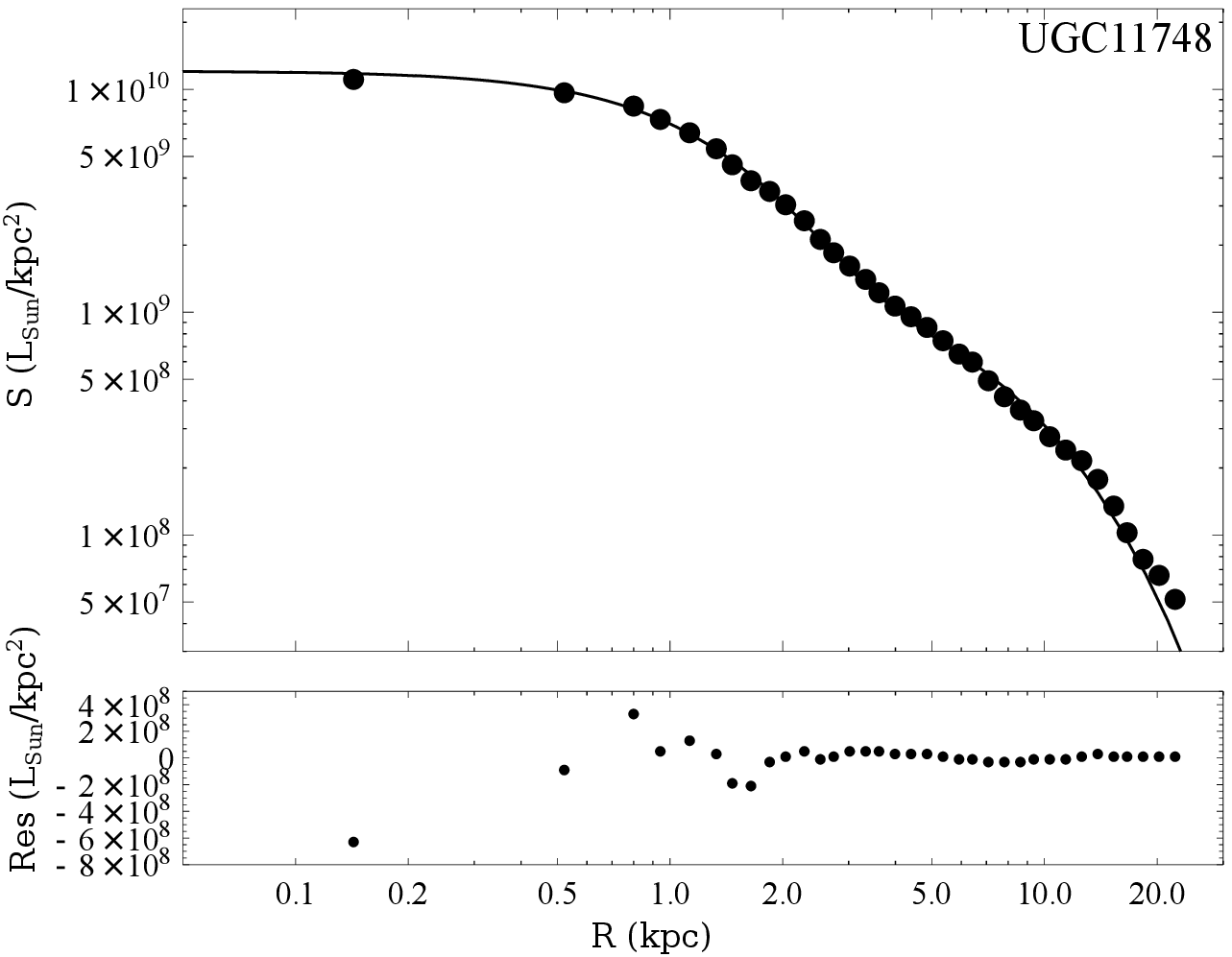}
\includegraphics[width=150pt,height=130pt]{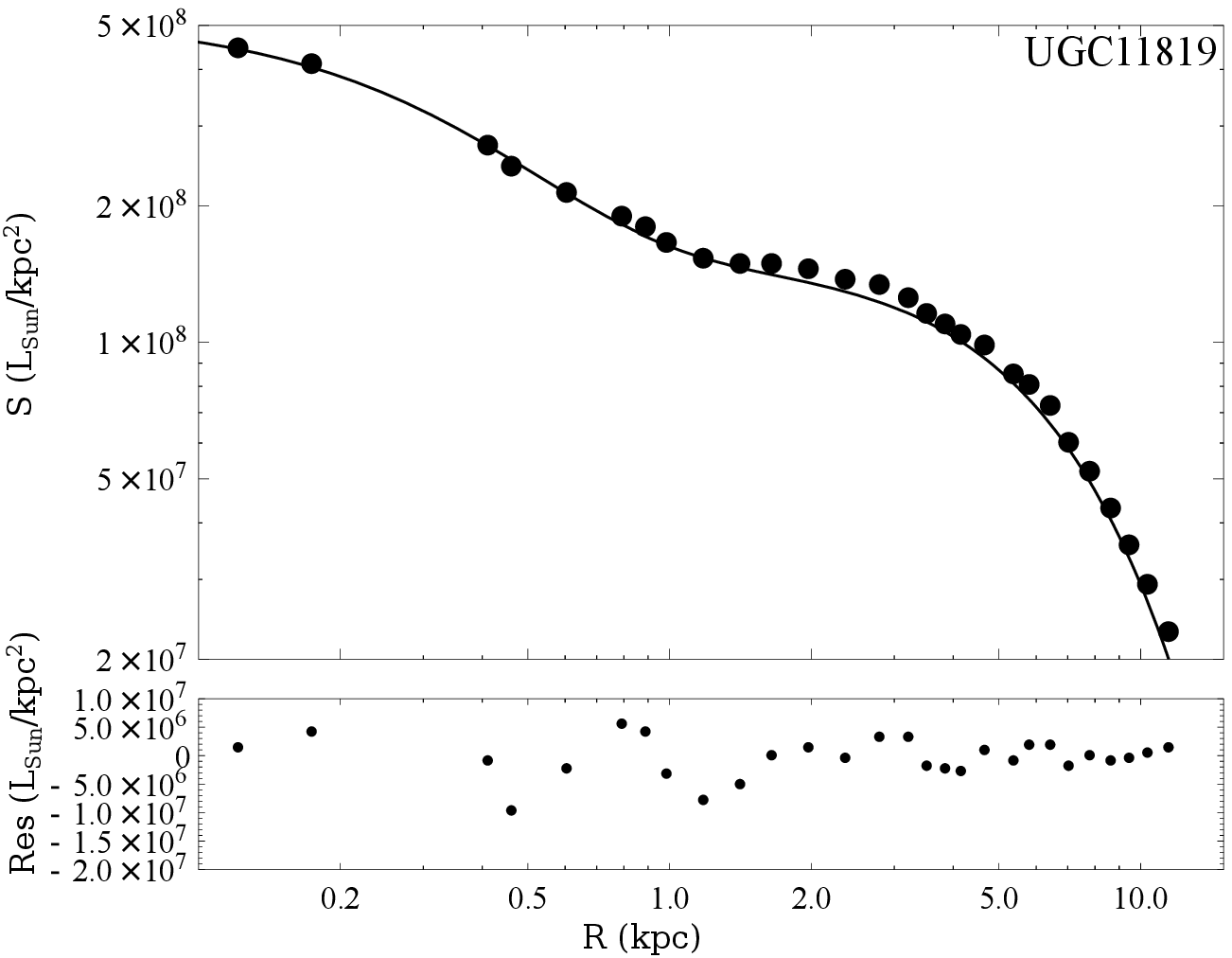}\newline
\caption{Best-fit 1D surface brightness models of the 15 LSB galaxies (upper panels) and their residuals (lower panels). The measurements are represented as larger black dots, the residuals by points, while the model predictions are shown as black continuous lines.}
\label{fig:bright_fit_plots2}
\end{figure*}

\end{document}